\documentclass[]{aa}
\usepackage{graphicx}
\usepackage{psfig}
\usepackage{txfonts}
\usepackage{longtable}

\usepackage[]{natbib}
\bibpunct{(}{)}{;}{a}{}{,} 
\def\lax    {\ifmmode{_<\atop^{\sim}}\else{${_<\atop^{\sim}}$}\fi}
\def\gax    {\ifmmode{_>\atop^{\sim}}\else{${_>\atop^{\sim}}$}\fi}
\def\kms    {\ifmmode{{\rm ~km~s}^{-1}}\else{~km~s$^{-1}$}\fi}

\newcommand{\mhtwo}{$M_{\rm H_2}$}
\newcommand{\mhi}{$M_{\rm HI}$}
\newcommand{\lfir}{$L_{\rm FIR}$}
\newcommand{\lb}{$L_{\rm B}$}
\newcommand{\lk}{$L_{\rm K}$}
\newcommand{\msun}{$M_\odot$}
\newcommand{\lsun}{$L_\odot$}

\begin{document}
	
\title{On the molecular gas content and SFR in Hickson Compact Groups: enhanced or deficient?\thanks{
  Full Tables~1, 2, 3 and 5  
 are available in electronic form at the CDS via anonymous ftp to {\tt cdsarc.u-
strasbg.fr (130.79.128.5)} or via {\tt http://cdsweb.u-strasbg.fr/cgi-bin/qcat?J
/A+A/vvv/ppp} and from {\tt http://amiga.iaa.es/}. 
}}

\titlerunning{Molecular gas content and SFR in Hickson Compact Groups}

\author{V. Martinez-Badenes
	\inst{1},
	U. Lisenfeld\inst{2},
	D. Espada\inst{1,3},
	L. Verdes-Montenegro\inst{1},
	S. Garc\'ia-Burillo\inst{4},
	S. Leon\inst{5},
	J. Sulentic\inst{1},
	\and
	M. S. Yun\inst{6}
	}

\institute{	
Instituto de Astrof\'isica de Andaluc\'ia (IAA/CSIC), Apdo. 3004, 18080, Granada, Spain.
\email{vicentm@iaa.es},
\and
Departamento de F\'isica Te\'orica y del Cosmos, Facultad de Ciencias, Universidad de Granada, Spain.
\email{ute@ugr.es},
\and
National Astronomical Observatory of Japan, 2-21-1, Osawa, Mitaka, Tokyo 181-8588, Japan
\and
Observatorio Astron\'omico Nacional (OAN)â Observatorio de Madrid, C/ Alfonso XII 3, 28014, Madrid, Spain
\and
Joint ALMA Observatory/ESO, Vitacura, Santiago, Chile.
\and
Department of Astronomy, University of Massachusetts, Amherst, MA 01003, USA.
}

 \abstract
  {}  
   {We  study the effect of the extreme environment  in Hickson Compact groups (HCGs) 
   on the molecular gas mass, \mhtwo , and the star formation rate (SFR) of galaxies as a function of 
  atomic hydrogen (HI)  content and evolutionary phase of the group.}
   {We have selected a redshift limited (D$<$100 Mpc) sample of  88 galaxies in 20 HCGs with available atomic hydrogen (HI) VLA maps, covering a wide range of
   HI deficiencies and  evolutionary phases of the groups, and containing at least one spiral galaxy. 
 We derived  the far-infrared (FIR) luminosity (${\it L_{\rm FIR}}$) from IRAS data and  used it as a tracer of the star formation rate (SFR).
   We calculated  the HI mass (${\it M_{\rm HI}}$),  ${\it L_{\rm FIR}}$  and ${\it M_{\rm H_{2}}}$ deficiencies. 
   }
   {
      The mean deficiencies of ${\it L_{\rm FIR}}$ and ${\it M_{\rm H_{2}}}$  of spiral galaxies in HCGs are close  to 0, indicating that their average SFR and 
   molecular gas content are similar   to those of  isolated galaxies. 
However, there are indications of an excess in   \mhtwo\   ($\sim$ 50\%) in spiral galaxies in HCGs which
   can be interpreted, assuming that there is no systematic difference in the CO-to-H$_2$ conversion factor, as either an enhanced molecular gas content or as a higher concentration  of the molecular component towards the center in comparison to
   galaxies in lower density environments.
In contrast, the mean  ${\it M_{\rm HI}}$ of spiral galaxies in HCGs is only 12\% of the expected value.
  The specific star formation rate (sSFR= SFR/stellar mass) tends to be lower for galaxies  with a higher \mhtwo\
   or \mhi\ deficiency.
   This trend is not seen for  the star formation efficiency (SFE=SFR/\mhtwo ), which
   is very similar to isolated galaxies.    We found tentative indications for an enhancement of \mhtwo\ in spiral galaxies in HCGs in an early evolutionary
   phase and a decrease in  later phases.

   We suggest that this might be due to
 an enhancement of the conversion from atomic to molecular gas due to on-going tidal interactions
 in an early evolutionary phase,
 followed  by 
 HI stripping  and a decrease of the  molecular gas content
 because of lack of replenishment.
 }
{The properties of \mhtwo\ and \lfir\ in galaxies in HCGs are surprisingly similar to those
of isolated galaxies, in spite of the much higher def(\mhi) of the former.
The  trends of the sSFR and def(\mhtwo) with def(\mhi) and the evolutionary state
indicate, however, that  the ongoing interaction might have some effect on the molecular gas and SF.
}

               
\authorrunning{Martinez-Badenes et al.}
\maketitle

\section{Introduction}
\label{Introduction}
Hickson Compact Groups (HCGs) \citep{1982ApJ...255..382H} are dense and relatively isolated groups of 4-8 galaxies in the nearby universe. The combination of high galaxy density \citep{1982ApJ...255..382H} and low density environment coupled with low systemic velocity dispersions ($< \sigma >$ = 200 km s$^{-1}$, \citealt{1992ApJ...399..353H}) make HCGs especially interesting systems to study how gas content and star formation activity in galaxies are influenced by the environment.

The most remarkable effect of multiple and strong interactions between galaxies in HCGs involves an atomic gas (HI) deficiency.
VLA measures of individual spiral galaxies in HCGs show them to have only  24\% of the atomic hydrogen (HI) mass, \mhi , expected from their optical luminosities and morphological types
 \citep{2001A&A...377..812V}. The inferred deficiency becomes even larger if one assumes that many of the lenticular galaxies, that are over-represented in HCGs, are stripped spirals.
 \citet{2001A&A...377..812V} proposed an evolutionary sequence for HCGs in which the HI is continuously removed from the galaxies, finally leading
 to groups where most of the HI is located outside of the galaxies. However, not only the individual galaxies in HCGs are HI deficient, but
also HCGs as a whole  \citep{2001A&A...377..812V}. This leads to the still open question of where the missing HI has gone and by which mechanism it was removed.
In order to investigate the role played by a hot intragroup medium (IGM), \citet{2008MNRAS.388.1245R} 
performed Chandra and XMM-Newton observations to study  eight
of the most HI deficient HCGs. They found bright X-ray emission in only 4 groups  suggesting that galaxy-IGM interactions are not the dominant
mechanism driving cold gas out of the galaxies.
\citet{2010ApJ...710..385B} found with new single-dish Green Bank Telescope (GBT) observation of HGCs  an important diffuse, low-column density
intragroup HI component, missed by VLA observations. Taking into account these components  reduced, but not completely eliminated, the  HI-deficiency
of the groups.

The effect of an extreme environment on molecular gas properties is controversial. An enhancement of molecular gas content  with respect to isolated
galaxies has been reported for strongly interacting
systems  \citep[][and references therein]{2004A&A...422..941C}, defined in that work as
galaxies appearing to be clearly interacting with nearby objects, presenting tidal tails or bridges, merging systems and galaxies with disturbed structures.
With respect to galaxies in clusters, no deficiency of the 
molecular gas content has been found in global studies of the Virgo cluster \citep{1986ApJ...301L..13K,2002A&A...384...33B} 
and the Coma Supercluster \citep{1991A&A...249..359C,1997A&A...327..522B} in spite of large HI deficiencies that some galaxies 
presented. 
The spatially resolved study of \citet{2009ApJ...697.1811F} found, however, that a significant number ($\sim$ 40\%) of HI-deficient spiral galaxies
was also depleted in molecular gas, {\it if} the HI was removed from within the optical disk.
Scott et al. (in prep.) found a trend for spirals in Abell 1367 in more evolved evolutionary states to be more depleted in  \mhtwo\  than those in
 less evolved evolutionary states.
Thus, there are indications that the cluster environment does affect the molecular gas content.

For galaxies in HCGs, observations of the molecular gas up to date are inconclusive.
 \citet{1998A&A...330...37L} found the ${\it M_{\rm H_{2}}}$/${\it L_{\rm B}}$ ratio of galaxies in HCGs to be enhanced with respect to a sample of field and interacting galaxies. 
On the contrary, \citet{1998ApJ...497...89V} found no evidence for an enhancement of  the molecular gas mass (${\it M_{\rm H_{2}}}$)  in HCG galaxies relative to a sample of isolated galaxies.  
Studying the relation between atomic and molecular gas for a sample of 32 spiral galaxies, \citet{2001A&A...377..812V} found
tentative evidence for a depressed molecular gas content in HI deficient galaxies in HCGs.

The level of star formation (SF) in HCGs has also been subject to considerable debate with original claims of a far-infrared (FIR) excess \citep{1989ApJ...341..679H} subsequently challenged \citep{1993ApJ...410..520S}. 
From the enhanced SF observed in galaxy pairs \citep{1991ApJ...374..407X},
an increase in  SF in HCGs might be expected as a consequence of the continuous encounters and tidal interactions which take place within such groups. Nevertheless, the Star Formation Rate (SFR) in HCGs, obtained from FIR \citep{1998ApJ...497...89V}, mid-infrared \citep{2010A&A...517A..75B} and H$\alpha$ luminosities \citep{1999ApJ...518...94I}  has been found to be similar to those of the control samples.  

There have been a few attempts to study the relation of ${\it M_{\rm H_{2}}}$ and ${\it L_{\rm FIR}}$  with  the HI properties of the HCG galaxies. 
\citet{2007ggnu.conf..349V}, based on  CO, FIR and HI single-dish data, together with VLA maps for 8 groups, found that the \mhtwo\
and \lfir\ are lower than expected for HI deficient galaxies, when compared to a well-defined sample of isolated galaxies \citep[AMIGA project, 
Analysis of the interstellar Medium of Isolated GAlaxies, http://amiga.iaa.es;][]{2005A&A...436..443V}. A possible explanation for this trend is that, as HI is needed to replenish the 
molecular clouds and molecular gas is necessary to fuel SF, a HI deficiency ultimately can lead to a decrease in the SFR. However, the result of 
 \citet{2007ggnu.conf..349V} was based on a small sample of galaxies that does not cover  the wide range of properties of HCGs and was therefore
 not statistically significant.
On the other hand,  while previous works studying ${\it M_{\rm H_{2}}}$ and 
 SFR of  galaxies in HCGs \citep{1998ApJ...497...89V,1998A&A...330...37L} were  based on larger samples, those
  did not have the HI mass of the individual galaxies to compare  with 
 ${\it M_{\rm H_{2}}}$ and ${\it L_{\rm FIR}}$. 
 Thus, up to date, no study of the relation between \mhtwo , \mhi\ and SFR for a statistically significant sample has been carried out.

To shed light on the relations between ${\it M_{\rm H_{2}}}$ and SFR with \mhi\  properties
 of the HCGs, we present here
a systematic study  for galaxies in a sample of 20 HCGs for which we have HI measurements for the entire groups, as well as 
for a large fraction of the individual galaxies. This enables us to
take into account ${\it M_{\rm HI}}$ of the galaxies as an additional parameter, as well as 
the  evolutionary phase of the group according to \citet{2001A&A...377..812V} (see Sect. 2). We compare the properties of galaxies in HCGs with those of 
isolated galaxies in  the AMIGA sample   \citep{2005A&A...436..443V}.
Our goal is to determine whether deviations in the HI content with respect to  isolated galaxies 
translate into anomalies in the ${\it M_{\rm H_{2}}}$ and the SFR.

The outline of this paper is as follows. We present the sample in Sec. \ref{sec:sample}. CO(1-0) and CO(2-1) data coming either from our observations or from the literature, 
together with reprocessed IRAS FIR data, are presented in Sec. \ref{sec:data}. In Sec. \ref{sec:results} we compare the 
${\it M_{\rm H_{2}}}$, ${\it L_{\rm FIR}}$ (as a tracer of the SFR) and ${\it M_{\rm HI}}$ of the galaxies, studying their deficiencies and their relation with the HI content and evolutionary 
phase of the group. A 
discussion of a possible evolutionary sequence for the molecular gas content in the HCGs is presented in Sec. \ref{sec:evol_mh2}. Finally, the conclusions
 of our work are summarized in Sec. \ref{Conclusions}.

\section{The Samples}\label{sec:sample}

\subsection{Galaxies in HCGs}
Our sample was selected from the revision of the original \citet{1982ApJ...255..382H} catalogue performed by \citet{1992ApJ...399..353H}. 
From the groups included in that work, we study 86 galaxies belonging  to 20 different HCGs: 7, 10, 15, 16, 23, 25, 30, 31, 37, 40, 44, 58, 67, 68, 79, 88, 92, 93, 97 and 100. The groups, which cover all  evolutionary stages and a wide range of HI deficiencies, satisfy the following criteria:
\begin{itemize}
\item Having at least four members, so triplets are excluded, according to the original \citet{1982ApJ...255..382H} criterion. We also exclude false groups, where a single knotty irregular galaxy has been confused with separated galaxies \citep{2001A&A...377..812V}.
\item Containing at least one spiral galaxy, since we are mainly interested in studying the relation between the SF process and ${\it M_{\rm H_{2}}}$, which are most clearly linked for spiral galaxies.
\item Being at a distance $D \leq$100 Mpc (assuming H$_{0}$=75 km s$^{-1}$ Mpc$^{-1}$), so that 
observations of  groups have a better sensitivity limit and minimize possible source confusion within the telescope beam. At 100 Mpc, the 30m beam would have a size of 10.7 kpc and the VLA beam (considering a size of 50$^{\prime\prime}\times$50$^{\prime\prime}$)  24.2 kpc.
\end{itemize}

The HCGs in our sample cover the full range of HI contents. Their deviation from normalcy
is  measured with respect to that of isolated galaxies, as given by \citet{1984AJ.....89..758H}. 
This deviation is 
usually referred to as deficiency and is defined as the decimal logarithm of the ratio between the sum of the
expected HI masses for all the galaxies in the group, based on their optical luminosity and morphology, 
and the HI mass of the entire group as derived from the single dish observations in \citet{2010ApJ...710..385B} \citep[see][and also Sec. \ref{subsec:Deficiencies}]{2001A&A...377..812V}. As a function of their total HI deficiency, the HCGs in our sample can be classified as follows:

 \begin{itemize}
\item HCGs with a normal HI content (at least 2/3 of its expected value): HCG 23, 25, 68 and 79.
\item HCGs with a slight HI deficiency (between 2/3 and 1/3 of the expected value): HCG 7, 10, 15, 16, 31, 37, 40, 58, 88, 92, 97 and 100.
\item HCGs with a large HI deficiency (under 1/3 of the expected value): HCG 30, 44, 67 and 93.

\end{itemize}

\citet{2001A&A...377..812V} proposed an evolutionary sequence model where the HI is continuously stripped from the galaxies. According to this model, HCGs can be classified into 3 phases as follows: in Phase 1 the HI is mainly found in the disks of galaxies. In Phase 2, 30\% to 60\% of the HI has been removed from the disks by tidal interaction. Finally, in Phase 3, almost all the HI is found out of the galactic disks, either forming tidal tails of stripped gas (Phase 3a) or, in a few cases, in a large HI cloud with a single velocity gradient in which the galaxies are embedded (Phase 3b).
 
According to the evolutionary phases defined in \citet{2001A&A...377..812V}, the HCGs in our sample were classified by \citet{2010ApJ...710..385B} as: 	
  
\begin{itemize}
\item Phase 1: HCG 7, 23, 67, 79 and 88.
\item Phase 2: HCG 10, 16, 25, 31, 40\footnote{While HCG 40 was classified in \citet{2001A&A...377..812V}  as Phase 3,  new VLA observations (Yun et al. in preparation) showed 
that a significant amount of HI was missed due to a narrow spectral window, and based on these data has been reclassified as Phase 2.}
, 58 and 100. 
\item Phase 3: HCG 15, 30, 37, 44, 68, 92, 93 and 97. 
\end{itemize}

The evolutionary state is an indicator of the evolution of the cold ISM of the group but it does not necessarily give the age of the group. E.g. HCG79 consists on 3 early-type galaxies and one intruding spiral galaxy.
Stellar halo data indicates that it is an old  group
\citep{2008AJ....135..130D}. However, since the main part of the HI is located within the disk of 
the intruder galaxy,  it is classified  in evolutionary phase 1. 

We revised the velocities of the individual galaxies in the HCGs of our sample. Two galaxies not considered in \citet{1992ApJ...399..353H} have been added: HCG100d, which had no velocity data in that work, and HCG31g, added to the catalogue of HCGs by \citet{1990ApJ...365...86R}. 

The basic properties of the galaxies in our sample are detailed in Table \ref{Sample_lb_1}. The columns are: 

\begin{enumerate}
\item Galaxy: galaxy  designation, following the notation of \citet{1982ApJ...255..382H}.
\item  $V$: heliocentric radial velocity in km s$^{-1}$(weighted average of optical measurements taken from the LEDA\footnote{http://leda.univ-lyon1.fr/intro.html} database) converted from the optical to the radio definition for comparison with the CO spectra.

\item $\rm \sigma_{V}$: velocity dispersion of the galaxies in the group.
\item $D$: distance to the corresponding HCG in Mpc, derived from the mean heliocentric velocity of the group as D = V/H$_{0}$, assuming a value of H$_{0}$ = 75 km s$^{-1}$ Mpc$^{-1}$. The mean velocity of the group is calculated averaging the velocity of the individual galaxies (column 2).
\item $T$: morphological type taken from LEDA, following the RC3 classification \citep{1991trcb.book.....D}.
\item $D_{\rm 25}$: optical major diameter in arcmin at the 25 mag \rm{arcsec$^{- 2}$} isophot taken from LEDA.
\item $B_{\rm c}^{\rm T}$: apparent blue magnitude taken from LEDA, corrected for Galactic dust extinction, internal extinction and K-correction.
\item log(\lb): decimal logarithm of the blue luminosity, derived from $B_{\rm c}^{\rm T}$ as:

\begin{equation}
{\log}\Big(\frac{L_{\rm B}}{L_{\odot}}\Big)=2\rm{log}D-0.4B_{\rm c}^{\rm T}+11.95.
\end{equation}

\noindent This definition provides an estimate of the blue luminosity ($\nu L_{\nu}$) at 4400 \AA{}.

\item 
log(\lk): decimal logarithm of the luminosity in the K-band in units of the solar luminosity in the 
$K_{\rm S}$-band ($L_{K,\odot} = 5.0735 \times 10^{32}$ erg s$^{-1}$), 
calculated from the extrapolated magnitude in the $K_{\rm S}$ (2.17 $\mu$m) band from the 2MASS Extended Source Catalogue \citep{2000AJ....119.2498J}. 
We calculated the $K_{\rm S}$ luminosity, ${\it L_{\rm K}}$, from the total (extrapolated) $K_{\rm S}$ flux, $f_{\rm K}$, as ${\it L_{\rm K}}$ = $\nu f_{\rm K}(\nu)$ (where $\nu$ is the frequency of the K-band, 1.38$\times$10$^{14}$ Hz) .

\item  log($M_{\rm HI}$): logarithm of the mass of the atomic hydrogen, in solar masses, for 66 of the galaxies in our sample observed with the VLA, using different combinations of the C and D configurations with beam sizes ranging from 16$^{\prime\prime}\times$14$^{\prime\prime}$ to 72$^{\prime\prime}\times$59$^{\prime\prime}$ \citep[][and Verdes-Montenegro, private communication]{2001A&A...377..812V}.
\end{enumerate}

\begin{table*}
\caption{Basic parameters of the galaxies in the HCG sample}
\label{Sample_lb_1}
\begin{tabular}{lccccccccc}
\hline\hline
Galaxy&V&$\sigma_{V}$&D& T(RC3)&D$_{25}$&B$_{c}^{T}$&log$({\it L_{\rm B}})$&log$({\it L_{\rm K})}$&log(\mhi)\\
  &(km $s^{-1}$)&(km $s^{-1}$)&(Mpc)& &(arcmin)&(mag)&($L_{\odot}$)&($L_{\odot}$)&($M_{\odot}$)\\
  (1) & (2) & (3) &   (4) & (5) & (6) &   (7) & (8) & (9) & (10)  \\ 
\hline
      7a   &    4141   &      117    &   53.4   &    1.0   &    2.06   &   12.96   &   10.22   &   11.09     &  9.12  \\
      7b   &    4175   &      117    &   53.4   &   -1.9   &    1.27   &   14.29   &    9.69   &   10.78     & $<$7.83  \\
      7c   &    4347   &      117    &   53.4   &    5.0   &    1.71   &   13.36   &   10.06   &   10.80     &  9.56  \\
      7d   &    4083   &      117    &   53.4   &   -1.4   &    0.94   &   14.04   &    9.79   &   10.11     &  9.00  \\
     10a   &    5104   &      269    &   65.6   &    3.1   &    2.92   &   12.53   &   10.57   &   11.27     &   ...  \\
.... & .... &.... &.... &.... &.... &.... &.... &.... &.... \\
\hline
\end{tabular}  

{\bf Notes.} The full table is available in electronic form at the CDS and from http://amiga.iaa.es.


\end{table*}

\subsection{Reference sample: Isolated galaxies}
\label{ref_sample}

We chose  the AMIGA sample of isolated galaxies \citep{2005A&A...436..443V},
 which is based on the CIG catalogue 
\citep{1973SoSAO...8....3K}, as a reference for the FIR and molecular gas properties.
The FIR properties of an optically complete subsample of this catalogue have been studied in \cite{2007A&A...462..507L}. The data 
 that we  are using for this subsample are  slightly different from \citet {2007A&A...462..507L}
 because we have taken into account a recent update of some basic properties
which affects the blue magnitude, distance, morphological type and isolation degree 
\citep[detailed information are provided in][]{fernandez2011}.
 We also use the  CO data  of a velocity restricted subsample (1500$<$V$<$5000 km s$^{-1}$) of 173 AMIGA galaxies \citep{2011lisenfeld}  to 
 compare to our galaxies.
 For the analysis of the HI  properties  we use  the work of \citet{1984AJ.....89..758H}, presenting the observations and analysis of CIG galaxies, as a reference. 
 
\begin{figure}[h]
\centering{
\includegraphics[scale=.5, angle=-90]{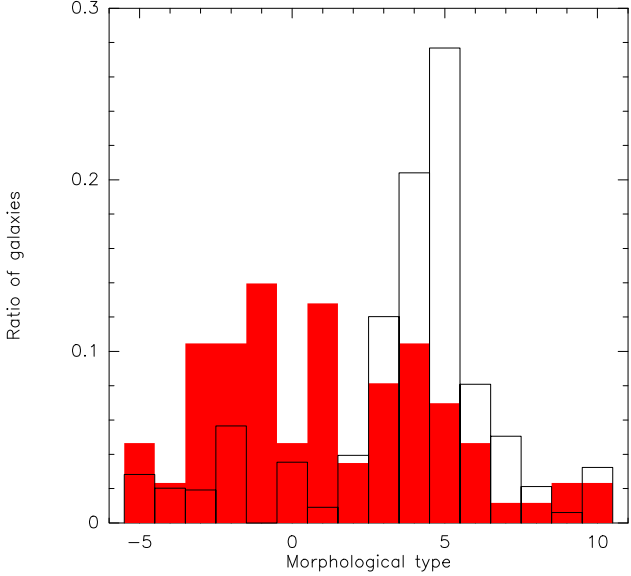}
}
\caption{Morphological types, T(RC3), for the AMIGA isolated (black line) and HCGs (red filled bars) galaxies.}
\label{histo_hubble_over}       
\end{figure} 

 There are two intrinsic differences between the AMIGA and the HCG samples that must be taken into account when performing a comparison: 
(i) the HCG sample has a larger rate of  early-type galaxies ($\sim$45\%), whereas 87\% of the AMIGA galaxies are spirals (Fig. \ref{histo_hubble_over}) and
%
(ii)  the sample of  AMIGA galaxies with ${\it M_{\rm H_{2}}}$ data  is 
restricted to a velocity range of 1500$<V<$5000 km s$^{-1}$, while the range of the HCGs extends to larger velocities.
Thus, the isolated galaxies are, on average, at a lower distance
(47 Mpc average distance versus 68 Mpc for the HCG sample) which can explain their lower average 
luminosities (see Table \ref{Medias_2muestras}).
However, the values of the deficiencies, ratios or correlations, that we are going to discuss in the following
 are not expected to be affected by the difference in distance.

\section{The data}\label{sec:data}
\subsection{CO data}\label{subs:COdata}
We obtained CO data, either by our observations or from the literature,  for 86 galaxies in the selected 20 HCGs. 
%
CO data are missing for only 2 galaxies in these 20 groups, HCG 67d and HCG 92f. The CO(1-0) line was detected for 45 galaxies.

\subsubsection{IRAM CO(1-0) and CO(2-1) observations and data reduction}\label{subsubs:COobs}
We observed 47 galaxies belonging to 14 different HCGs. The observations of the CO rotational transition lines J=1$\rightarrow$0 and J=2$\rightarrow$1 (at 115.271 and 230.538 GHz, respectively) were carried out with the IRAM 30m radio telescope at Pico Veleta\footnote{IRAM is supported by CNRS/INSU (France), the MPG (Germany) and the IGN (Spain).} during June, October and December 2006. We performed single-pointing observations using the wobbler switch mode, with a switch frequency of 0.5 Hz and a throw of 200$^{\prime\prime}$. 
 We checked 
for all the objects that the off-position did not coincide with a neighbor galaxy.

The dual polarization receivers A100 and B100 were used to observe in parallel the CO(1-0) and CO(2-1) lines. The median system temperature was 231 K for the CO(1-0) observations, with $\sim$80\% of the galaxies observed with system temperatures between 150 and 350 K. In the case of CO(2-1), the median system temperature was 400 K, with a temperature range between 230 and 800 K for 85\% of the galaxies. For CO(1-0) the 1 MHz filterbank was used, and for CO(2-1) the 4 MHz filterbank. The corresponding velocity resolutions were 2.6 km s$^{-1}$ and 5.3 km s$^{-1}$ for CO(1-0) and CO(2-1), respectively. The total bandwidth was 1 GHz. The Half Power Beam Width (HPBW) is 22$^{\prime\prime}$ and 11$^{\prime\prime}$ for 115 and 230 GHz, respectively. All CO spectra and intensities are presented on the main beam temperature scale ($T_{\rm mb}$) which is defined as $T_{\rm mb} = (F_{\rm eff}/B_{\rm eff})\times T_{\rm A}^*$. The IRAM forward efficiency, $F_{\rm eff}$, was 0.95  and 0.91 at 115 and 230 GHz,
and the beam efficiency, $B_{\rm eff}$, was 0.75 and 0.54, respectively.

The data reduction and analysis was performed using CLASS, while further analysis was performed using GREG, both part of the GILDAS\footnote{http://www.iram.fr/IRAMFR/GILDAS} package developed by IRAM. First we visually inspected the spectra and discarded bad scans. Then, spikes were removed and
a constant or linear baseline was subtracted from  each spectrum.   The scans were then averaged to achieve a single spectrum for each galaxy and transition. These spectra were smoothed to a velocity resolution of 21 to 27 km s$^{-1}$, depending on the rms. 
A total of 24 galaxies were detected in CO(1-0) (2 of them marginal), 22 in CO(2-1) (4 of them marginal) and 18 in both transitions. The spectra are shown in Appendix A, 
Fig. ~\ref{spec-co10} displays the spectra detected in CO(1-0) and  Fig. ~\ref{spec-co21} those  detected in CO(2-1).

For each spectrum, we integrated the intensity along the velocity interval where emission was detected. In the case of nondetections we set an upper limit as:

\begin{equation}
I_{\rm CO} < 3 \times rms \times \sqrt{\delta \rm{V} \ \Delta V},
\end{equation}

\noindent where $\delta \rm{V}$ is the channel width and $\Delta$V is the total line width. We used a value of $\Delta$V = 300 km $s^{-1}$ when there was no detection in CO(1-0) or  CO(2-1).
 In those cases where the source was detected in only one transition, this line width was used to calculate the upper limit in the other transition.

The results of our CO(1-0) and CO(2-1) observations are displayed in Table~\ref{Sample_CO_1}.
The columns are: 

\begin{enumerate}
\item Galaxy: galaxy designation.
\item  $I_{\rm CO(1-0)}$: velocity integrated intensity of the CO(1-0) emission in K km s$^{-1}$.
\item rms : root-mean-square noise of the CO(1-0) spectrum (if available) in mK.
\item Ref.: reference of the CO(1-0) data, detailing whether data come from our observations or from the literature (see \ref{COliterature}).
\item Beam: HPBW of the telescope in arc second.
\item $\rm \Delta \rm{V}_{CO(1-0)}$: line width of the CO(1-0) emission (if detected) in km s$^{-1}$.
\item $I_{\rm CO(2-1)}$: velocity integrated intensity of the CO(2-1) emission (if observed) in K km s$^{-1}$.
\item rms: rms of the CO(2-1) spectrum (if observed) in mK.
\item $\rm \Delta \rm{V}_{CO(2-1)}$: line width of the CO(2-1) emission (if detected) in km s$^{-1}$.
\item log($M_{\rm H_{2}~obs}$): \rm logarithm of the H$_{2}$ mass (in solar masses) calculated from the observed  central $I_{CO}$ (see Sec.\ref{subs:MolGas}).
\item log(\mhtwo): \rm logarithm of the H$_{2}$ mass (in solar masses) extrapolated to the emission from the total disk (see Sec.~\ref{subs:MolGas}).
\end{enumerate}

\begin{table*}
\caption{Observed and derived molecular gas properties}
\label{Sample_CO_1}
\begin{tabular}{lcccccccccc}
\hline\hline
Galaxy&$\it I_{\rm CO(1-0)}$&rms&Ref.&HPBW&$\Delta$V&$I_{CO(2-1)}$&rms&$\Delta$V&log(${\it M_{\rm H_{2}{\rm obs}}}$)&log(${\it M_{\rm H_{2}}}$)\\
 &(K km $s^{-1}$)&(mK)&&(arcsec) &(km $s^{-1}$)&(K km $s^{-1})$&(mK)&(km s$^{-1}$)&($M_{\odot}$)&($M_{\odot}$)\\
  (1) & (2) & (3) &   (4) & (5) & (6) &   (7) & (8) & (9) & (10) & (11) \\ 
\hline
      7a   & 7.20   &   &    3   &   43   &     500   &   &   &   &    9.51   &    9.71  \\
      7b   &   $<$ 0.70   &   &    3   &   55   & ...     &   &   &   &$<$8.71   &    $<$8.80  \\
      7c   & 1.40   &   &    3   &   55   &     183   &   &   &   &    9.01   &    9.17  \\
      7d   &   $<$ 0.60   &   &    3   &   43   & ...     &   &   &   &$<$8.43   &    $<$8.52  \\
     10a   & 2.72$\pm$0.49   &   &    2   &   22   &     339   &   &   &   &    8.79   &    9.51  \\
.... & .... &.... &.... &.... &.... &.... &.... &.... &....&.... \\
\hline
\end{tabular}  

\footnotesize{
\noindent{$^{(1)}$CO reference: 1: Our observations. 2: \citet{1998A&A...330...37L}. 3: \citet{1998ApJ...497...89V}.}\\
{\bf Notes.} The full table is available in electronic form at the CDS and from http://amiga.iaa.es.
}
\end{table*}

\subsubsection{CO(1-0) data from the literature}\label{COliterature}
We have searched in the literature for  available CO(1-0) data for the 20 HCGs of our sample and have compiled data for the velocity integrated CO(1-0) intensities 
and line widths (also listed in Tab.~\ref{Sample_CO_1}) from the following sources:  
\begin{itemize}

\item \citet{1998ApJ...497...89V}: 24 galaxies from 9 different HCGs. 20 of these galaxies were observed with the NRAO 12 m telescope at Kitt Peak with a beam size of 55$^{\prime\prime}$. The data from the other 4 galaxies are from \citet{1996A&A...314..738B}, observed with the SEST 15m telescope, with a 43$^{\prime\prime}$ beam.
Two of these galaxies (68d and 88c) were also observed by us, but we chose the \citet{1998ApJ...497...89V} data because of their better quality.

\item \citet{1998A&A...330...37L}: 17 galaxies corresponding to 10 different HCGs, observed with the IRAM 30m telescope with a similar setting as in our observations (see Sec. \ref{subsubs:COobs}). 

\end{itemize}

There are 16 galaxies that were observed both by us and by \citet{1998ApJ...497...89V} or  \citet{1998A&A...330...37L}.
Furthermore,  14  galaxies were observed by both \citet{1998ApJ...497...89V} and \citet{1998A&A...330...37L}. In order to choose 
between the different existing spectra (either from our observations or from the literature), we first checked that they were consistent and then applied the following criteria: 
if available, we chose the spectrum with detected emission. If more than one detected spectrum existed, we chose the one with the lower rms or --in case of comparable rms- the spectrum observed with a larger beam, in order to probe a larger fraction of the disk. Except for the two galaxies mentioned above, 
we always selected our data due to their better quality in case of duplication.
In total, we have CO(1-0) spectra for 86 galaxies (45 from our own observations and 41 from the literature) for our statistical analysis.

\subsubsection{Molecular gas mass}\label{subs:MolGas}

We calculate the molecular gas mass, ${\it M_{\rm H_{2}}}$ using the following equation:
\begin{equation}
M_{H_{2}}=75 \times D^{2}I_{CO(1-0)}\Omega
\end{equation}

\noindent where $\Omega$ is the area covered by the observations in arcsec$^{2}$ (i.e. $\Omega$ = 1.13 $\theta^{2}$ for a single pointing with a gaussian beam where $\theta$ is the HPBW). This equation assumes a CO-to-H$_2$ conversion factor 
 X=$N_{\rm H_{2}}/I_{\rm CO}$ = $2\times 10^{20}\rm cm^{-2}$ (K km s$^{-1})^{-1}$ \citep[e.g.] []{1986ApJ...309..326D}. 
No correction factor for the fraction of helium and other heavy metals is included. 
 The molecular gas masses of the AMIGA galaxies are calculated in the same way.

In both the observations that we carried out and the data from the literature, a single position at the center of the galaxy was observed. Because of this and the 
different beams used by us and in the literature we need to correct for possible emission outside the beam. To extrapolate  the observed CO intensities to
the total value within the disk we 
need to know the distribution and extension of the CO emission. Different authors \citep{2001PASJ...53..757N, 2001ApJ...561..218R, 2008AJ....136.2782L} 
found that the velocity integrated CO intensity in spiral galaxies follows an exponential distribution as a function of radius with a scale length $r_{\rm e}$:

\begin{equation}
I_{\rm CO}(r) = I_{0}\propto \exp(r/r_{\rm e})
\label{Ico_r}
\end{equation}

We adopt a scale length of $r_{\rm e}$ = 0.2$\times$$r_{\rm 25}$, where $r_{\rm 25}$ is the major optical 25 mag arcsec$^{-2}$ isophotal radius, 
following \citet{2011lisenfeld}, who derive this scale length from studies of the mentioned authors and from their own CO data. 
We have used this distribution to calculate the expected CO emission from the entire disk, taking into account the galaxy inclination
 \citep[see][for more details]{2011lisenfeld}. This approach assumes that the distribution of the molecular gas in galaxies in HCGs is the
 same as in field spiral galaxies. The implications of this approach will be discussed. 

\begin{figure}
\centering
\includegraphics[width=6.5cm, angle=-90]{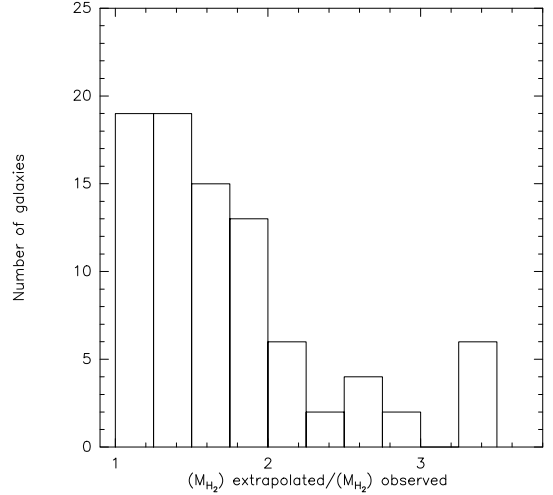}
\caption{ Distribution of the aperture correction factor for ${\it M_{\rm H_{2}}}$.}
\label{Histo_correccion}       
\end{figure}

The resulting aperture correction factor  for  \mhtwo\ (defined as the ratio between \mhtwo\ observed in the central pointing and  \mhtwo\
extrapolated to the entire disk) is shown in Figure \ref{Histo_correccion}. The ratio between the extrapolated and central 
intensities is below 2 for most galaxies (66 out of 86, or 77\%), with an average value of 1.78. To check the consistency of the extrapolation, 
we have also performed the analysis presented in this paper for a sample restricted to galaxies  with a small (less than a factor 1.6) aperture correction ($n=45$), 
finding no significant differences with respect to the full sample. Thus,  we conclude that the aperture correction does not introduce any bias in the results.

\begin{figure}
\includegraphics[width=7.cm, angle=-90]{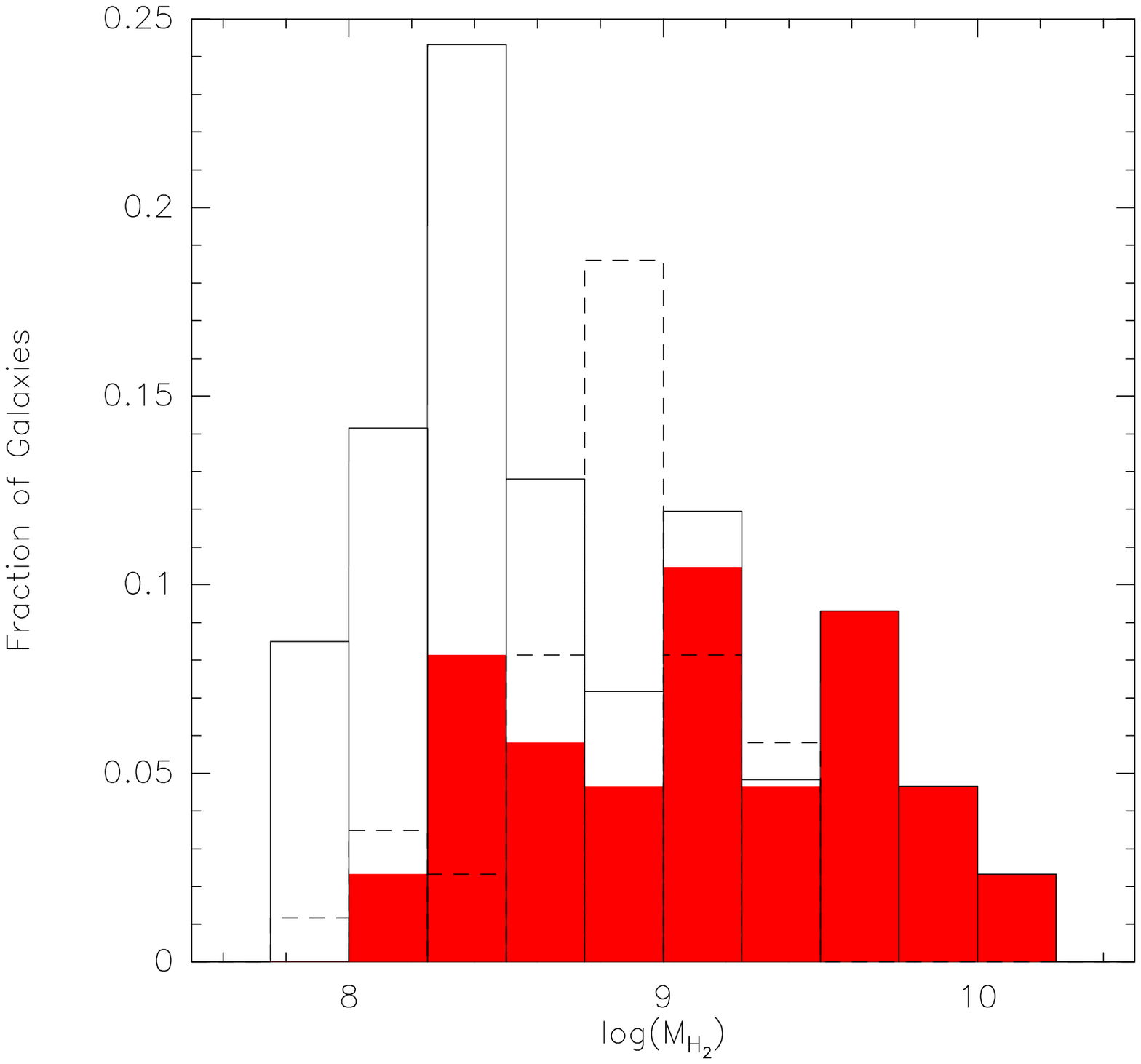}
\quad
\includegraphics[width=7.cm, angle=-90]{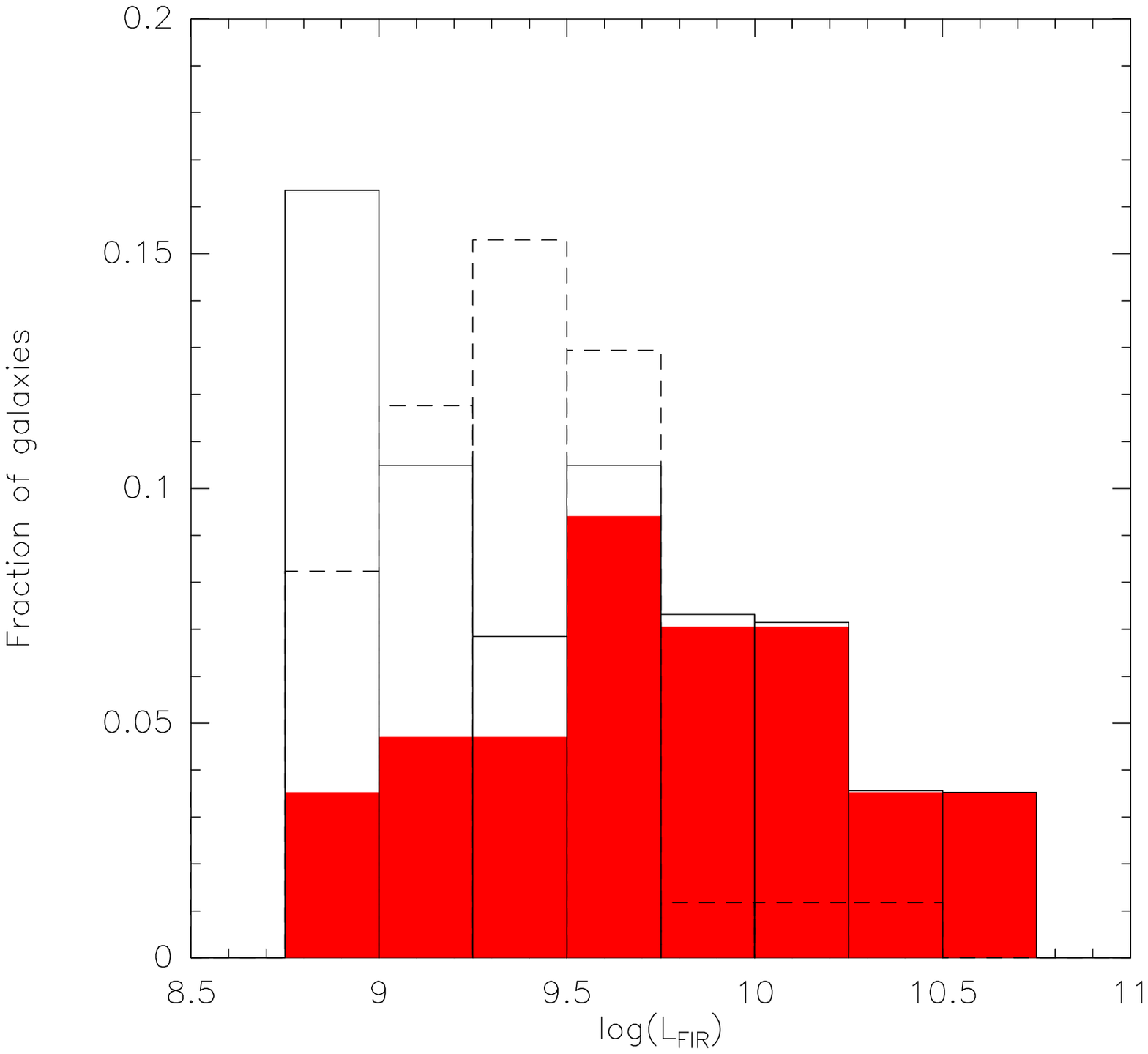}
\caption{${\it M_{\rm H_{2}}}$ and ${\it L_{\rm FIR}}$ distribution  of the HCG galaxies. The red filled bins show the distribution of the detected galaxies, the dashed line gives the distribution of upper limits and the full line shows the distribution calculated  with  ASURV.}  
\label{histo_h2_fir_hcgs}
\end{figure}

The values for the molecular gas mass in the central pointing and the extrapolated molecular gas mass are listed in Table \ref{Sample_CO_1}. 
Here, and in the following, we always use the extrapolated molecular gas mass and denote it as 
\mhtwo\ for simplicity. The \mhtwo\ distribution is shown in Fig. \ref{histo_h2_fir_hcgs}. The average value for   spiral galaxies
($T\ge 1$)  is listed in Table \ref{Medias_2muestras}. 
The distribution and average values of \mhtwo , as well as the statistical distributions and average values throughout this work, have been calculated using the Kaplan-Meier estimator implemented in ASURV\footnote{Astronomical Survival Analysis (ASURV) Rev. 1.1 \citep{1992BAAS...24..839L} is a generalised statistical package that implements the methods presented by \cite{1985ApJ...293..192F}}, to take into account the upper limits in the data.

\subsection{Far-Infrared Data}\label{subs:FIRdata}

FIR fluxes were obtained from ADDSCAN/SCANPI, a utility provided by the Infrared Processing and Analysis Center (IPAC)\footnote{http://scanpi.ipac.caltech.edu:9000/}. This is a one-dimensional tool that coadds calibrated survey data of the Infrared Astronomical Satellite (IRAS). It makes use of all scans that passed over a specific position and produces a scan profile along the average scan direction. It is 3-5 times more sensitive than the IRAS Point Source Catalog (PSC) since it combines all survey data and it is therefore more suitable for detection of the total flux from slightly extended objects. 

We have compiled from \citet{1998ApJ...497...89V}  the FIR data (also derived using ADDSCAN/SCANPI) for 63 galaxies in our sample. In the case of the remaining 23 galaxies, we derived FIR fluxes directly using ADDSCAN/SCANPI. To choose the best flux estimator we have followed the guidelines given in the IPAC website\footnote{http://irsa.ipac.caltech.edu/IRASdocs/scanpi\_interp.html}, which are also detailed in \citet{2007A&A...462..507L}. 
%
%
As a consistency check,  we also applied this procedure to 14 galaxies in the list of \citet{1998ApJ...497...89V}. We found no significant differences, 
with an average difference of 15\% between our reprocessed fluxes and those in \citet{1998ApJ...497...89V}. 

From  the fluxes at 60 and 100 $\mu$m the FIR luminosity, ${\it L_{\rm FIR}}$,
 is computed as:
\begin{equation}
 \log(L_{\rm FIR}/L_{\odot}) = \log(FIR)+2\log(D)+19.495
\end{equation}

\noindent where FIR is defined as \citep{1988ApJS...68..151H}:
\begin{equation}
FIR=1.26\times10^{-14} (2.58\,F_{\rm 60} + F_{\rm 100}){\rm W m^{-2}}.
 \label{FIR_eq}
\end{equation}

The computed ${\it L_{\rm FIR}}$, together with the 60 and 100 $\mu$m fluxes compiled from 
ADDSCAN/SCANPI, are detailed in Table \ref{Sample_FIR}. 
The distribution of ${\it L_{\rm FIR}}$ is shown in Fig. \ref{histo_h2_fir_hcgs}. 
The average value of ${\it L_{\rm FIR}}$  for  spiral galaxies is given in Table \ref{Medias_2muestras}.

For galaxies HCG 31a and HCG 31c the FIR fluxes could not be separated. Therefore, we use the sum of both. 
When comparing $L_{\rm FIR}$ to other magnitudes (${\it L_{\rm B}}$, ${\it M_{\rm H_{2}}}$ or $M_{HI}$), we also use the sum of both galaxies.

In order to check the accuracy of the low resolution IRAS data, we compared them  to 
 24$\mu$m data from Spitzer
for  the 12 groups for which Spitzer data is available. We compared the SFR derived from  $L_{\rm FIR}$ (calculated from eq. 10) to that
derived from the Spitzer 24$\mu$m luminosity, $L_{\rm 24\mu m}$,  \citep[from][]{2010A&A...517A..75B,2011bitsakis}  using the equation 
$SFR$(\msun yr$^{-1}) = 8.10 \times 10^{-37} (L_{\rm 24\mu m} ({\rm erg\,s^{-1}}))^{0.848}$
\citep[from][]{2010ApJ...714.1256C}.
For most of the objects the agreement was satisfactory: the values of the SFR derived in both ways agreed to better than a factor 2.5 or,
in case of IRAS upper limits, the  resulting upper limits for the SFR were above those derived 
 from $L_{\rm 24\mu m}$. 
There were only  3 galaxies in 2 groups with  a larger discrepancy: for HCG 79b and for HCG 37b we obtained a  value
 of the SFR derived from IRAS that was  a factor of 6  higher than the SFR from the 24$\mu$m data and 
for HCG 37a the difference was a factor of 10.
After checking the Spitzer images and IRAS data, we found that 
in the case of HCG 79 the reason for the discrepancy was the  blending of HCG 79a and 79b in the IRAS beam. We thus assumed that
the value of $L_{\rm FIR}$  given for HCG 79b in \citet{1998ApJ...497...89V} arises 
from  both galaxies and 
assigned to each object a fraction of the IRAS fluxes and \lfir\  such that SFR(IRAS)=SFR(24$\mu$m).
A similar situation occurred in the case of HCG 37, where 3 objects (HCG 37a, HCG 37b and HCG 37c) are blended in the IRAS beam.
Here, we assume that the  value of $L_{\rm FIR}$  given for HCG 37b in \citet{1998ApJ...497...89V} was emitted from  all three galaxies and correctedin the
same way as for HCG 79.

\begin{table*}
\caption{FIR, SFR, SFE and sSFR}
\label{Sample_FIR}
\begin{tabular}{llccccccc}
\hline\hline
Galaxy&Ref$^{(1)}$&$I_{60}$&$I_{100}$&log(${\it L_{\rm FIR}}$)&SFR&log(SFE)$^{(2)}$&log(sSFR)\\
& & (Jy) & (Jy) &($L_{\odot}$)&($M_{\odot}$ yr$^{-1}$)&(yr$^{-1}$)&($M_{\odot}$ yr$^{-1}$)\\
\hline
      7a   &   2   &    3.32   &    6.61   &   10.23   &		  3.75  &        -9.14   &      -10.33     \\
      7b   &   2   & $<$ 0.18   & $<$ 0.32   & $<$ 8.95   &		$<$0.20 &                &      $<$-11.30 \\            
      7c   &   2   &    0.61   &    2.35   &    9.65   &   		  0.99  &        -9.18   &      -10.62  \\              
      7d   &   2   & $<$ 0.15   & $<$ 0.39   & $<$ 8.95   &		$<$0.20 &                &      $<$-10.63 \\            
     10a   &   2   &    0.50   &    1.81   &    9.72   &   		  1.16  &        -9.45   &      -11.02  \\  
.... & .... &.... &.... &.... &.... &.... &....  \\
\hline
\end{tabular}  

\footnotesize{
\noindent
$^{(1)}$ Reference code (see \ref{subs:FIRdata}): 1: Our data analysis. 2: \citet{1998ApJ...497...89V}.\\
$^{(2)}$The value of the SFE is not displayed for the galaxies with upper limits in both ${\it L_{\rm FIR}}$ and ${\it M_{\rm H_{2}}}$.\\
{\bf Notes.} The full table is available in electronic form at the CDS and from http://amiga.iaa.es.
}
\end{table*}


\section{Results}\label{sec:results}

In this section, we aim to 
study the relation between \mhtwo\ and the SFR in HCG galaxies and compare them
to isolated galaxies. Furthermore, we search for relations with the atomic gas deficiency of the galaxies and the groups and with the evolutionary phase of the groups.
We furthermore investigate the ratio between the two CO transitions, CO(1-0) and CO(2-1).

In order to search for differences  to isolated galaxies,
we used two methods: (i) We normalized \mhtwo\ and \lfir\ to the blue luminosity, \lb , or the
luminosity in the K-band, \lk , and compared the ratios to those of isolated galaxies, and
(ii) we  calculated the deficiency parameters of ${\it M_{\rm H_{2}}}$, ${\it L_{\rm FIR}}$ and ${\it M_{\rm HI}}$ of the galaxies 
(see Sect.~\ref{subsec:Deficiencies}). 
We obtained in general very consistent results for
 \lb\ and  \lk .

 
 We carry out this analysis separately for early-type galaxies and spirals because of the following
 reasons:
 (i) the morphological distribution is very different for both samples, with a much larger
 fraction of early-type galaxies among HCG galaxies,
(ii) the number of early-type galaxies in the AMIGA reference sample is very small so that no statistically
significant comparison sample is available. In particular, no deficiency parameter can be derived. 
(iii) Early-type galaxies tend to have a significantly lower molecular gas content than late-type galaxies, and their FIR emission 
is not as clearly related to their SFR as it is in late-type galaxies, as a result of the lack of  strong SF. 
Therefore, the use of \lfir\  as a SF tracer is more questionable.

\subsection{Relation between ${\it M_{\rm H_{2}}}$, ${\it L_{\rm FIR}}$, ${\it M_{\rm HI}}$ and ${\it L_{\rm B}}$ }

\begin{table*}
\caption{Correlation analysis of ${\it M_{\rm H_{2}}}$ vs ${\it L_{\rm B}}$, ${\it L_{\rm FIR}}$ vs ${\it L_{\rm B}}$ and ${\it M_{\rm H_{2}}}$ vs ${\it L_{\rm FIR}}$}
\begin{tabular}{lcccccc}
\hline\hline
\centering
Magnitude&	Sample	&&	Slope&	Intercept&	Slope&	Intercept\\
		&			&&(bisector)&(bisector)&	(${\it L_{\rm B}}$ indep.)&(${\it L_{\rm B}}$ indep.)\\	
\hline
${\it M_{\rm H_{2}}}$ vs ${\it L_{\rm B}}$	&	HCGs	&All&	1.37$\pm$0.15	&	-4.74$\pm$1.48&	0.81$\pm$0.14&	0.73$\pm$1.35\\
&									&T$>$0&	1.40$\pm$0.16	&	-4.94$\pm$1.61&	0.95$\pm$0.20&	-0.43$\pm$1.97\\
		&	AMIGA	&&	1.45$\pm$0.08&	-5.61$\pm$0.77&	1.12$\pm$0.08&		-2.43$\pm$0.83\\
\hline
${\it L_{\rm FIR}}$ vs ${\it L_{\rm B}}$	&	HCGs	&All&	1.47$\pm$0.16&	-5.29$\pm$1.54&	0.79$\pm$0.15&	1.43$\pm$1.49\\
										&	&T$>$0&	1.31$\pm$0.16&	-3.37$\pm$1.99&	0.77$\pm$0.16&	2.00$\pm$1.58\\
		&	AMIGA	&&	1.35$\pm$0.04&	-4.06$\pm$0.37&	1.12$\pm$0.04&	-1.73$\pm$0.38\\
\hline
${\it M_{\rm H_{2}}}$ vs ${\it L_{\rm FIR}}$
 &	HCGs	&All&	0.90$\pm$0.09&	0.41$\pm$0.83	&	0.75$\pm$0.09& 	1.82$\pm$0.86\\   					
 &			&T$>$0&	1.21$\pm$0.11&	-2.63$\pm$1.11	&	1.04$\pm$0.11& 	-1.00$\pm$1.08\\   					
					&AMIGA&&	1.16$\pm$0.08&	-2.14$\pm$0.72&	0.98$\pm$0.06&	-0.46$\pm$0.61\\
\hline
\end{tabular}
 \label{Slopes}
{\footnotesize \\ The slope and intercept are defined as log(${\it M_{\rm H_{2}}}$) = log($L_{B})\times slope$ + intercept, log(${\it L_{\rm FIR}}$) = log($L_{B})\times slope$ + intercept and log(${\it M_{\rm H_{2}}}$) = log($L_{FIR})\times slope$ + intercept.
 The fits on ${\it L_{\rm FIR}}$ vs ${\it L_{\rm B}}$ for the AMIGA sample are slightly different from the values in \citet{2007A&A...462..507L}
because we have taken into account a recent update of the basic properties of the galaxies 
 \citep[e.g. distance and morphological type; see][for more details]{fernandez2011}.
The AMIGA fits involving ${\it M_{\rm H_{2}}}$ are taken from 
\citet{2011lisenfeld}.}
\end{table*}

\begin{figure}
\includegraphics[width=8.5cm, angle=-90]{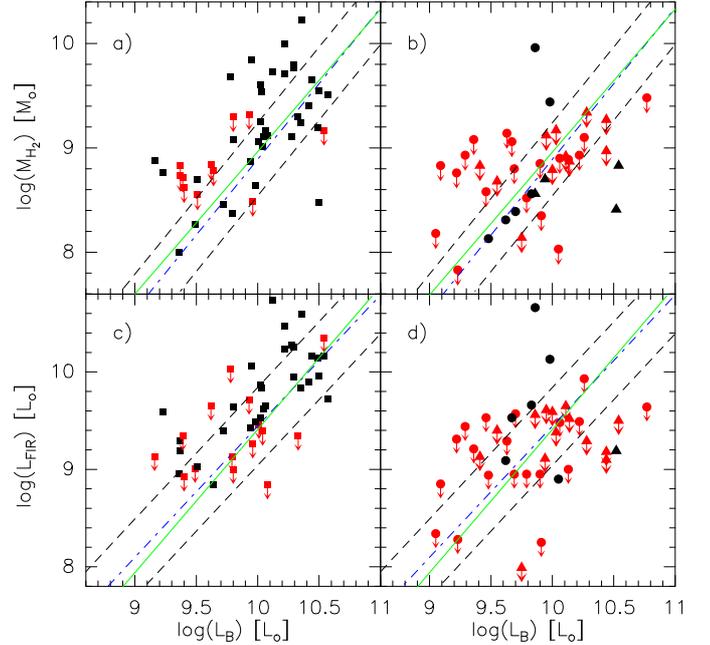}
\caption{\mhtwo\ vs \lb\ for  {\it a) } spiral galaxies (T$\ge$1)  and  {\it b)}  elliptical (circles) and S0 galaxies 
(triangles).  \lfir\ vs \lb\ for  {\it c) } spiral galaxies (T$\ge$1) and  {\it d)}   elliptical (circles) and S0 galaxies 
(triangles).  
%
%
The full green line corresponds to the bisector fit found for HCG galaxies (fit parameters are
given in Table \ref{Slopes}), while the blue dashed-dotted line corresponds to the bisector fit found for the AMIGA isolated galaxies. 
Both fits are done for the entire range of morphological types.
The dashed black lines are offset by the standard deviation of the correlation for the isolated galaxies, 
which is $\pm$0.35 for the ${\it M_{\rm H_{2}}}$ and $\pm$0.4 for ${\it L_{\rm FIR}}$. 
Black symbols denote detections and red symbols upper limits.}
\label{H2_FIR_LB}
\end{figure}

%
%
Fig. \ref{H2_FIR_LB} shows ${\it M_{\rm H_{2}}}$ (top) and  ${\it L_{\rm FIR}}$ (bottom)  versus  ${\it L_{\rm B}}$  for
spirals galaxies (left) and early-type galaxies (right). 
For spiral galaxies good correlations exist between both   ${\it M_{\rm H_{2}}}$, respectively 
${\it L_{\rm FIR}}$,  and  ${\it L_{\rm B}}$.
A linear fit to  the total sample of HGCs is plotted, together with the corresponding fit to the AMIGA sample. The coefficients are listed in Table \ref{Slopes}. 
 A  slightly shift  towards higher values  in \mhtwo\  seems to be present in comparison to the best-fit line of  isolated galaxies.
The linear   regressions between ${\it L_{\rm FIR}}$ and 
${\it L_{\rm B}}$, ${\it M_{\rm H_{2}}}$ and 
${\it L_{\rm B}}$ or ${\it M_{\rm H_{2}}}$ and ${\it L_{\rm FIR}}$ (Table \ref{Slopes}) show no significative differences between HCGs and isolated galaxies.
 For early-type galaxies no clear correlation is visible and for   log$(L_{\rm B}) \gtrsim 10$, the values of both ${\it M_{\rm H_{2}}}$ and ${\it L_{\rm FIR}}$ 
are below those of spiral galaxies.

 %

\begin{figure}[h]
\centering
\includegraphics[scale=.5, angle=-90]{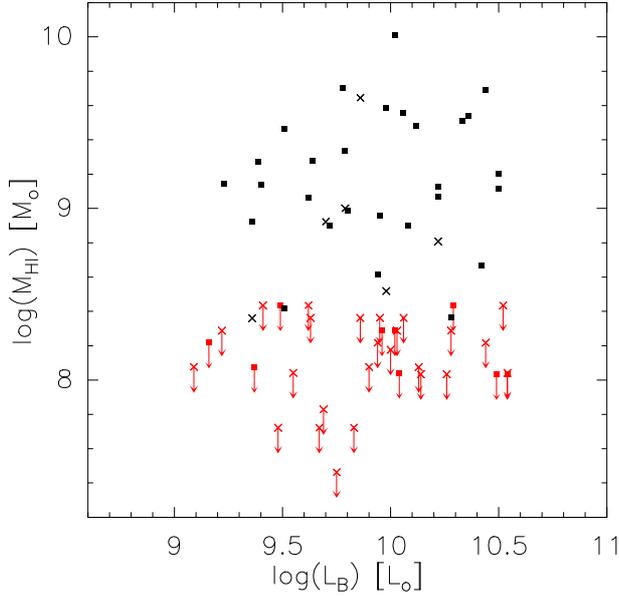}
\caption{\mhi\ vs \lb\ for  late-type (T$\ge$1,  squares) 
and early-type (crosses) galaxies. Black symbols denote detections and red symbols upper limits. }
\label{MHI_LB}       
\end{figure}

We note that, in contrast to ${\it M_{\rm H_{2}}}$ and ${\it L_{\rm FIR}}$, ${\it M_{\rm HI}}$ shows no correlation with ${\it L_{\rm B}}$ (Fig. \ref{MHI_LB})
reflecting the fact that HI is very strongly affected by the interactions and in many galaxies of our evolved groups largely removed from the galaxies.
	
\begin{figure}[h]
\centerline{
\includegraphics[scale=.5, angle=-90]{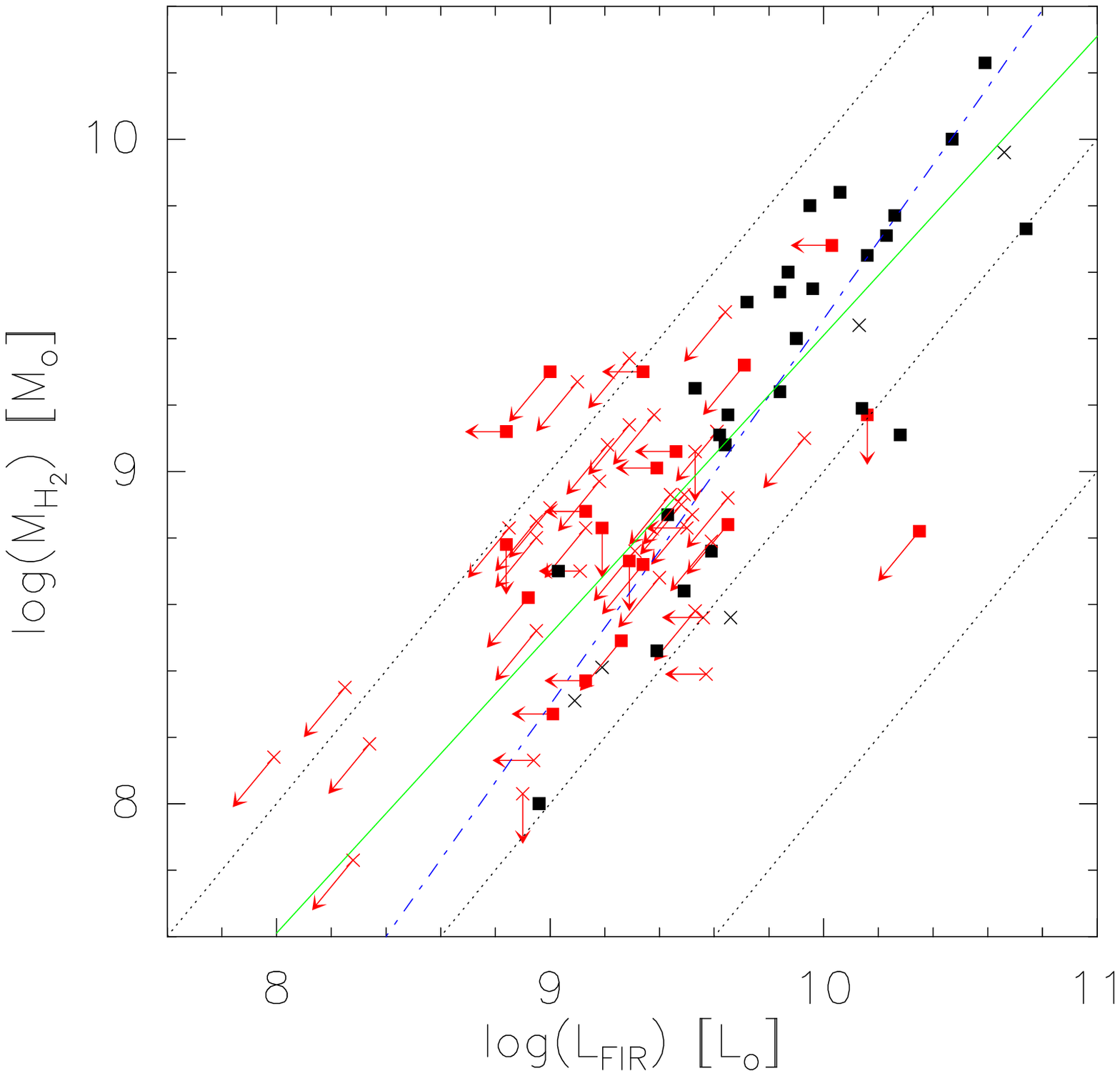}
}
\caption{\mhtwo\ vs \lfir\ for  late-type (T$\ge$1,  squares)  and early-type (E+S0, crosses) galaxies. The green line corresponds to the bisector fit found for HCGs galaxies, while the blue dashed-dotted
line corresponds to the bisector fit found for the AMIGA isolated galaxies from \citet{2011lisenfeld}. The fits are detailed in Table \ref{Slopes}. The dotted black lines correspond to the ${\it L_{\rm FIR}}/{\it M_{\rm H_{2}}}$ ratios 1 (left), 10 (middle) and 100 (right) $L_{\odot}$/$M_{\odot}$.
Black symbols denote detections and red symbols upper limits.}
\label{h2_fir}       
\end{figure}

Previous surveys \citep[see e.g.][]{1991ARA&A..29..581Y} have found a linear correlation between ${\it M_{\rm H_{2}}}$ and ${\it L_{\rm FIR}}$. 
A linear correlation can also be seen in our sample (Fig. \ref{h2_fir}). 
We include in this figure the lines for constant ${\it L_{\rm FIR}}/{\it M_{\rm H_{2}}}$ values equal to 1, 10 and 100 $L_{\odot}/M_{\odot}$. 
Practically all of our galaxies lie in the range
of ${\it L_{\rm FIR}}/{\it M_{\rm H_{2}}}= 1 -10\, L_{\odot}$/$M_{\odot}$, typical for normal, quiescent galaxies  \citep{1991ARA&A..29..581Y}.

\begin{figure}
\includegraphics[width=7.cm, angle=0]{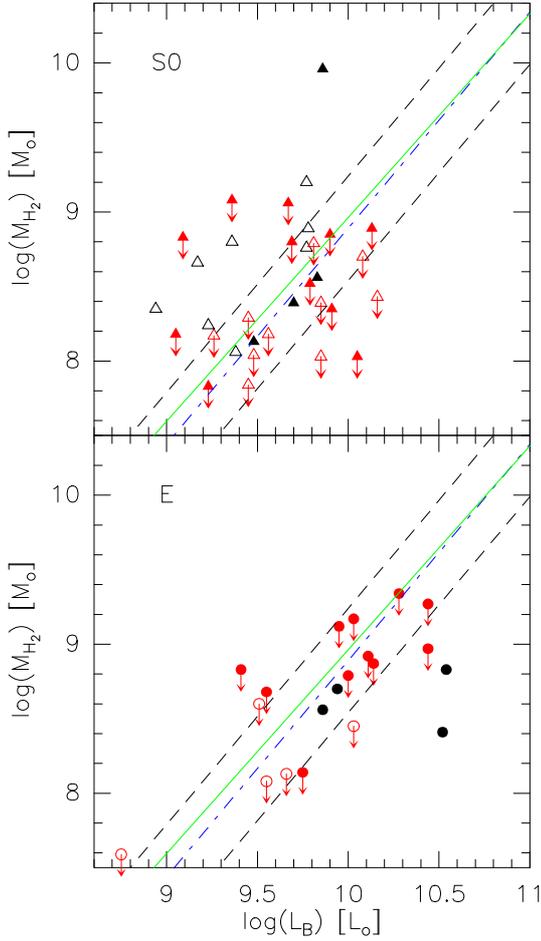}
\caption{${\it M_{\rm H_{2}}}$ vs $ L_{\rm B} $ for early-type galaxies in HCGs (full symbols) 
and from the AMIGA sample of isolated galaxies (open symbols) with distances
between 20 and 70 Mpc. The lines are the same as in Fig. 5a and b.
Black symbols denote detections and red symbols upper limits.
 {\it Top:} S0 galaxies (triangles), {\it Bottom:} elliptical galaxies (circles).
 }  
\label{compare_early_amiga}
\end{figure}

Finally,  we have directly compared E and S0 galaxies in HCGs to galaxies of the same types in the AMIGA sample.  
In the  case of lenticular galaxies we have limited the sample in HCGs to the same distance range as the AMIGA sample (40 - 70 Mpc) since for the largest distances the rate of upper limits is very high in HCGs and does not provide any further information. In Fig.~\ref{compare_early_amiga} (top) we show the relation between ${M_{\rm H_{2}}}$  
and $L_{\rm B}$ for the S0s in  HCGs and from the AMIGA sample.
Even  though the number of data points is low, a  trend 
seems  to be present for S0s in isolated galaxies to have a higher ${M_{\rm H_{2}}}$  for the same $L_{\rm B}$.  A similar result is found for $L_{\rm FIR}$ (not shown here), where most lenticular isolated galaxies present higher values than expected for their optical luminosity, while most of the objects in  HCGs  show upper limits excluding any excess.
If S0 galaxies in these dense environments originate from stripping of spirals, this might indicate
that  molecular gas has also been removed in the process. Although this interpretation is
speculative due to the low statistics, it provides hints for further research in future works.

Concerning the elliptical galaxies, none of the isolated galaxies is detected in CO, while among the four detections in HCGs two have a mass similar to the expected for spiral isolated galaxies (HCG 15d and HCG 79b) while the other two show significantly lower masses (HCG 37a and HCG 93a), pointing to an external 
origin (Fig.~\ref{compare_early_amiga}, bottom).
The FIR luminosity of the Es in HCGs (not shown here) is similar to that expected for spiral galaxies.
It is also noticeable that while the range of $L_{\rm B}$ values for the S0s in HCGs covers about the same range as for isolated galaxies, Es in HCGs are 
up to half an order of magnitude more luminous than isolated Es.



\subsection{Deficiencies}\label{subsec:Deficiencies}
We have calculated the ${\it M_{\rm H_{2}}}$, ${\it L_{\rm FIR}}$ and ${\it M_{\rm HI}}$ deficiencies following the definition of \citet{1984AJ.....89..758H} as

\begin{equation}
{\rm Def}(X)=\rm log(X_{predicted})-\rm log(X_{observed})
\end{equation}

\noindent where we calculated the predicted value of the variable X from ${\it L_{\rm B}}$. Following this definition, a negative deficiency implies an excess with respect to the predicted value.

The expected ${\it M_{\rm H_{2}}}$ for each galaxy is calculated from its \lb\ using the fit to the AMIGA sample in \citet{2011lisenfeld}. Note that the fit, which is given in Table \ref{Slopes}, was calculated without distinguishing morphological types.  Due to the dominance of spiral galaxies in the AMIGA sample,
the fit is only adequate for spiral galaxies. Because of the low number of early-type galaxies
in the AMIGA sample it is not possible to derive a meaningful deficiency parameter for them.
In addition, we calculated  the deficiency derived from the relation between \mhtwo\ and \lk\
of the AMIGA sample \citep{2011lisenfeld}, 
log(\mhtwo) = -2.27+1.05$\times$ log(\lk)).
In a similar way, the expected ${\it L_{\rm FIR}}$ is calculated from the fit between \lfir\ and \lb\
obtained for the AMIGA isolated galaxies (Table \ref{Slopes}) for  the 
sample presented in \citet{2007A&A...462..507L}.

 The correlations between \mhtwo\ (respectively \lfir ) and \lb , or \lk, have a considerable scatter
with standard deviations of 0.35 dex for \mhtwo\ and 0.4 dex for \lfir. 
These standard deviations are much higher than the observational measurement errors. In this case, the error of the mean
values  are completely dominated by the statistical errors and therefore we
neglect the observational errors in our calculations.
The high standard deviation  means that
  individual galaxies with deficiencies up to about these values can  just represent normal
  deviations from the mean.
  However, the much smaller error of the {\it mean} deficiency allows to compare samples
  of galaxies (here: galaxies in HCGs and isolated galaxies) with a higher precision.
 %


The HI deficiency of the galaxies is calculated following the morphology-dependent fit between 
\mhi\ and \lb\ in \citet{1984AJ.....89..758H}. We have 
considered h = $H_{0}$/100 = 0.75. 
We adapted their results which were based on mag$_{zw}$ to our use of $B_{\rm c}^{\rm T}$ with 
the relation found by \cite{2005A&A...436..443V} (mag$_{zw}$ = $B_{\rm c}^{\rm T}$+0.136).
%
 Taking furthermore into account that 
 we express ${\it L_{\rm B}}$ as a function of the solar bolometric luminosity (mag = 4.75), we introduce the following correction:

\begin{equation}
(\rm log\it L_{\rm B})_{Haynes} = (\rm log\it L_{\rm B})_{ours} + 0.14
\end{equation}

\noindent to express ${\it L_{\rm B}}$ in the terms we assume (Sec. \ref{sec:sample}) to calculate the expected content of HI. 
The deficiencies in ${\it M_{\rm H_{2}}}$, ${\it L_{\rm FIR}}$ and ${\it M_{\rm HI}}$  derived from \lb\   are listed in Table \ref{Tab_defs_1}.

\begin{table}
\caption{Deficiencies  of ${\it M_{\rm H_{2}}}$, ${\it L_{\rm FIR}}$, and $M_{\rm HI}$ derived from \lb}
\label{Tab_defs_1}
\begin{tabular}{l|ccc}
\hline\hline
Galaxy&Def($\it M_{\rm H_{2}})$ &Def(${\it L_{\rm FIR}}$) &Def($M_{\rm HI}$)\\
\hline
     7a   &   -0.50   &   -0.49   &    0.67 \\		                          	
      7b   & $>$-0.36   & $>$0.07   & $>$1.38 \\	                          
      7c   &   -0.19   &   -0.13   &    0.29 \\		                          
      7d   & $>$-0.06   & $>$0.21   &    0.28 \\	                          
     10a   &    0.21   &    0.49   &   ... \\		                          
....&....&....&....\\
\end{tabular}

{\bf Notes.} The full table is available in electronic form at the CDS and from http://amiga.iaa.es.

\end{table}

\subsubsection{${\it M_{\rm H_{2}}}$ and ${\it L_{\rm FIR}}$ deficiencies\label{subsub:def_h2_fir}}

 \begin{table*}
\caption{Mean values  for
spiral galaxies ($T\ge 1$) in HCGs and from the AMIGA sample.
The mean values and their errors are calculated with ASURV, taking upper limits into account.
We neglect observational errors since the data is dominated by statistical errors. 
The quoted errors represent the error of the mean values, not the standard deviation.
}
\begin{tabular}{l|cc|cc}
\hline\hline
&\multicolumn{2}{c}{\textbf{HCGs}}&\multicolumn{2}{c}{\textbf{AMIGA}$^{(1)}$}\\
& Mean  &$n_{UL}/n$&Mean &$n_{UL}/n$\\
\hline
log(${\it L_{\rm B}}$)	   (\lsun)                    &9.95$\pm$0.06            &0/46                        &9.75$\pm$0.04            &0/150\\
log(${\it M_{\rm H_{2}}}$)	  (\msun)            &9.02$\pm$0.09            &11/46     &8.38$\pm$0.09            &64/150\\
log(${\it L_{\rm FIR}}$)	(\lsun)  		&9.53$\pm$0.09&15/45 &9.16$\pm$0.05&58/150\\
\hline
Def(${\it M_{\rm H_{2}}}$) (from \lb\ )        &-0.14$\pm$0.09            &11/46	                      &0.06$\pm$0.04            &64/150\\
Def(${\it M_{\rm H_{2}}}$) (from \lk\ )        &-0.15$\pm$0.06            &10/45	                      &-0.01$\pm$0.05           &58/149 \\
Def(${\it L_{\rm FIR}}$)		&-0.11$\pm$0.08&15/45	           &-0.09$\pm$0.04&58/150\\
Def(HI)                                  &0.93$\pm$0.13&9/37	& -& -\\
\hline
log(\mhtwo/\lb), all \lb            &-0.96$\pm$0.08            &11/46	                      &-1.25$\pm$0.04            & 64/150 \\
(\msun/\lsun) & & & & \\
 log(\mhtwo/\lb), low \lb$^{(2)}$      &-1.04$\pm$0.10            &10/22             &        -1.36$\pm$0.05          & 56/103  \\
(\msun/\lsun) & & & & \\
log(\mhtwo/\lb), high  \lb$^{(3)}$       &-0.88$\pm$0.09            &1/24                      &-1.06$\pm$0.05            &8/47 \\
(\msun/\lsun) & & & & \\
 log(\mhtwo/\lk)            &  -1.58$\pm$0.05            &10/45	                      &-1.76$\pm$0.05            & 50/135 \\
(\msun/L$_{K,\odot}$) & & & & \\
log(\lfir/\lb)                  &    -0.45$\pm$0.07 & 15/45	&  -0.52$\pm$0.03 & 58/149  \\
\hline
\end{tabular}\\
\\ {\footnotesize $^{(1)}$The mean values of the AMIGA galaxies are calculated 
  for the subsample of galaxies with ${\it M_{\rm H_{2}}}$ data. For each subsample, $n$ is the number of galaxies and $n_{UL}$ is the number of upper limits. ${\it L_{\rm FIR}}$ and \lb\ of the AMIGA galaxies  are from the new data release  (see Sec. \ref{ref_sample}),  
   while ${\it M_{\rm H_{2}}}$ and \lk\  are from  \citet{2011lisenfeld}.

$^{(2)}$ for \lb $<10^{10}$ \lsun

$^{(3)}$ for \lb $>10^{10}$ \lsun
}
\label{Medias_2muestras}
\end{table*}

The mean ${\it M_{\rm H_{2}}}$ 
and ${\it L_{\rm FIR}}$ deficiencies  for spiral galaxies in HCGs 
are similar (see Table \ref{Medias_2muestras}). 
Galaxies showing an excess in ${\it M_{\rm H_{2}}}$ or ${\it L_{\rm FIR}}$ have values spanning over the full range  of ${\it L_{\rm B}}$, as can be seen in Fig. \ref{H2_FIR_LB}. Thus, the excess in ${\it M_{\rm H_{2}}}$ or ${\it L_{\rm FIR}}$ is not associated with the brightest objects \textit{per se}.
We have checked in detail the properties of the 9 galaxies showing the  largest $M_{\rm H_{2}}$ excess
(HCG 10c, HCG 16a, HCG 16c, HCG 16d, HCG 23b, HCG 23d, HCG 40c, HCG 58a, HCG 88c), and we find that half of them present strong signs of distortion (tidal tails in 
the optical and/or HI, kinematical perturbations, etc).

\begin{figure}[h]
\centerline{
\includegraphics[scale=.5, angle=0]{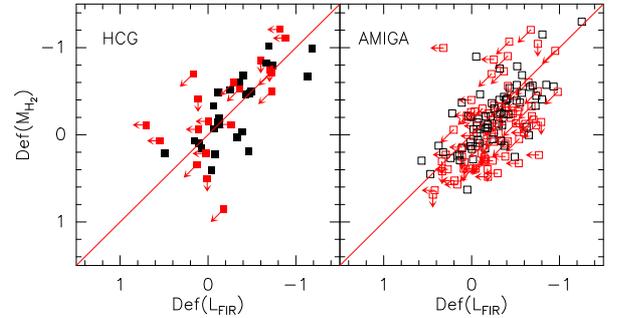}
}
\caption{${\it M_{\rm H_{2}}}$ deficiency vs ${\it L_{\rm FIR}}$ deficiency for late-type (T$\ge$1)
galaxies in HCGs (left) and  from the AMIGA sample (right).
Red symbols represent upper limits in either \mhtwo\ or \lfir , and black symbols detections.
The y=x line is plotted as reference and does not represent a fit to the data.
}    
\label{Def_h2_def_fir} 
\end{figure}


Fig. \ref{Def_h2_def_fir} (left) shows Def(${\it M_{\rm H_{2}}}$) (from \lb)  vs Def(${\it L_{\rm FIR}}$) for each galaxy. Both are strongly correlated, which 
can be understood as due to the causal relation between the  molecular gas and SFR, leading to a lower SFR if the 
molecular gas as the fuel for SF decreases. 
For comparison, 
Fig. \ref{Def_h2_def_fir} (right)
displays Def(${\it M_{\rm H_{2}}}$) of the isolated galaxies versus their Def(${\it L_{\rm FIR}}$). 
The behavior of the isolated galaxies does not show a significant difference 
compared to galaxies in HCGs with a very similar range covered by both samples.
However, for the isolated galaxies, Def(${\it M_{\rm H_{2}}}$) extends to slightly lower values  for a given  Def(${\it L_{\rm FIR}}$).
This is also reflected in the mean values of Def(${\it M_{\rm H_{2}}}$) and Def(${\it L_{\rm FIR}}$) of AMIGA and HCG galaxies (Table \ref{Medias_2muestras}): while the values of Def($ L_{\rm FIR}$) for spiral galaxies are almost the same for both samples, Def(${\it M_{\rm H_{2}}}$)
in spirals is larger by 0.15-0.20 for HCG than for AMIGA galaxies (corresponding to a 40-60\% larger ${\it M_{\rm H_{2}}}$ than expected for isolated galaxies).

\begin{figure}
\includegraphics[width=7.cm, angle=-90]{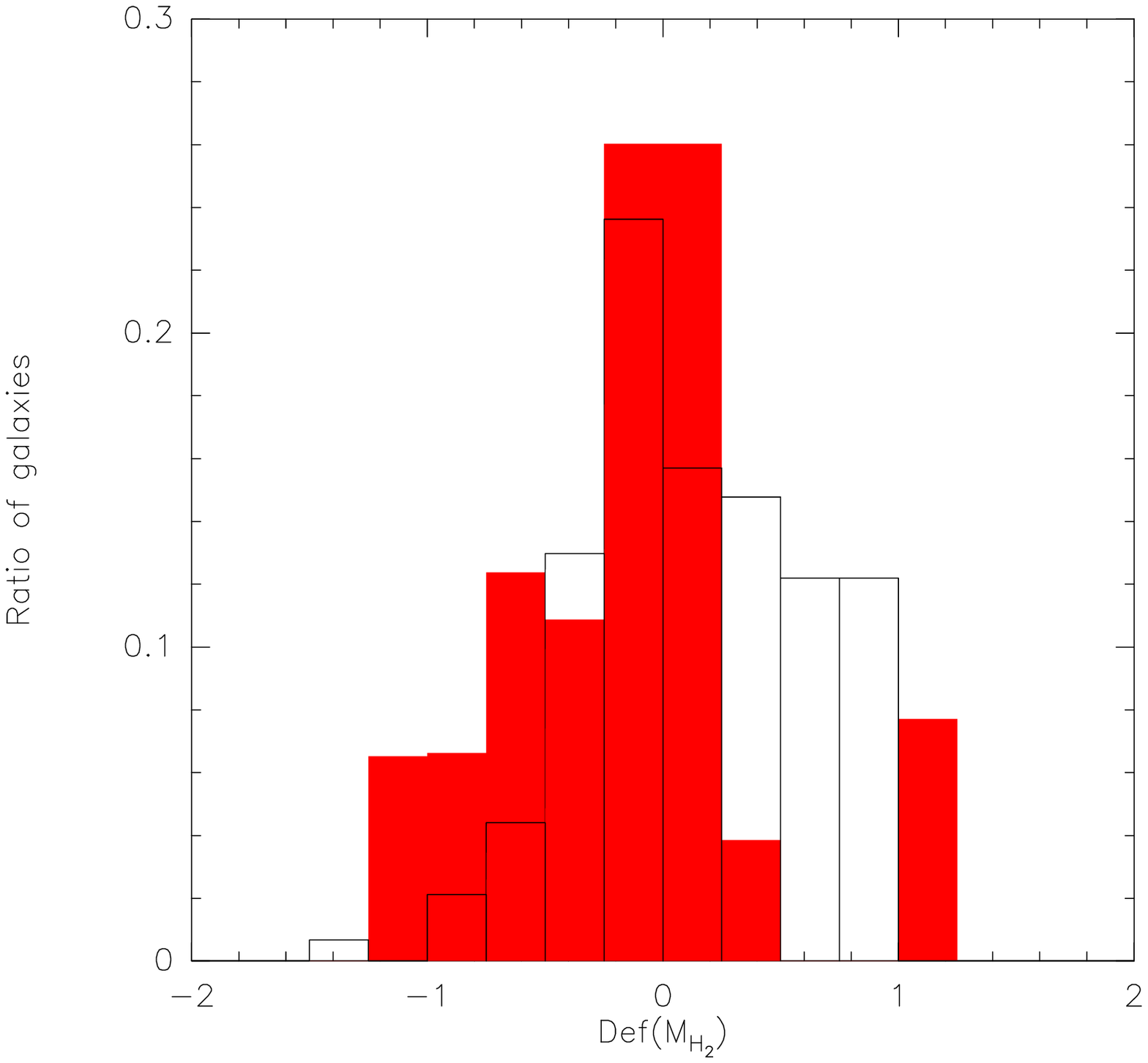}
\quad
\includegraphics[width=7.cm, angle=-90]{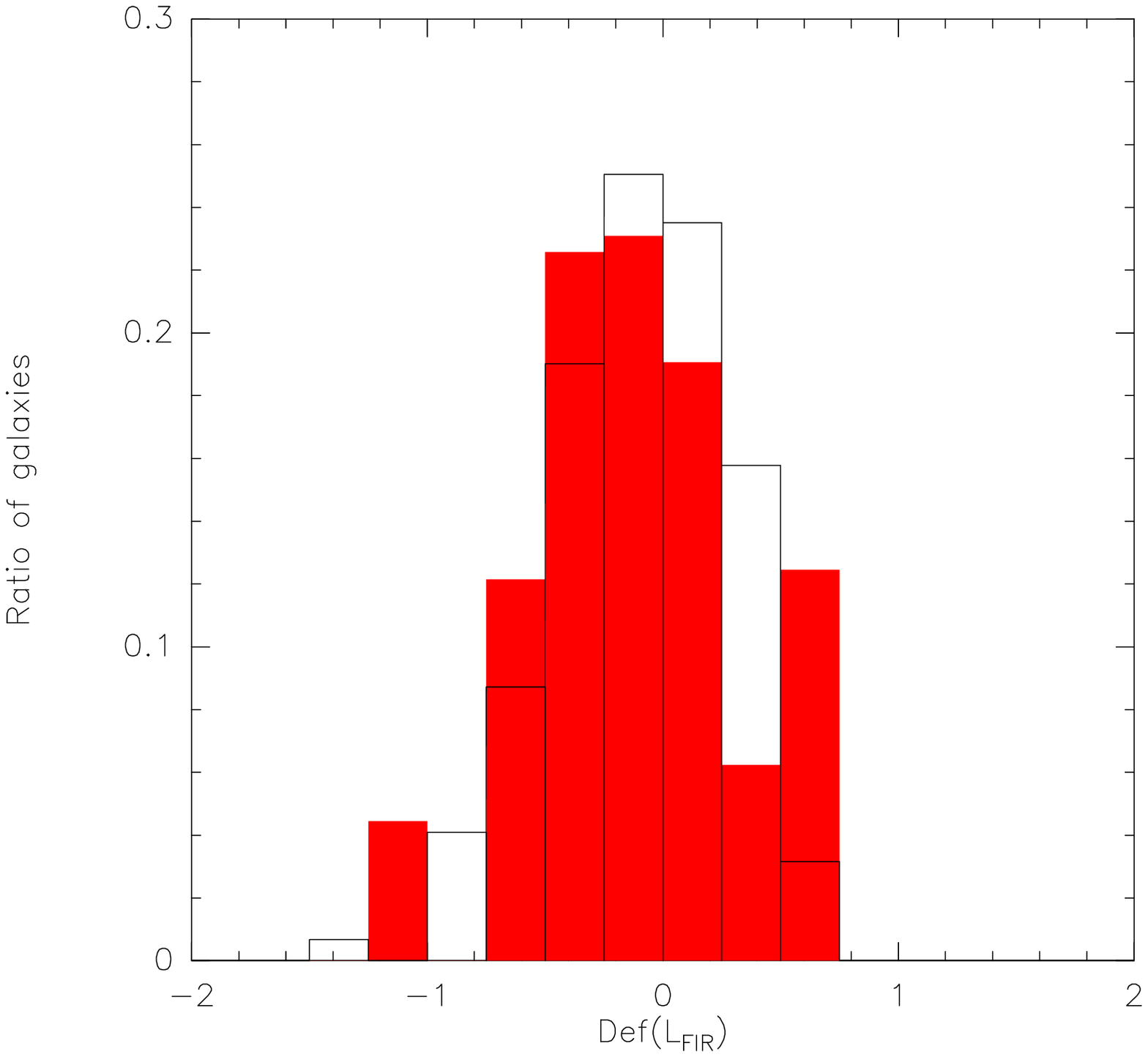}
\caption{Def(${\it M_{\rm H_{2}}}$) (top) and Def(${\it L_{\rm FIR}}$) (bottom) distribution of  spiral galaxies in AMIGA (black line) and in HCGs (red filled bars), calculated with ASURV in order to take the upper limits into account. }
\label{Histo_DefH2_DefFIR}
\end{figure}

The histograms shown in Fig. \ref{Histo_DefH2_DefFIR} underline these findings: whereas the distribution of Def(${\it M_{\rm H_{2}}}$)
for spiral galaxies in HCGs is shifted to negative deficiencies (i.e. an excess) compared to AMIGA galaxies, the distribution of Def(${\it L_{\rm FIR}}$) is very similar for spiral galaxies in HCGs and in the AMIGA sample. 
Two sample tests (Gehan's Generalized Wilcoxon Test and Logrank Test)   confirm that the distributions of Def(${\it M_{\rm H_{2}}}$)  
are different with a probability of $>96\%$ , whereas the distributions Def($L_{\rm FIR}$) are identical with a 
non-negligible probability. 

As an additional test, we have compared the ratios \mhtwo/\lb\ and  \mhtwo/\lk\ of HCG galaxies to those of isolated galaxies (values are listed in Table \ref{Medias_2muestras}).
In the case of  \mhtwo/\lb\ we have derived the ratios both for the entire luminosity range and for low (\lb $\le 10^{10}$ \lsun) and high
(\lb $> 10^{10}$ \lsun)  luminosity galaxies in order not to be affected   by the  nonlinearity of the \mhtwo-\lb\ relation. 
In all cases we found a lower ratio (by $\sim$ 0.2-0.3 dex) for the isolated galaxies, confirming our findings from the deficiency parameter.

The larger  \mhtwo\  for a given  $L_{\rm B}$ found for spiral galaxies in HGCs could be explained in three ways:
a) a real excess of the total molecular gas mass (and will  be further discussed as such in the following section),
b) a higher concentration towards the center of the molecular gas in HCG galaxies compared to  isolated galaxies, so that the
extrapolation of the flux based on a similar extent (see Sec.~\ref{subs:MolGas})  would lead to an overestimate of ${\it M_{\rm H_{2}}}$, or
 c) a systematic difference in the CO-to-H$_2$ conversion factor between the AMIGA and HCG sample.
Although we cannot exclude this  possibility, we do not consider it very likely. 
The CO-to-H$_2$ conversion factor is known to depend on a number of galactic properties as the metallicity, gas temperture, gas density and velocity 
dispersion \citep[e.g.][]{1988ApJ...325..389M,2011MNRAS.418..664N}. These properties 
are likely similar in both samples because of 
the similar ranges in \lb\ and \lfir\ (tracing SFR) that they cover.
The first two effects (a and b) could both be at work at the same time. In fact, as indicated in e.g. \cite{2008A&A...485..475L}, galaxies in the AMIGA sample are dominated by disk SF while surveys of compact groups \citep{1995MNRAS.274..845M} show that most
radio detections involve compact nuclear emission. This can be explained since nuclear emission is
thought to be enhanced by interactions that produce a loss of angular momentum of the molecular
gas, that subsequently falls towards the center of the galaxy. These dissipative effects are 
 likely near minimum in isolated galaxies. This result was also proposed by \cite{1998ApJ...497...89V}, where the enhanced $I_{\rm 25}/\it I_{\rm100}$ ratio in HCGs was suggested
to be caused by local starbursts, presumably in the nuclear
region. This result is still compatible with the
conclusion of a normal level of FIR emission among HCG
galaxies that we find here, if the activity responsible for enhanced 24$\mu$m emission and
enhanced/more concentrated molecular gas is localized compared to the overall distribution of gas and
dust in the galaxies.


\subsubsection{Comparison to the  $M_{\rm HI}$ deficiency }\label{subsub:def_HI}	

\begin{figure}
\includegraphics[width=8.5cm, angle=0]{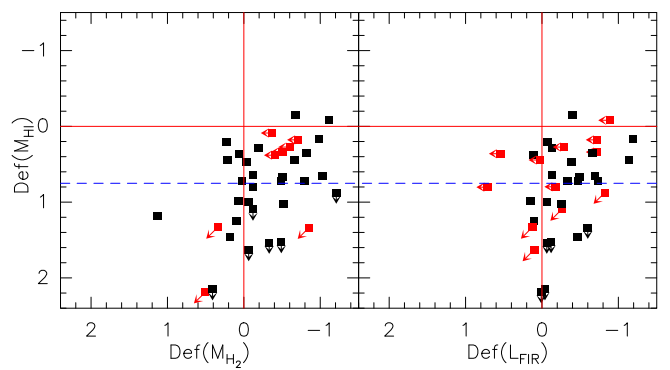}
\caption{${\it M_{\rm HI}}$ vs ${\it M_{\rm H_{2}}}$  deficiencies (left) and  ${\it M_{\rm HI}}$ vs ${\it L_{\rm FIR}}$ deficiencies (right)
for  spiral  galaxies (T$\ge 1$). The red lines show Def(${\it M_{\rm HI}}$) = 0, Def(${\it M_{\rm H_{2}}}$) = 0 and
 Def(${\it L_{\rm FIR}}$) = 0, and 
the dashed lines give Def(${\it M_{\rm H_{2}}}$)=0.75, separating low and highly HI-deficient galaxies in our analysis.
 Red symbols denote upper limits in \mhtwo\ or \lfir .}  
\label{Def_H2_FIR_HI}
\end{figure}

In Fig.~\ref{Def_H2_FIR_HI} we display Def(${\it M_{\rm HI}}$) vs Def(${\it M_{\rm H_{2}}}$) (left) and Def(${\it L_{\rm FIR}}$) (right).
The mean value of Def(${\it M_{\rm HI}}$) of the galaxies with available HI data is $0.93\pm0.13$ (12\% of the expected value)
for spiral galaxies and $1.31\pm0.11$ (5\% of the expected value)
for all morphological types, which is one order of magnitude larger than 
Def(${\it M_{\rm H_{2}}}$) and Def(${\it L_{\rm FIR}}$). 
We stress here that the samples used in the present paper and in \citet{2001A&A...377..812V} are not the same. This earlier study 
concentrated on the set of data available at that time, 
which was biased towards HI bright groups. Later, more groups with  higher HI deficiencies have been observed with the VLA 
\citep{2007ggnu.conf..349V}, and are part of the present sample. Therefore, the mean HI deficiency of the galaxies in  \citet{2001A&A...377..812V}  (25\% of the expected value
for spiral galaxies) is less than
the mean HI deficiency of the present sample. We have checked that the HI deficiencies calculated in this paper are consistent with the values 
for the groups in common with \citet{2001A&A...377..812V}.



\begin{table*}
\centering
\caption{Mean values of deficiencies and ratios of  \mhtwo\ and \lfir\    for different samples. Only spiral
galaxies (T$\ge$1) are considered.  Mean values are calculated as explained in
Table \ref{Medias_2muestras}.
}
\begin{tabular}{lc|ccccc}
\hline\hline
&  &  Def(\mhtwo )&  log(\mhtwo/\lb) & $n_{UL}/n$    &   log(\mhtwo/\lk)  & $n_{UL}/n$\\
&   &                        &  (\msun/\lsun)   &    & (\msun/L$_{K,\odot}$)  &  \\
\hline
\textbf{Total}	   &	&-0.14$\pm$0.09 &  -0.96$\pm$0.08  &	11/46 &  -1.58$\pm$0.05  & 10/45\\
\hline
\textbf{HI content}			&Def(HI)$<$0.75&	  	-0.34$\pm$0.10  &  -0.82$\pm$0.10  & 5/21 &  -1.40$\pm$0.07  & 4/20 \\ 	
\textbf{of galaxies}			&Def(HI)$>$0.75&	 	-0.07$\pm$0.16   & -1.15$\pm$0.13  &  3/16 &  -1.77$\pm$0.08  & 3/16\\	
\hline
\textbf{HI content}	&Normal			 &-0.38$\pm$0.20&   -0.76$\pm$0.15  & 1/6 &   -1.52$\pm$0.11  & 0/5\\
\textbf{of the group}	&Slightly deficient	&-0.08$\pm$0.11 &   -0.99$\pm$0.10 & 8/32 &   -1.59$\pm$0.07  & 8/32\\  
				&Very deficient		&-0.21$\pm$0.08 &   -0.95$\pm$0.10 & 2/8&   -1.60$\pm$0.12  & 2/8 \\ 
\hline
\textbf{Evolutionary}&Phase 1   &-0.35$\pm$0.14& -0.76$\pm$0.11 & 2/11 &   -1.46$\pm$0.09  & 1/10\\
\textbf{Phase}		&Phase 2  &-0.16$\pm$0.13 & -0.92$\pm$0.12 & 5/21 &   -1.55$\pm$0.07  & 5/21\\
				&Phase 3	 & -0.04$\pm$0.09 & -1.07$\pm$0.08 & 4/14 &   -1.71$\pm$0.11  & 4/14\\	
				\hline


\hline
&  & Def(\lfir )&  log(\lfir/\lb) & $n_{UL}/n$  &log(\lfir/\lk) & $n_{UL}/n$ \\
&   &                &    &   &  (\lsun/L$_{K,\odot}$)   &  \\
\hline 
\textbf{Total}	& 	&-0.11$\pm$0.08 & -0.45$\pm$0.07  & 15/45 & -1.14$\pm$0.09  & 14/44\\
\hline           
\textbf{HI content}     &Def(HI)$<$0.75 &-0.32$\pm$0.11 &  -0.28$\pm$0.11 & 6/20  & -0.84$\pm$0.10  &  5/19  \\                    
\textbf{of galaxies}     &Def(HI)$>$0.75  & 0.03$\pm$0.11  &  -0.60$\pm$0.11 &6/16    & -1.38$\pm$0.13 &   6/16    \\

\hline                                      
\textbf{HI content}     &Normal    &-0.19$\pm$0.19 &   -0.36$\pm$0.11 & 3/6  &-1.07$\pm$0.17 &   2/5\\ 
\textbf{of the group}   &Slightly deficient   &-0.08$\pm$0.09 &   -0.45$\pm$0.09 & 10/31 &-1.15$\pm$0.11 &    10/31\\           
                 &Very deficient       &-0.23$\pm$0.08 &  -0.37$\pm$0.07   & 2/8 & -1.03$\pm$0.14  &  2/8\\

\hline                                  
\textbf{Evolutionary}&Phase 1         & -0.17$\pm$0.15 &  -0.36$\pm$0.13  & 4/11 &  -1.03$\pm$0.13 & 3/10  \\     
\textbf{Phase}          &Phase 2        &-0.12$\pm$0.13 &  -0.43$\pm$0.13 &7/20 & -1.11$\pm$0.14  &  7/20 \\    
			       &Phase 3   &-0.12$\pm$0.05 &  -0.45$\pm$0.04 & 4/14 &-1.11$\pm$0.09 &   4/14\\       
\hline
\end{tabular}

\label{More_deficiencies}

\centering
{\footnotesize For each subsample, $n$ is the number of galaxies and $n_{UL}$ is the number of upper limits.}
\end{table*}

Most noticeable in Fig.~\ref{Def_H2_FIR_HI} is that even very HI-deficient galaxies have a rather normal 
\mhtwo\ or \lfir .
There is no clear correlation between Def(${\it M_{\rm HI}}$) and Def(${\it M_{\rm H_{2}}}$) or 
Def(${\it L_{\rm FIR}}$). 
There  might be a weak trend in the sense that a larger  ${\it M_{\rm HI}}$ deficiency 
leads to larger  ${\it M_{\rm H_{2}}}$ and ${\it L_{\rm FIR}}$ deficiencies. 
This trend is also seen
when calculating the mean deficiencies and ratios 
separately 
%
for low and highly \mhi\ deficient galaxies,
here chosen as galaxies with def(\mhi) $<$ 0.75 and 
def(\mhi) $>$ 0.75  in order 
 to obtain two groups of roughly the same size (Table \ref{More_deficiencies}).
 However, the differences are small and fall below significance when changing the
 separation to def(\mhi) =  0.50. Thus, the statistics in our sample is not sufficient
 to firmly conclude whether this trend is real.


\subsection{Comparison to the HI content and evolutionary stage of the group}\label{subsec:Def_content_phase}




To study the influence of the global HI content of the group on ${\it M_{\rm H_{2}}}$ and SFR of the individual galaxies we have classified the groups as a function of their Def(${\it M_{\rm HI}}$) as listed in Sec. \ref{sec:sample}. The average Def(${\it M_{\rm H_{2}}}$) and Def(${\it L_{\rm FIR}}$) of the galaxies belonging to these groups are detailed in Table \ref{More_deficiencies}. We find no clear relation between the Def(${\it M_{\rm H_{2}}}$) of the galaxies, nor the Def(${\it L_{\rm FIR}}$), with the global Def(${\it M_{\rm HI}}$) of the groups.




In a similar way, we calculated the average Def(${\it M_{\rm H_{2}}}$) and Def(${\it L_{\rm FIR}}$) of the galaxies belonging to HCGs in different evolutionary states, as defined by \cite{2010ApJ...710..385B} (see Sec. \ref{sec:sample}), which are also detailed in Table \ref{More_deficiencies}. The Def(${\it M_{\rm H_{2}}}$) of the galaxies increases slightly as the group evolves along the evolutionary sequence. This trend is also visible in the ratios \mhtwo/\lb\ and \mhtwo/\lk.
In the case of Def(${\it L_{\rm FIR}}$), there is no clear relation for spiral galaxies with the evolutionary state, we only find a trend when considering the total sample, most likely due to a changing fraction of ellipticals.

\begin{figure}[h!]
\includegraphics[width=7.cm, angle=0]{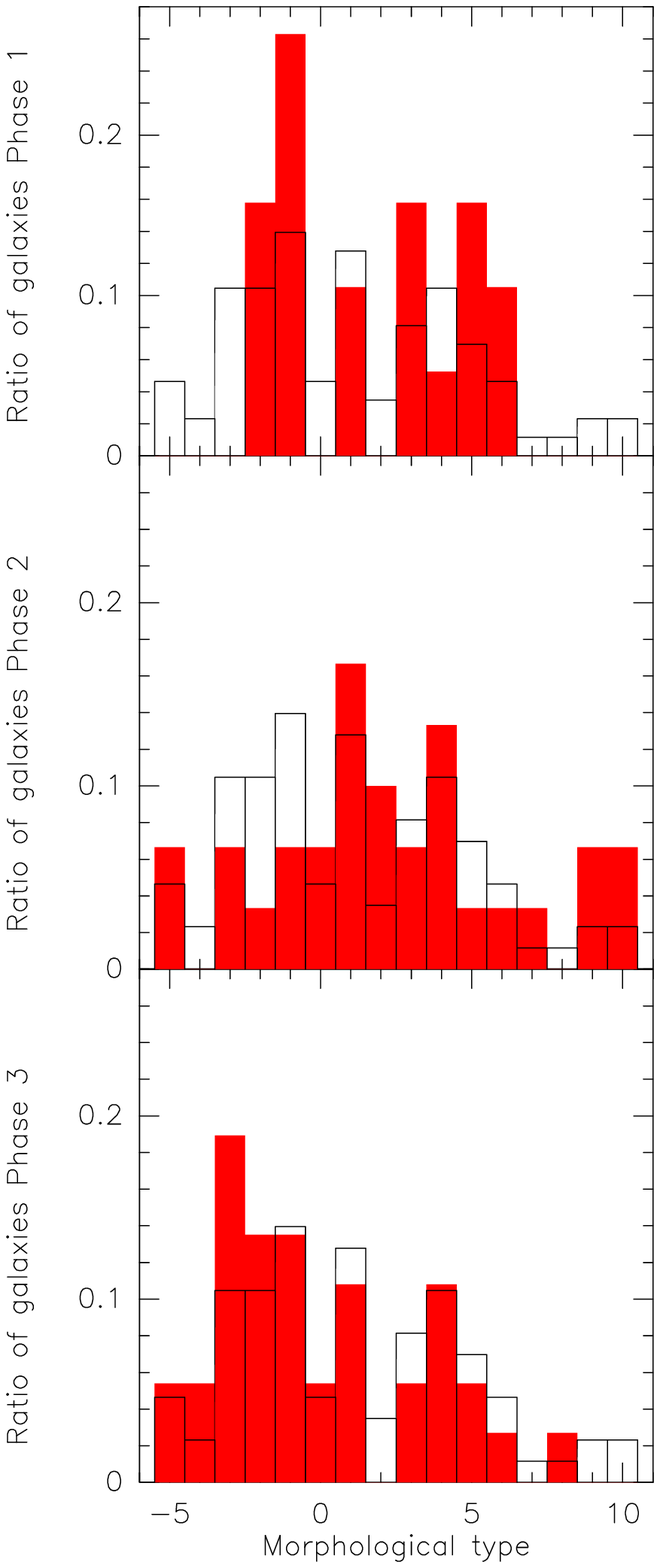}
\caption{Morphological type distribution for different evolutionary phases. From top to bottom, the morphological type distribution of galaxies in HCGs in evolutionary phases
 1, 2 and 3 are plotted. The filled red bins correspond to the distribution for the groups in each evolutionary state, while the black line bins correspond, for comparison, to galaxies of all phases.}
\label{Histo_evphase}
\end{figure}

A very pronounced variation with evolutionary phase is shown by the 
morphological types  (Fig. \ref{Histo_evphase}). 
 The ratio of elliptical and S0 galaxies increase strongly  in groups in phase 3. 
It has been proposed \citep[e.g.][]{2001A&A...377..812V,2011MNRAS.415.1783B} that S0  galaxies in HCGs might be stripped spirals. 

 \subsection[SFR, SFE and sSFR]{Star Formation Rate, Star Formation Efficiency and specific Star Formation Rate}\label{subsec_SFR_SFE}
 
We calculate the SFR from \lfir\ following the prescription of 
\citet{1998ARA&A..36..189K}:

\begin{equation}
SFR(M_{\odot}/\rm yr) = 4.5 \times 10^{-44}\, L_{\rm IR} (\rm erg s^{-1})
\label{eq.SFR}
\end{equation}

\noindent where ${\it L_{\rm IR}}$ refers to the IR luminosity integrated over the entire  mid- and far-IR spectrum (10-1000 $\mu$m). This expression
is based on a Salpeter IMF. We convert it to  the \citet{2001MNRAS.322..231K}
 IMF by dividing by a factor 1.59 \citep{2008AJ....136.2782L}.
In our analysis we use  ${\it L_{\rm FIR}}$   (eq. \ref{FIR_eq}), which estimates
the FIR emission in the wavelength range of 42.5-122.5 $\mu$m. We estimate  ${\it L_{\rm IR}}$ from
 ${\it L_{\rm FIR}}$ using the 
result of \citet{2003ApJ...586..794B} that on average ${\it L_{\rm IR}} \sim 2 \times {\it L_{\rm FIR}}$. Taking this into account, we can calculate the SFR from
${\it L_{\rm FIR}}$ as:

\begin{eqnarray}
SFR(M_{\odot}/\rm yr) &=& 4.5 \times 2 \times \frac{1}{1.59} \times 10^{-44} L_{\rm FIR} (\rm erg s^{-1}) \\ \nonumber
&=&  2.2 \times 10^{-10}~ L_{\rm FIR} (L_{\odot})
\end{eqnarray}

The values of the SFR of the galaxies in our sample are listed in Table \ref{Sample_FIR}.

We define the  SFE as the  ratio between the SFR and the molecular gas mass, SFE = SFR/\mhtwo .
The SFE of the individual galaxies in our sample are listed  in Table \ref{Sample_FIR}. 
Fig.~\ref{h2_fir}  shows a good, roughly linear correlation between \lfir\ and \mhtwo\  and indicates that the SFE in our sample 
is expected to show a rather narrow range.
To calculate the average SFE of our sample we must take 
into account that ASURV can only handle data showing upper or lower limits, but not both. Thus, we have first calculated the average SFE 
considering only galaxies detected in CO with an upper limit in FIR, together with the ones detected in both bands. Separately, 
we considered only those detected in FIR but not detected in CO and the ones detected in both bands. 
The average values are listed in Table \ref{SFEmedias}.

We have calculated the average SFE for the AMIGA sample of isolated galaxies, taking  ${\it M_{\rm H_{2}}}$  from 
\citet{2011lisenfeld} and  ${\it L_{\rm FIR}}$ from \cite{2007A&A...462..507L}, for comparison with the SFE in  
HCG galaxies. The values are listed in Table~\ref{SFEmedias}. We furthermore list the SFE
derived from a spatially resolved analysis for 30 nearby galaxies from
the HERACLES survey \citep{2011ApJ...730L..13B}. All values are adjusted to our CO-to-H$_2$ conversion factor, Kroupa IMF,
and no consideration of helium in the molecular gas mass. 
Tab.~\ref{SFEmedias}  shows a slightly lower  SFE in HCGs than
in AMIGA galaxies, in line with the previous results of  an excess in \mhtwo\, but
a normal value of   \lfir . 
In comparison to the galaxies from the
HERACLES survey there is no noticeable difference. 
Thus, overall there are no strong indications that the process of SF occurs in a different manner in the
different environment of HCGs.

\begin{table}
\caption{Mean log(SFE) for different samples and measurements (only spiral galaxies, T$\ge$1).
 Mean values are calculated as explained in
Table \ref{Medias_2muestras}.
}
\begin{centering}
\begin{tabular}{lc}
\hline\hline
Sample& $<$log(SFE) (yr$^{-1}$)$>$\\
\hline
HCGs        &-9.06$\pm$0.05$^{(1)}$/-9.22$\pm$0.06$^{(2)}$\\
CIGs          &-8.94$\pm$0.03$^{(1)}$/-9.07$\pm$0.04$^{(2)}$	\\
HERACLES   &-9.23$^{(3)}$\\
\hline
\end{tabular}
\\
\end{centering}
{\footnotesize$^{(1)}$ Values obtained with galaxies detected in both CO and FIR and galaxies detected in FIR but not detected in CO. $^{(2)}$ Values obtained with galaxies detected in both CO and FIR and galaxies detected in CO but not detected in FIR. $^{(3)}$ from  \cite{2011ApJ...730L..13B}.
The 1$\sigma$ standard deviation is 0.24dex.
}
\label{SFEmedias}
\end{table}

The specific SFR, sSFR, is defined as the ratio between the SFR and the stellar mass of a galaxy.  
We calculated the stellar mass from the K band luminosity  since  the light in this band is dominated by the emission of low-mass stars, which are responsible for the bulk of stellar mass in galaxies. 
From \lk\ we derived the stellar mass, $M_*$, by adopting a mass-to-luminosity ratio of M$_{\odot}$/ ${\it L_{\rm K,\odot}}$ = 1.32 \citep{2001MNRAS.326..255C} for the Salpeter Initial Mass Function (IMF), and applying a correction
factor of 0.5 \citep[from][]{2003ApJS..149..289B}  
to change to the \citet{2001MNRAS.322..231K}  IMF 
used in this paper.
The values for the individual galaxies are listed in Table \ref{Sample_FIR}.
The average sSFR for spiral galaxies in our sample is log(sSFR) =-10.61 $\pm$ 0.07 yr$^{-1}$.

\subsubsection{SFE and sSFR as a function of the deficiencies of the galaxies}\label{subsec:Def_SFE_SSFR}

\begin{table}
\caption{Mean log(sSFR) and log(SFE) as a function of  Def(${\it M_{\rm HI}}$) and Def(${\it M_{\rm H_{2}}}$) for spiral galaxies ($T\ge1$).
Mean values are calculated as explained in
Table \ref{Medias_2muestras}.}
{\small
\begin{centering}
\begin{tabular}{l|cc}
\hline\hline
&\multicolumn{2}{c}{log(sSFR)(yr$^{-1}$)}\\
&Mean &	n$_{UL}$/n		\\
\hline
Def(${\it M_{\rm HI}}$) $<$0.75			&-10.31$\pm$0.10&	(5/19)\\	
Def(${\it M_{\rm HI}}$) $>$0.75			&-10.85$\pm$0.13&	(6/16)\\

\hline
Def(${\it M_{\rm H_{2}}}$) $<$ -0.25		&-10.33$\pm$0.07&	(6/22)\\
Def(${\it M_{\rm H_{2}}}$) $>$ -0.25		&-10.81$\pm$0.12&	(8/22)\\
\hline
\end{tabular}
\\
\begin{tabular}{l|cccc}
&\multicolumn{2}{c}{log(SFE)(yr$^{-1}$)}\\
&Mean &	n$_{UL}$/n		\\
\hline
Def(${\it M_{\rm HI}}$) $<$0.75           &-9.08$\pm$0.07&        (5/19)\\
Def(${\it M_{\rm HI}}$) $>$0.75          &-9.16$\pm$0.12&        (6/16)\\        
                                                                                                          
\hline                                                                                                    
Def(${\it M_{\rm H_{2}}}$) $<$ -0.25     &-9.05$\pm$0.07&        (6/22)\\
Def(${\it M_{\rm H_{2}}}$) $>$ -0.25     &-9.04$\pm$0.13&        (8/22)\\
\hline
\end{tabular}

\end{centering}
}
{\footnotesize For each subsample, $n$ is the number of galaxies and $n_{UL}$ is the number of upper limits.
}
\label{sSFR_defs_medias}
\end{table}

\begin{figure}
\centering
\includegraphics[width=7.cm, angle=-90]{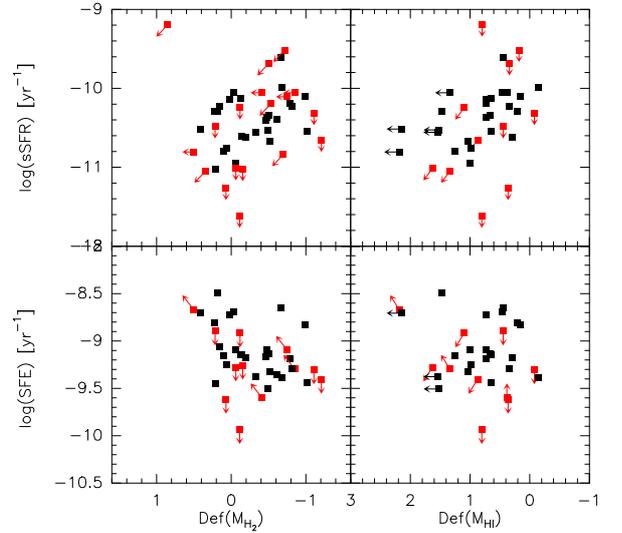}
\caption{Specific SFR (sSFR) (top) and star formation efficiency  (SFE) (bottom),  vs \mhtwo\  and  \mhi\  deficiencies of
spiral galaxies ($T\ge 1$) in HCGs. Red symbols denote upper
limits in \mhtwo\ or \lfir .}  
\label{SFE_SSFR_Def}
\end{figure}

In Fig. \ref{SFE_SSFR_Def} we display the SFE and the sSFR of the spiral galaxies in 
our sample as a function of their Def(${\it M_{\rm HI}}$) and Def(${\it M_{\rm H_{2}}}$). 
There is no  clear trend of the SFE with the gas deficiency of the galaxies, neither atomic nor molecular. This is confirmed by the mean values listed in 
Table~\ref{sSFR_defs_medias}. 
This result indicates that SF  proceeds with the same efficiency, independently of whether it occurs in a galaxy with a low or
high \mhi\ deficiency. 

On the other hand,  galaxies with a lower def(\mhtwo)  or def(\mhi) 
tend to have a higher sSFR (see Fig.\ref{SFE_SSFR_Def}, as well as   Table \ref{sSFR_defs_medias} for the quantitative trends). 
In particular, the trend with def(\mhi) is interesting as it suggests that, although the Def(${\it M_{\rm HI}}$) of a galaxy has no  influence 
on the absolute SFR or SFE, it has a noticeable effect on the SFR per stellar mass.


\subsection{Line Ratio}\label{res_lineratios}
\begin{figure}[h]
\centerline{
\includegraphics[width=6.5cm,angle=-90]{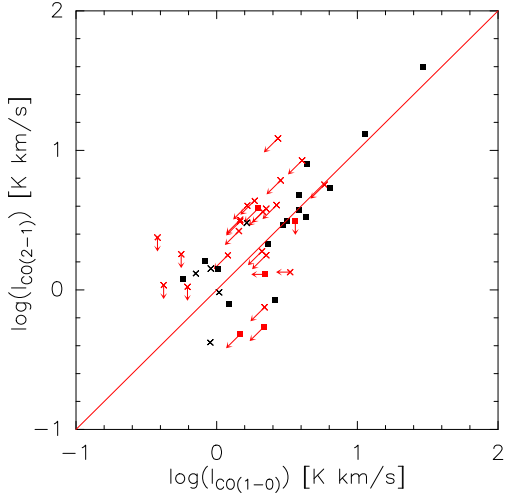}
}
\caption{log($I_{\rm CO(2-1)}$) versus log($I_{\rm CO(1-0)}$) for the galaxies observed by us. Spiral galaxies
(T$\ge$1) are shown as  filled squares and early-type (T$\le$0) as crosses.   Red symbols indicate upper limits in either
$I_{\rm CO(2-1)}$ or $I_{\rm CO(1-0)}$, and black symbols are detections.
The y=x line is plotted as reference and does not represent a fit to the data. }
\label{Comparacion_Ico}       
\end{figure}

Fig. \ref{Comparacion_Ico} shows the CO(1-0) versus the CO(2-1) intensity for the galaxies we observed (Sec. \ref{subsubs:COobs}). The plotted intensities are not aperture corrected. The mean ratio between both intensities is  $I_{\rm CO(2-1)}/I_{\rm CO(1-0)} =1.13\pm$0.11 for the full sample and  1.13$\pm$0.12 for spiral galaxies only.
To calculate this mean ratio with ASURV, we have taken into account galaxies with detections in both CO transitions as well as those detected only in 
CO(1-0). 
These values are slightly higher than those found by \citet{2009AJ....137.4670L} from  CO(2-1) and CO(1-0) maps for nearby galaxies from the SINGS sample
 ($I_{\rm CO(2-1)}/I_{\rm CO(1-0)} \sim$0.8)  and  than those from \citet{1993A&AS...97..887B} who obtained a mean 
 line ratio of $I_{\rm CO(2-1)}/I_{\rm CO(1-0)} = 0.89\pm0.06$ for a sample of nearby spiral galaxies. Both values are, in contrast to ours, corrected for beam-size effects.

In order to interpret the ratio of $I_{\rm CO(2-1)}/I_{\rm CO(1-0)}$ one has to consider two main parameters: the source distribution and the opacity. For optically thick, thermalized emission 
with a point-like distribution we expect a ratio $I_{\rm CO(2-1)}$/$I_{\rm CO(1-0)}$ = ($\theta_{CO(1-0)}$/$\theta_{CO(2-1)}$)$^{2}$ = 4 (with $I_{\rm CO}$ in $T_{\rm mb}$ and $\theta$ being the FWHM of the beams). On the other hand, for a uniform source brightness distribution we expect ratios larger than 1 for optically thin gas, and ratios  between about 0.6 and  1 for optically thick gas (with excitation temperatures above 5 K).

Due to the different beam sizes  of CO(1-0) and CO(2-1) in our observations we cannot distinguish these two cases.
However, we can conclude that our values  are  consistent with optically thick, thermalized gas with an extended distribution. Our mean value is slightly higher than the (beam-corrected) values  of \citet{2009AJ....137.4670L} and \citet{1993A&AS...97..887B} which might indicate that the molecular gas is not completely uniform over the CO(1-0) beam, but slightly concentrated towards the center.



\section{A possible evolutionary sequence of the molecular gas content and SFR in HCGs}\label{sec:evol_mh2}

In contrast to the HI content which can be highly deficient, the mean deficiencies for  
both \mhtwo\ and \lfir\  are low, close to the values found for isolated galaxies. 
In the case of \mhtwo\ we even find  indications for an 40-60\% excess
compared to isolated galaxies.
The difference in deficiency between the atomic and molecular gas is most likely due to the larger extent
of the HI gas, which can thus be  removed more efficiently from the galaxies, while the molecular gas, which is typically more 
concentrated in the inner regions, is presumably less affected by the environment. 
Subsequently, the lower HI mass might cause a lower ${\it M_{\rm H_{2}}}$ which leads to a lower SFR.
It is, however, remarkable that galaxies with a high HI deficiency can still contain a considerable amount of molecular gas and continue to form stars 
with a normal SFE.  This SF is not expected to last very long because once the molecular gas is used up, no HI is available to provide fuel for future SF.

Within this general picture of relative normality of \mhtwo\ and \lfir , we have 
found a relation between  def{\mhi) and the sSFR, and a tentative  trend
with def{\mhtwo) and def(\lfir).
%
Furthermore, there is a trend of 
 Def(${\it M_{\rm H_{2}}}$) with the evolutionary phase (Tab.~\ref{More_deficiencies}),
  in the sense that galaxies in HCGs belonging to phase 1 have the highest excess in ${\it M_{\rm H_{2}}}$. 
These trends might suggest that two mechanisms are at play. First, an increasing ${\it M_{\rm HI}}$ deficiency can be interpreted within a scenario
in which galaxies  in HCGs lose part of their HI as a result of mostly tidal stripping
during the initial  evolutionary phase, as suggested in the evolutionary model of \citet{2001A&A...377..812V}.
 On the other hand, in an early evolutionary phase the HI-to-H$_{\rm 2}$ conversion rate might be enhanced as a result 
 of the continuous interactions between galaxies, leading to the
enhancement in \mhtwo\ that we observe in evolutionary phase 1. This enhancement 
 of \mhtwo\  can not explain the high HI deficiencies observed in most galaxies, in agreement with the conclusions
 of  \citet{2008MNRAS.388.1245R},  but it could partly explain the lack of HI, 
 especially in the galaxies with the lowest HI deficiencies.
%

Based on our results, we thus suggest the following scenario which is speculative but compatible with our observations.
Galaxies in a HCG start with a normal content in \mhtwo\ and \mhi , i.e. they have 
Def(${\it M_{\rm HI}}$) = 0 and Def(${\it M_{\rm H_{2}}}$) = 0. 
Then, during the early evolutionary phase tidal interactions  enhance the conversion from atomic to molecular gas at the same time as they strip the HI from the galaxies, 
which leads to Def(${\it M_{\rm HI}}$) $>$ 0 and Def(${\it M_{\rm H_{2}}}$) $<$ 0. 
Finally, the multiple interactions within the group strip the main part of the HI in the 
disks, resulting in  Def(${\it M_{\rm HI}}$) $>>$ 0 and, as a consequence also  increase in   Def(${\it M_{\rm H_{2}}}$). 
This last effect could have contributed to  an increase of the fraction of lenticular
galaxies along the evolutionary sequence due to HI and H$_2$ stripping of spirals.



\section{Summary and Conclusions}\label{Conclusions}

We analyzed data for ${\it M_{\rm H_{2}}}$, obtained from observations with the IRAM 30m telescope and from the literature,
${\it L_{\rm FIR}}$ from IRAS and ${\it M_{\rm HI}}$ for 86 galaxies in 20 HCGs in order to study the relation between
atomic gas, molecular gas and SFR, traced by ${\it L_{\rm FIR}}$, in these galaxies.
We compared these properties to those
of isolated galaxies from the AMIGA project  \citep{2005A&A...436..443V}.
We adopted the same
CO-to-H$_2$ conversion factor for both samples.
The main conclusions of our study can be summarized as follows:

\begin{itemize}

\item The relation between ${\it M_{\rm H_{2}}}$, ${\it L_{\rm FIR}}$ and ${\it L_{\rm B}}$ in galaxies in HCGs is not significantly different from
the one found in isolated galaxies.  
The values of \lfir\ for spirals galaxies in HCGs are similar to those of the AMIGA galaxies for the same \lb .
For  ${\it M_{\rm H_{2}}}$ we find, however, a slight, but statistically significant, excess ($\sim$ 50\%) of HCGs spiral galaxies relative to AMIGA galaxies.
This could alternatively be explained by
a higher radial concentration of the molecular gas in HCG galaxies to
the center when compared with  isolated galaxies, so that the extrapolation of
the flux based on a similar extent (see Sec. 3.1.3) would lead
to an overestimate of \mhtwo\ for the group galaxies.
Another possible explanation  for this difference could be a systematically lower
CO-to-H$_2$ conversion factor for spirals in HCGs.

\item For elliptical and S0 galaxies the large number of upper limits do not allow strong conclusions about their \mhtwo\ or \lfir .
We note however that, while for S0s the \lb\ range is comparable to isolated S0 galaxies, 
Es in HCGs are up to half an order magnitude more luminous in \lb\ than isolated Es.

\item 
Practically all of our galaxies lie in the range of
\lfir/\mhtwo =1-10 \lsun/\msun, typical for normal, quiescent galaxies.
The deficiencies in ${\it M_{\rm H_{2}}}$ and ${\it L_{\rm FIR}}$ are  tightly correlated
and span about the same range as in isolated galaxies.

\item   The ${\it M_{\rm HI}}$ deficiency, calculated from the VLA data for individual galaxies,
 is much larger than the other deficiencies  with a mean value of 0.93$\pm$0.13 
(12\% of the expected value) for spiral galaxies,
and 1.31$\pm$0.11 (5\% the expected value) for all morphological types, 
   and represents the largest difference with respect to isolated galaxies.
Those values are significantly larger than those obtained in \citet{2001A&A...377..812V}  since the sample
in that study  was biased towards HI bright galaxies  while here we present a redshift selected sample.

\item The SFE of the spiral galaxies in the HCGs 
is slightly lower than in isolated galaxies, but in the range of values found for
nearby spiral galaxies (Bigiel et al. 2011). 
We have found no relation of the SFE with neither Def(${\it M_{\rm HI}}$) nor Def(${\it M_{\rm H_{2}}}$).

\item There is a trend of the sSFR to increase with decreasing Def(${\it M_{\rm HI}}$) and  Def(${\it M_{\rm H_{2}}}$).
This suggests that, although the Def(\mhi)
of a galaxy has only a weak influence on the absolute SFR, it has
a stronger influence on the SFR per stellar mass.



\item There is a trend of decreasing molecular gas deficiency with evolutionary phase,
with galaxies in groups in an early phase showing an excess in \mhtwo . This excess
goes to 0 in later phases.
A similar trend might exist with def(\mhi), but is statistically only marginally significant
in our sample.
%
This is interpreted as an initial
enhancement of the conversion from atomic to molecular gas due to on-going tidal interactions, later followed by
stripping of most of their HI. In these later phases, evolution of spiral to lenticular galaxies, would both explain the overabundance of those morphological types as well as the 
\mhi\  deficiency and decrease in  \mhtwo\ content of the galaxies.

\item No trend with the global HI deficiency of the  groups is found, which suggest that the molecular gas content and
SF are more driven by one-to-one interaction than directly by the local environment.

\end{itemize}

\begin{acknowledgements}
This work has
been supported by the research projects  AYA2008-06181-C02 and
 AYA2007-67625-C02-02  from the Spanish Ministerio de Ciencia y
Educaci\'on and the Junta
de Andaluc\'\i a (Spain) grants P08-FQM-4205, FQM-0108 and TIC-114.
DE was supported by a Marie Curie International Fellowship within the 6th European Community Framework Programme (MOIF-CT-2006-40298).
UL  warmly thanks IPAC (Caltech),  where this work was finished during a sabbatical stay,  for their hospitality.
We also thank T. Bitsakis and V. Charmandaris for letting us use their Spitzer data prior to publication,
and the anonymous referee for critical comments helping to put our conclusions on a firmer ground and
improving the quality of the paper.
This work is   based on observations with the  Instituto de Radioastronomia Milim\'etrica IRAM 30m and the
  Five College Radio Astronomy (FCRAO) 14m. The FCRAO is supported by NSF grant AST 0838222.
This research has made use of the NASA/IPAC Extragalactic Database (NED) which is operated by the Jet Propulsion Laboratory, California Institute of Technology, under contract with the National Aeronautics and Space Administration. We also acknowledge the usage of the HyperLeda database (http://leda.univ-lyon1.fr).
\end{acknowledgements}

\bibliography{Biblio_Hickson.bib} 
\bibliographystyle{aa}

\appendix

\section{CO spectra}

Figure~\ref{spec-co10}
shows the CO(1-0) spectra of the detections and tentative detections observed by us
and Figure~\ref{spec-co10} the CO(2-1) spectra.

\begin{figure*}[h!]
\centerline{
\psfig{file=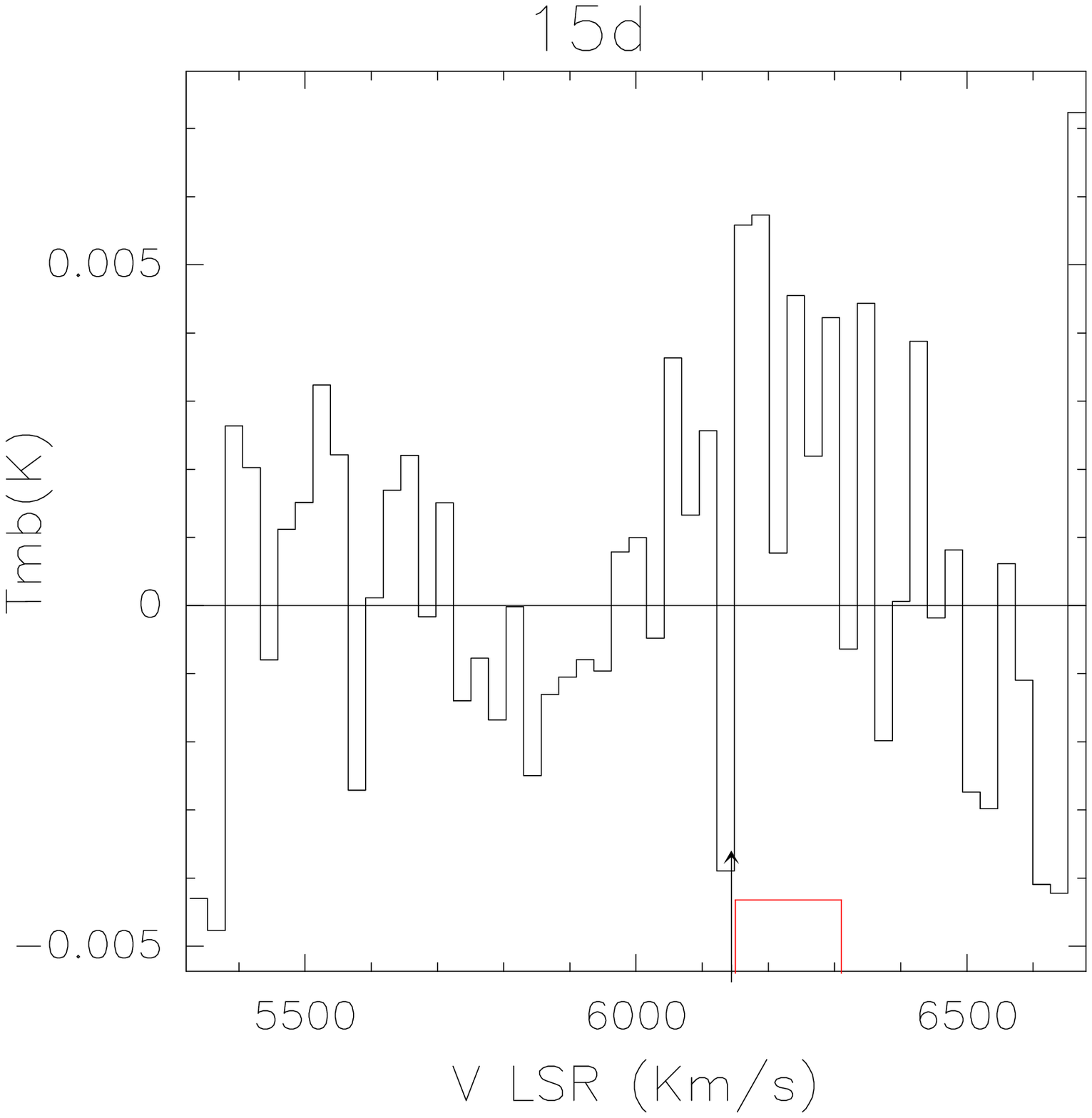,width=3.cm}
\quad
\psfig{file=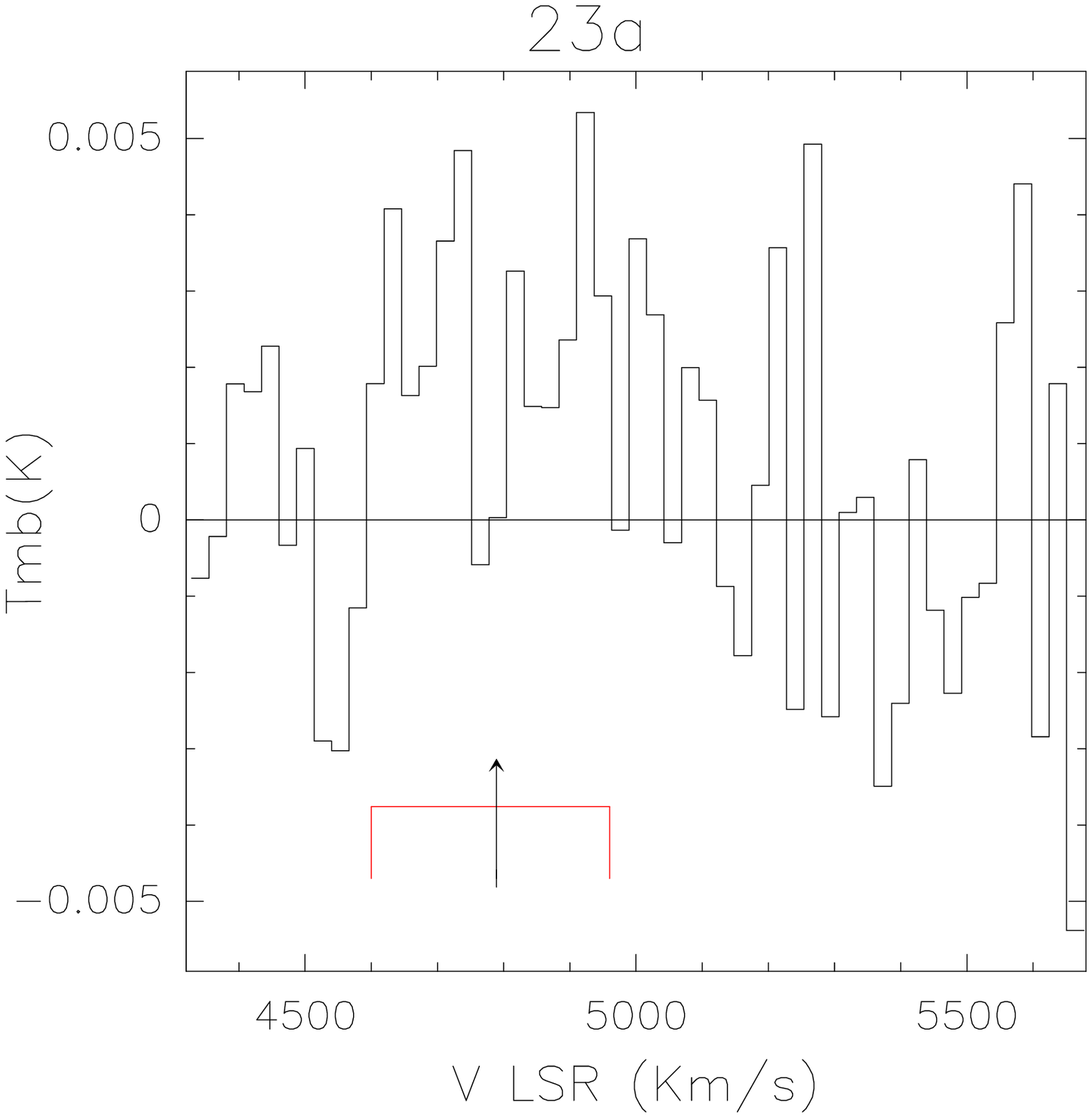,width=3.cm}
\quad
\psfig{file=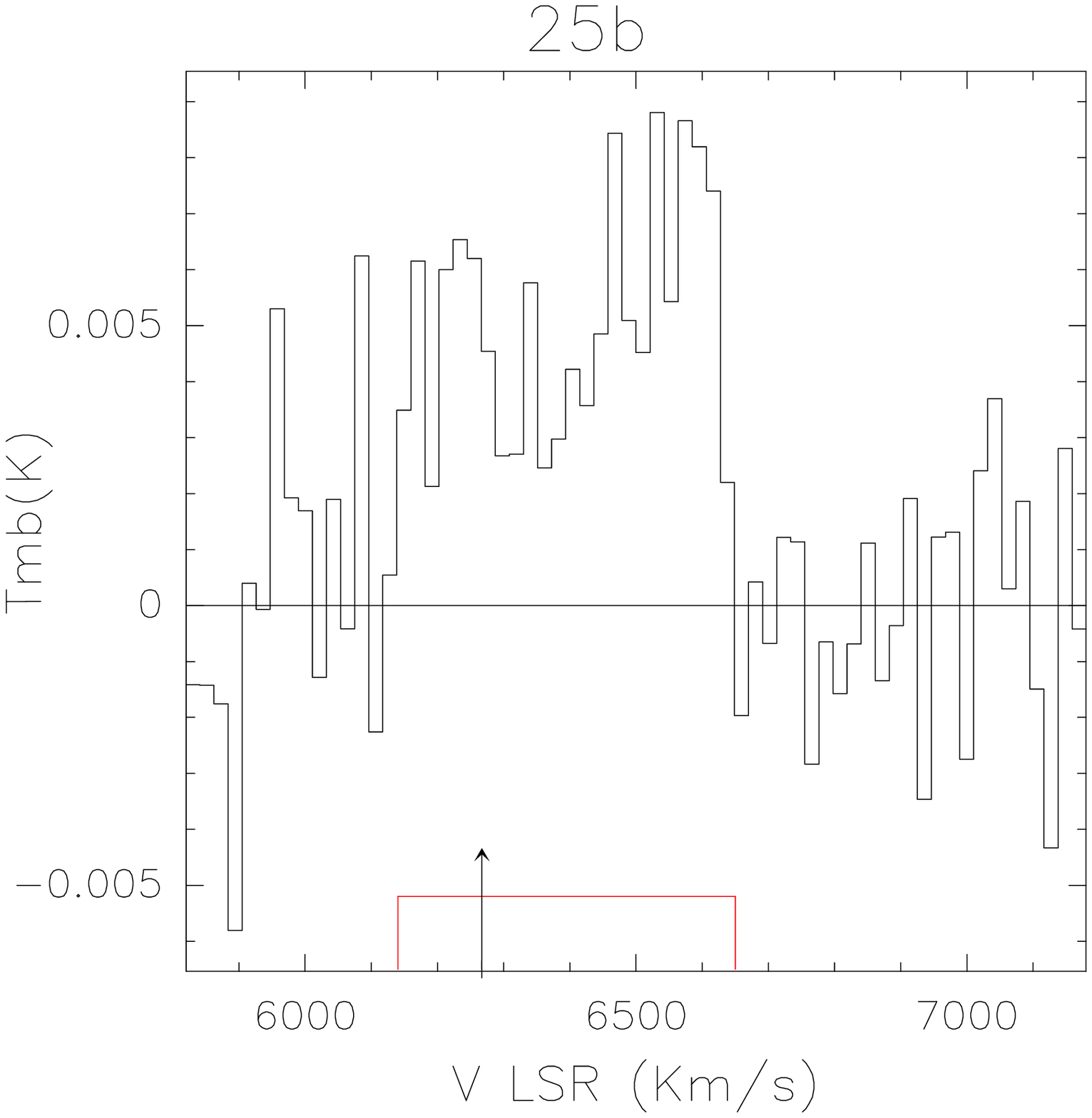,width=3.cm}
\quad
\psfig{file=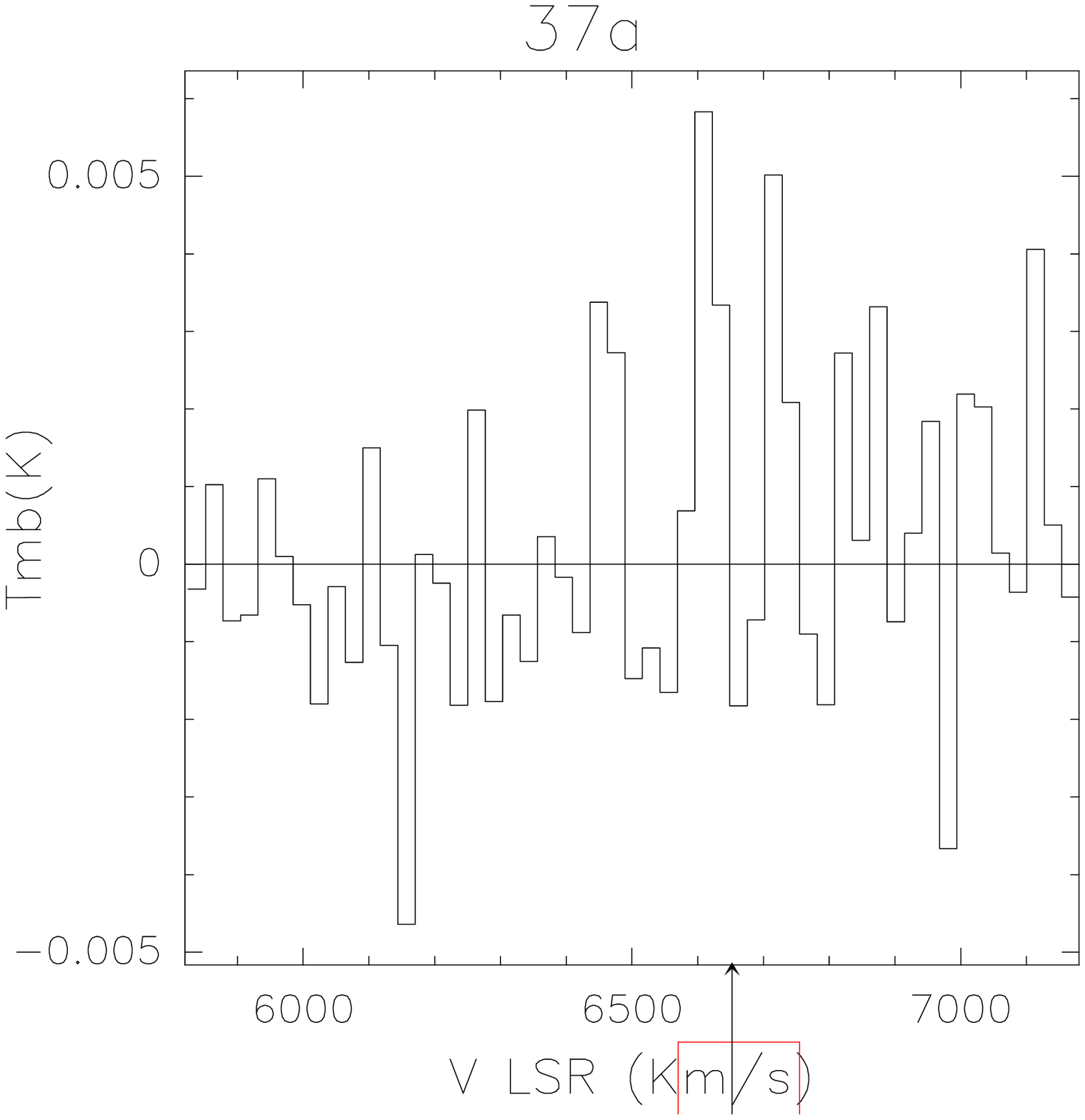,width=3.cm}
\quad
\psfig{file=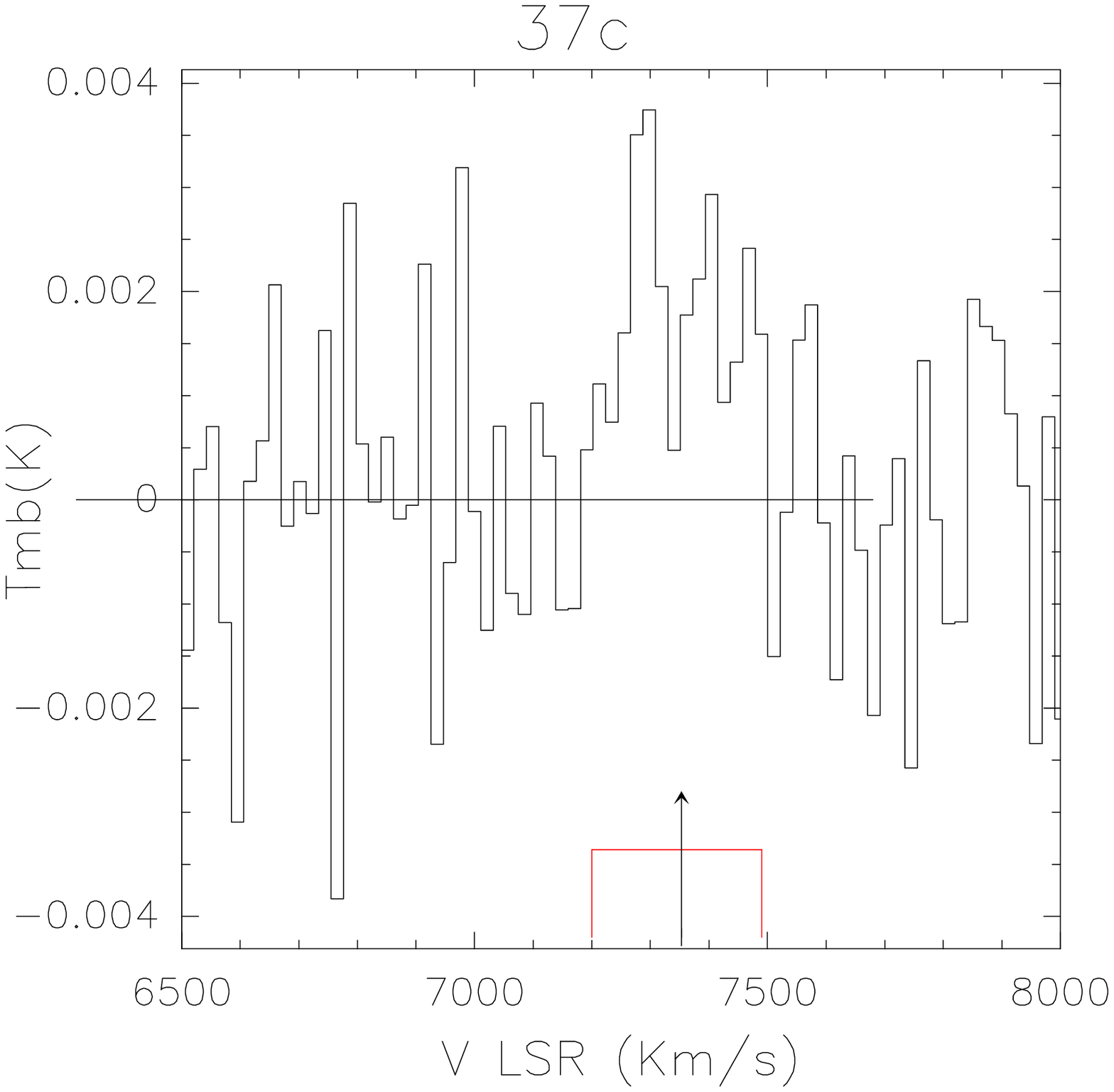,width=3.cm}
}

\centerline{
\psfig{file=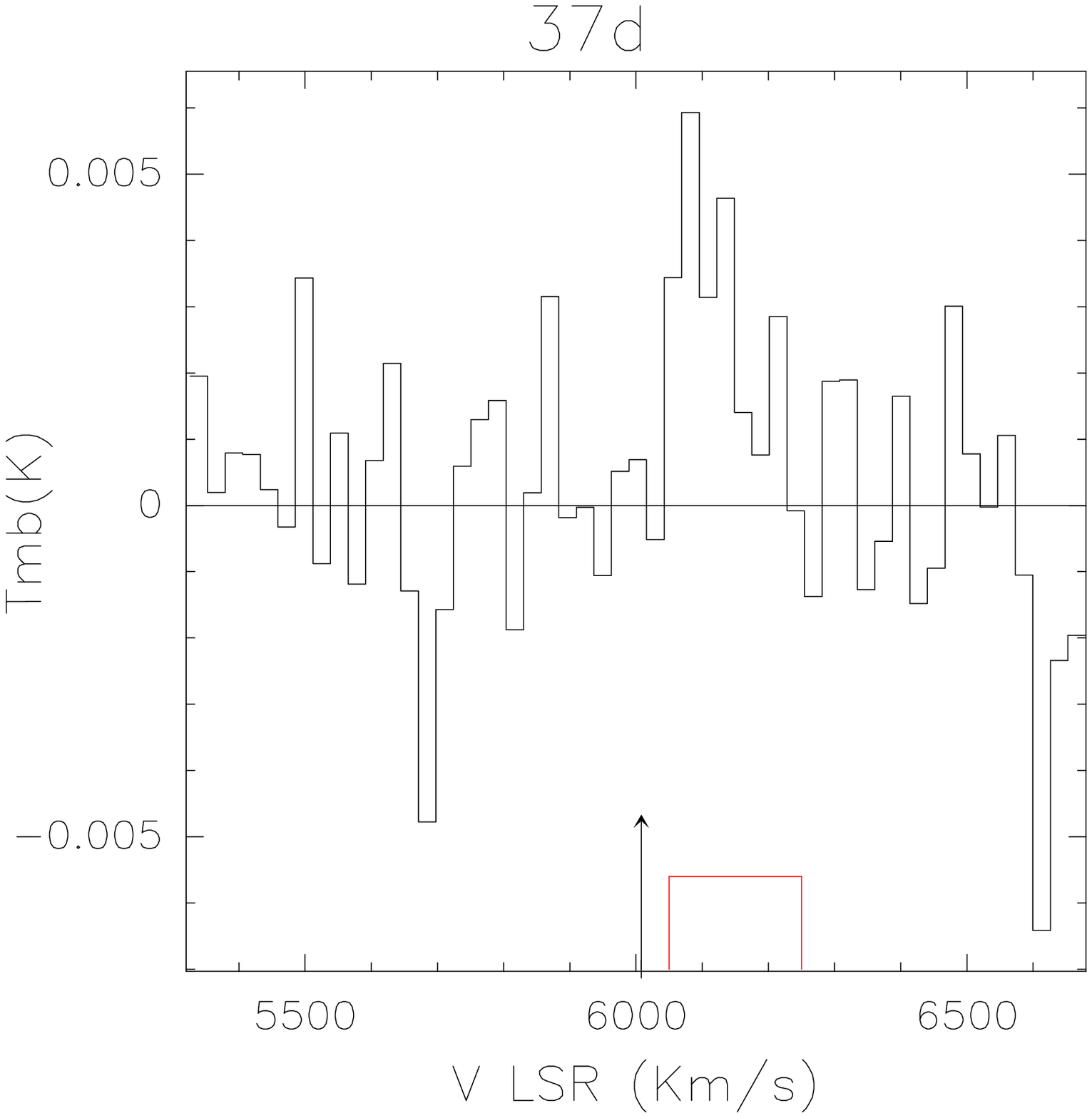,width=3.cm}
\quad
\psfig{file=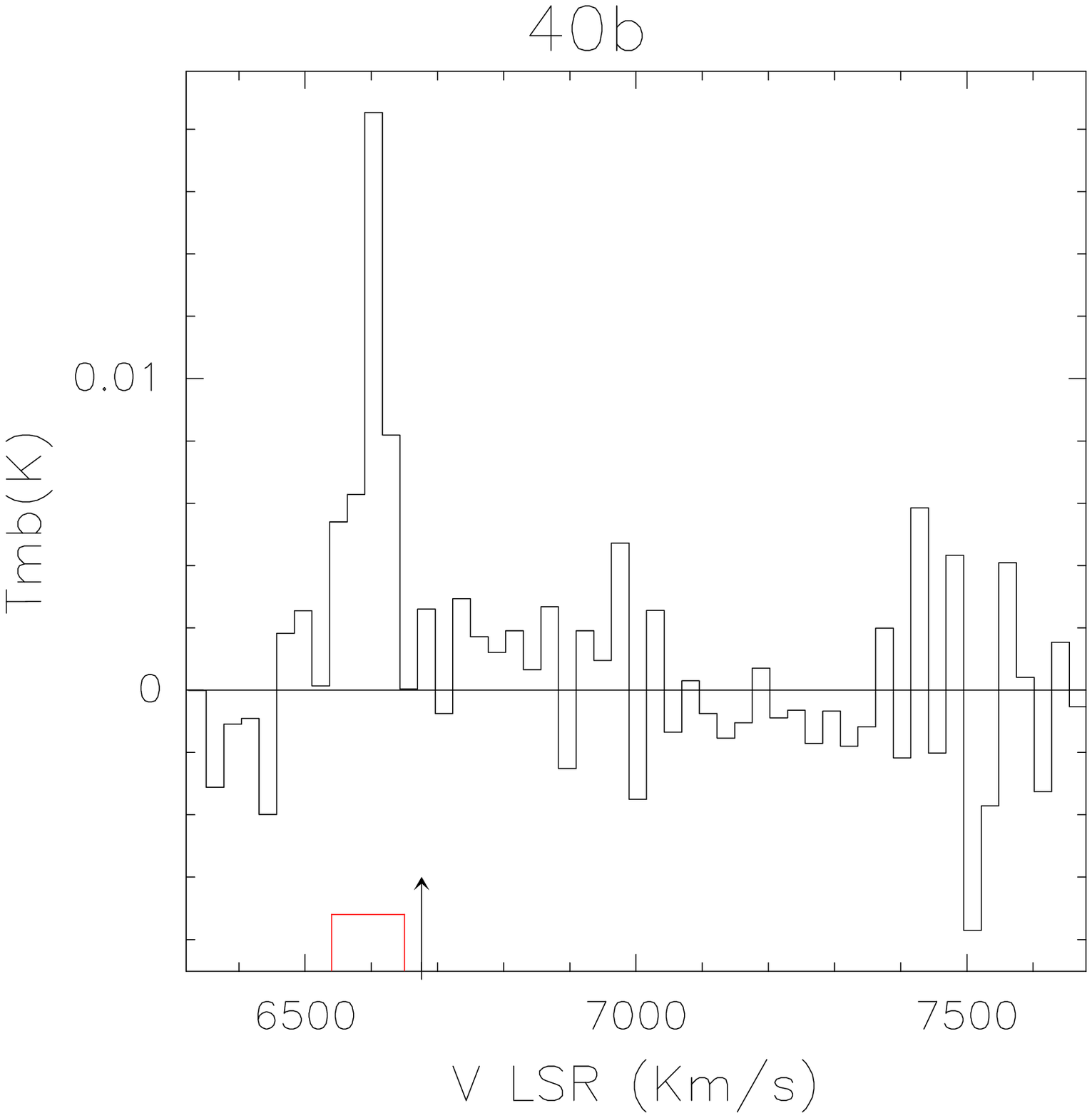,width=3.cm}
\quad
\psfig{file=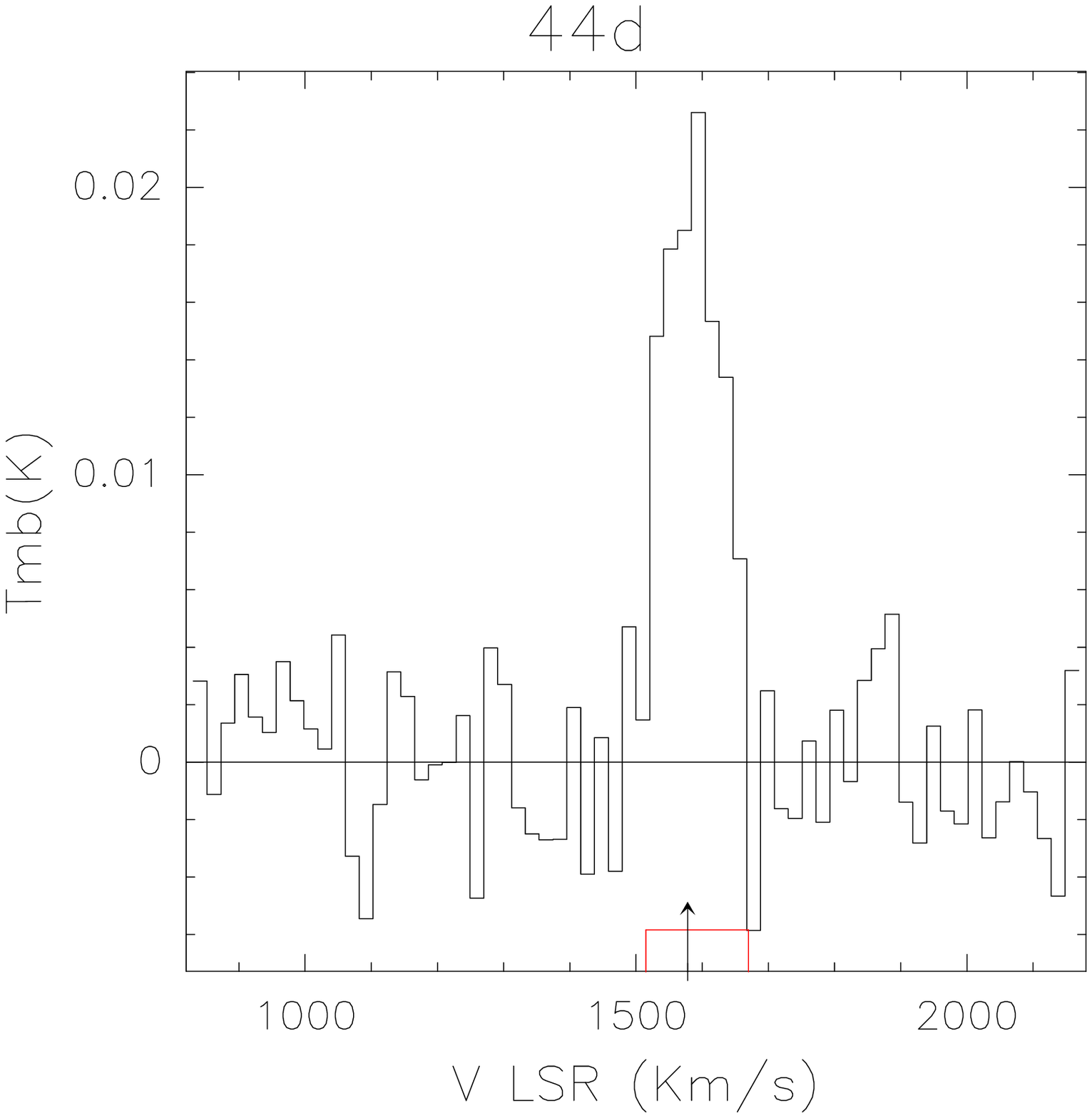,width=3.cm}
\quad
\psfig{file=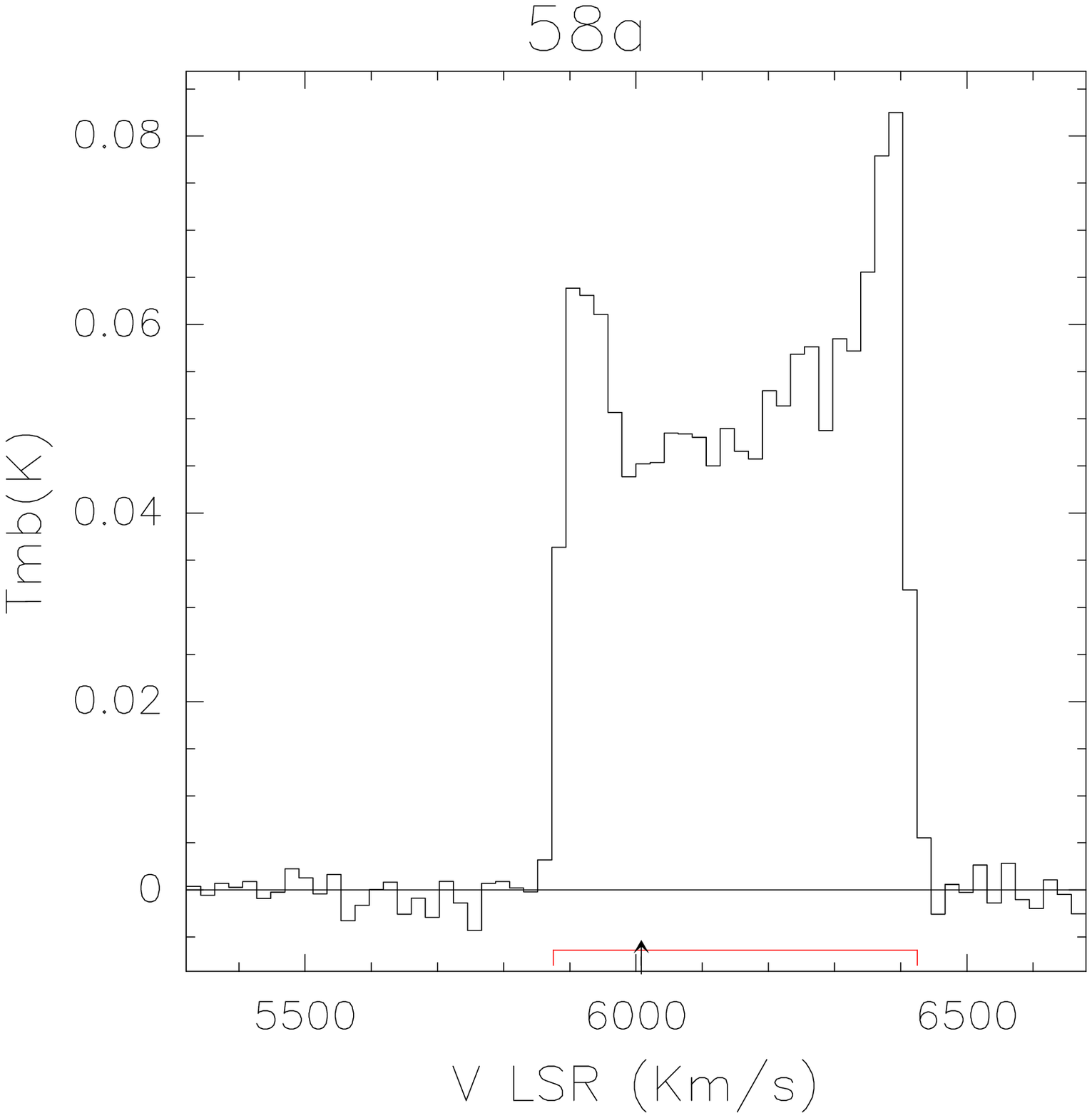,width=3.cm}
\quad
\psfig{file=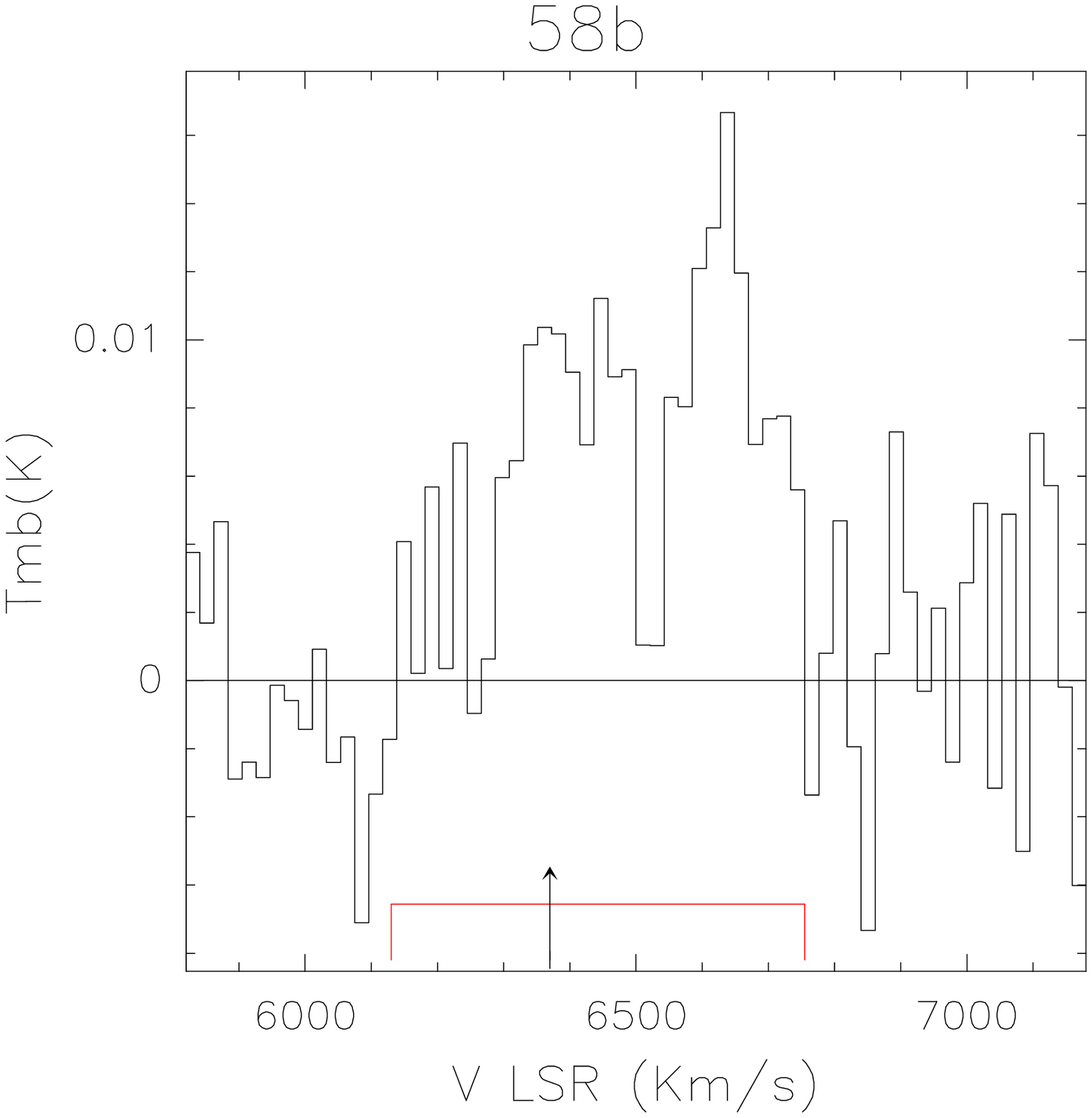,width=3.cm}
\quad
}
\centerline{
\psfig{file=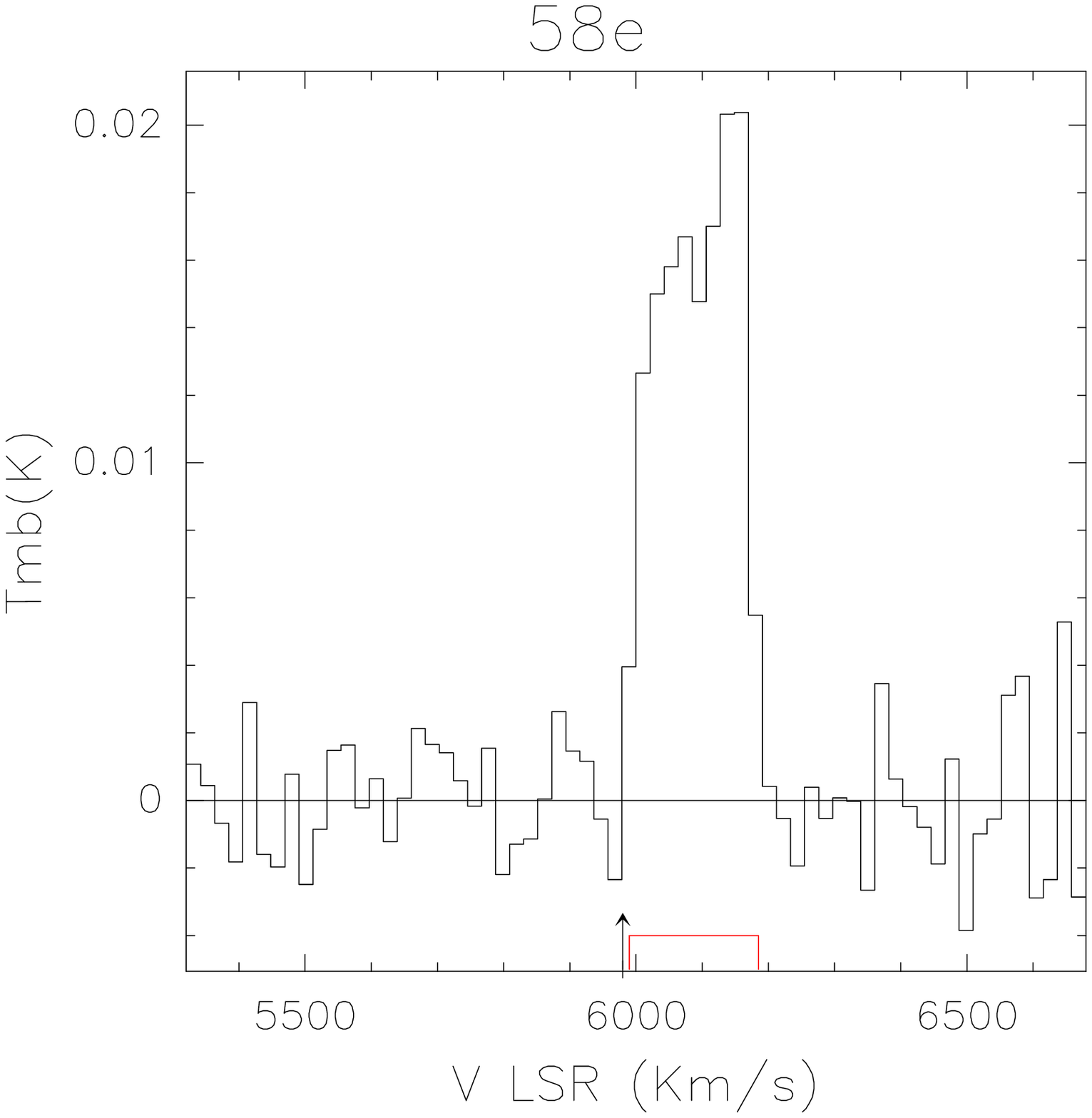,width=3.cm}
\quad
\psfig{file=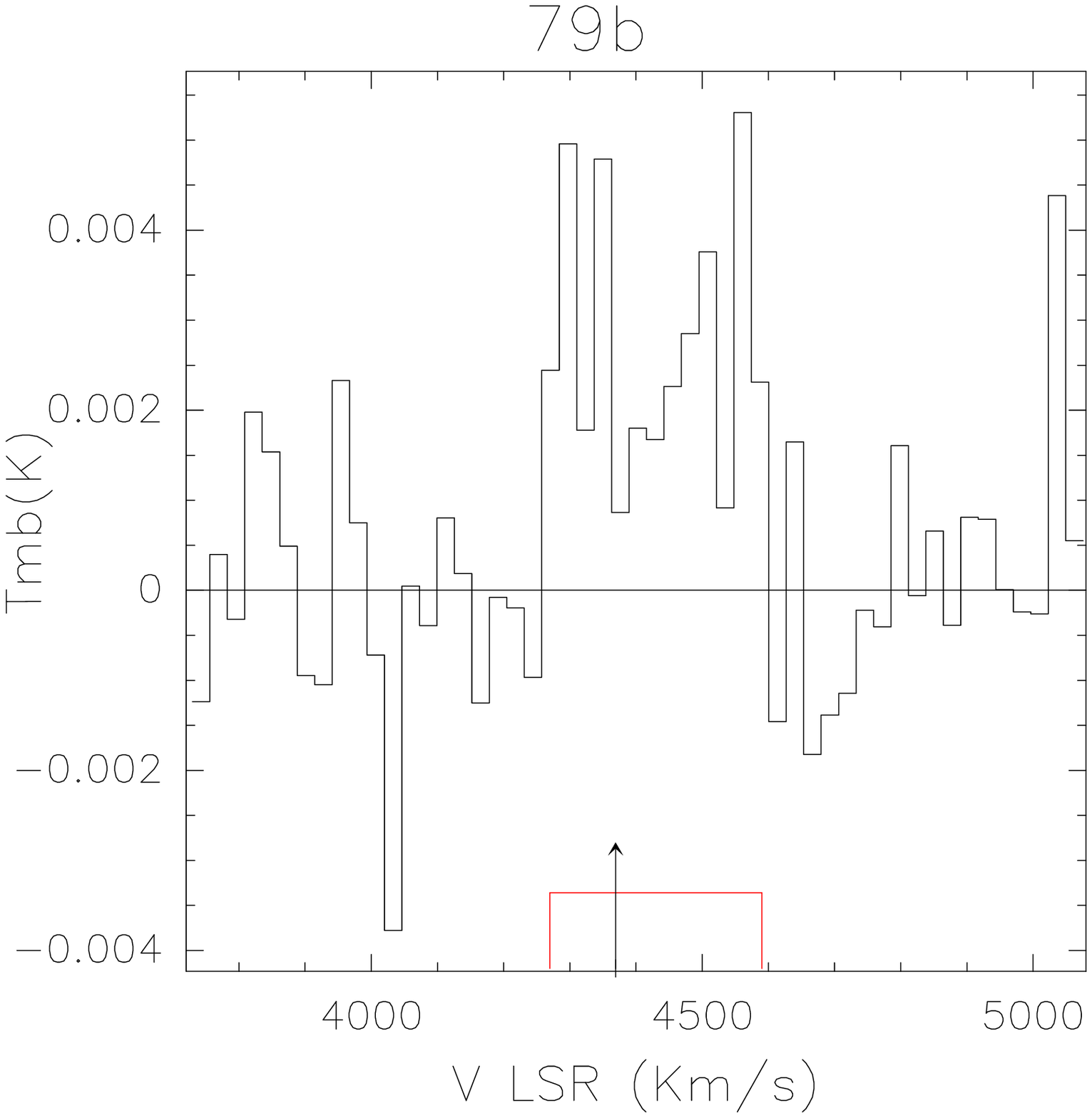,width=3.cm}
\quad
\psfig{file=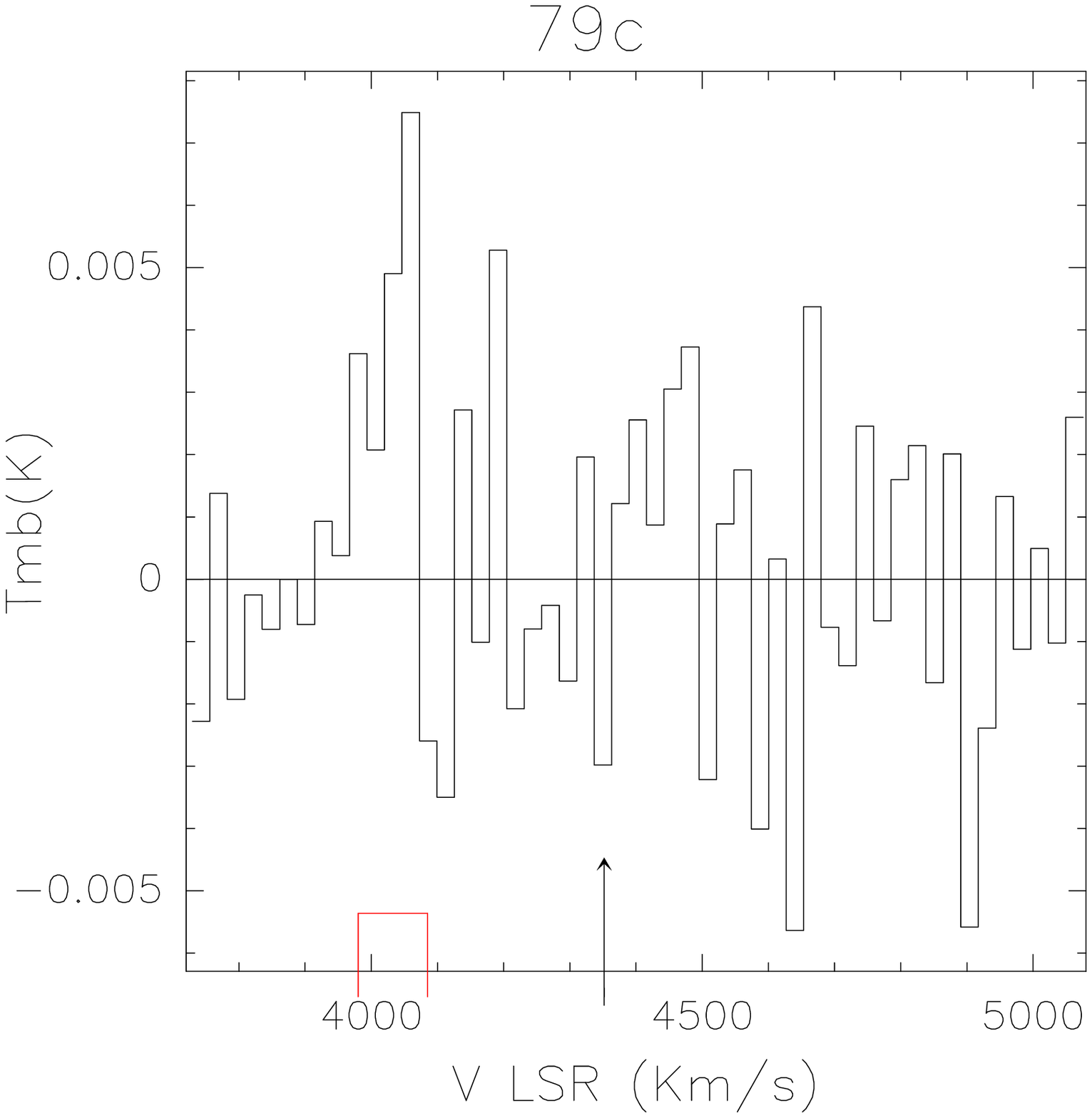,width=3.cm}
\quad
\psfig{file=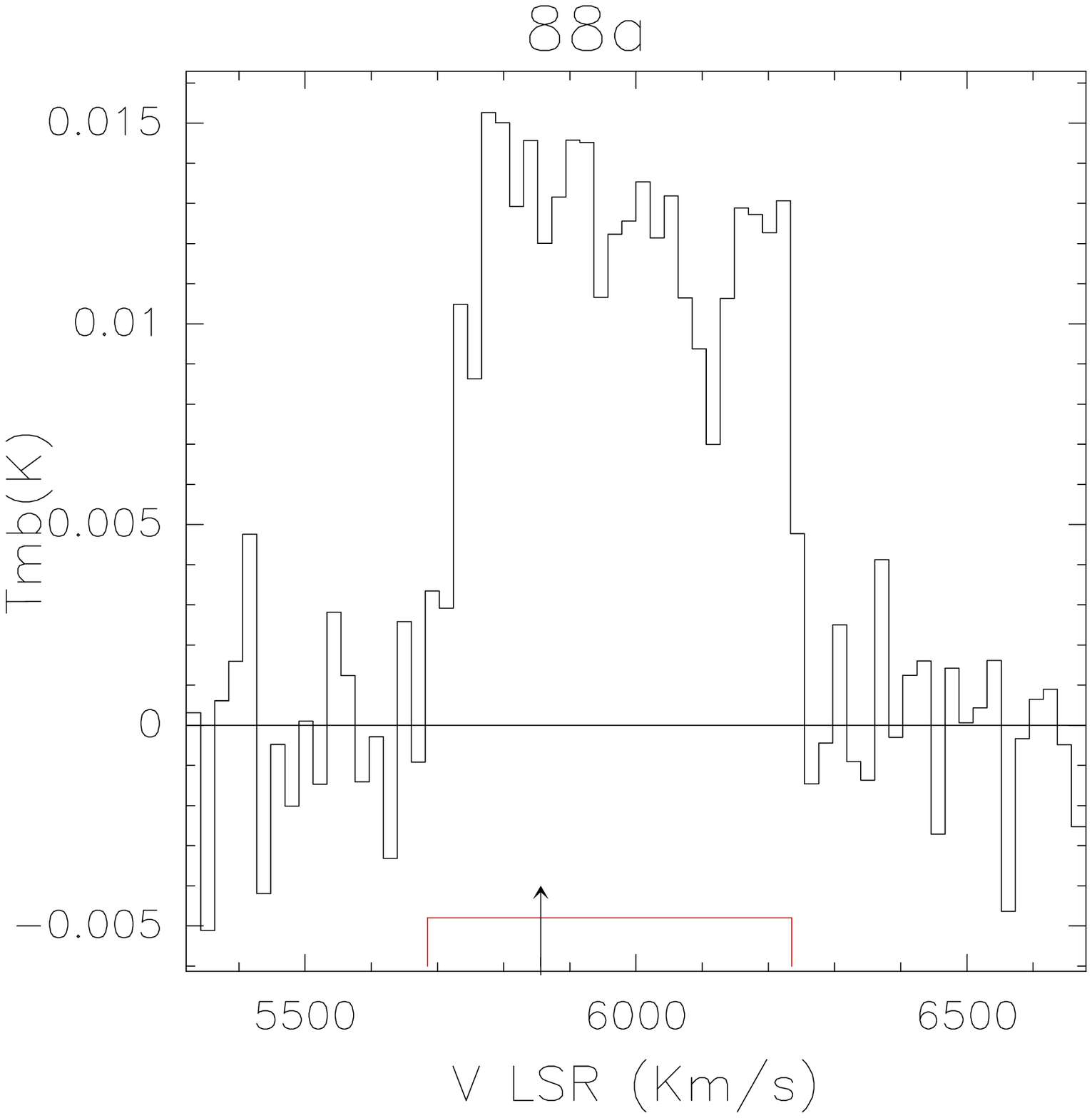,width=3.cm}
\quad
\psfig{file=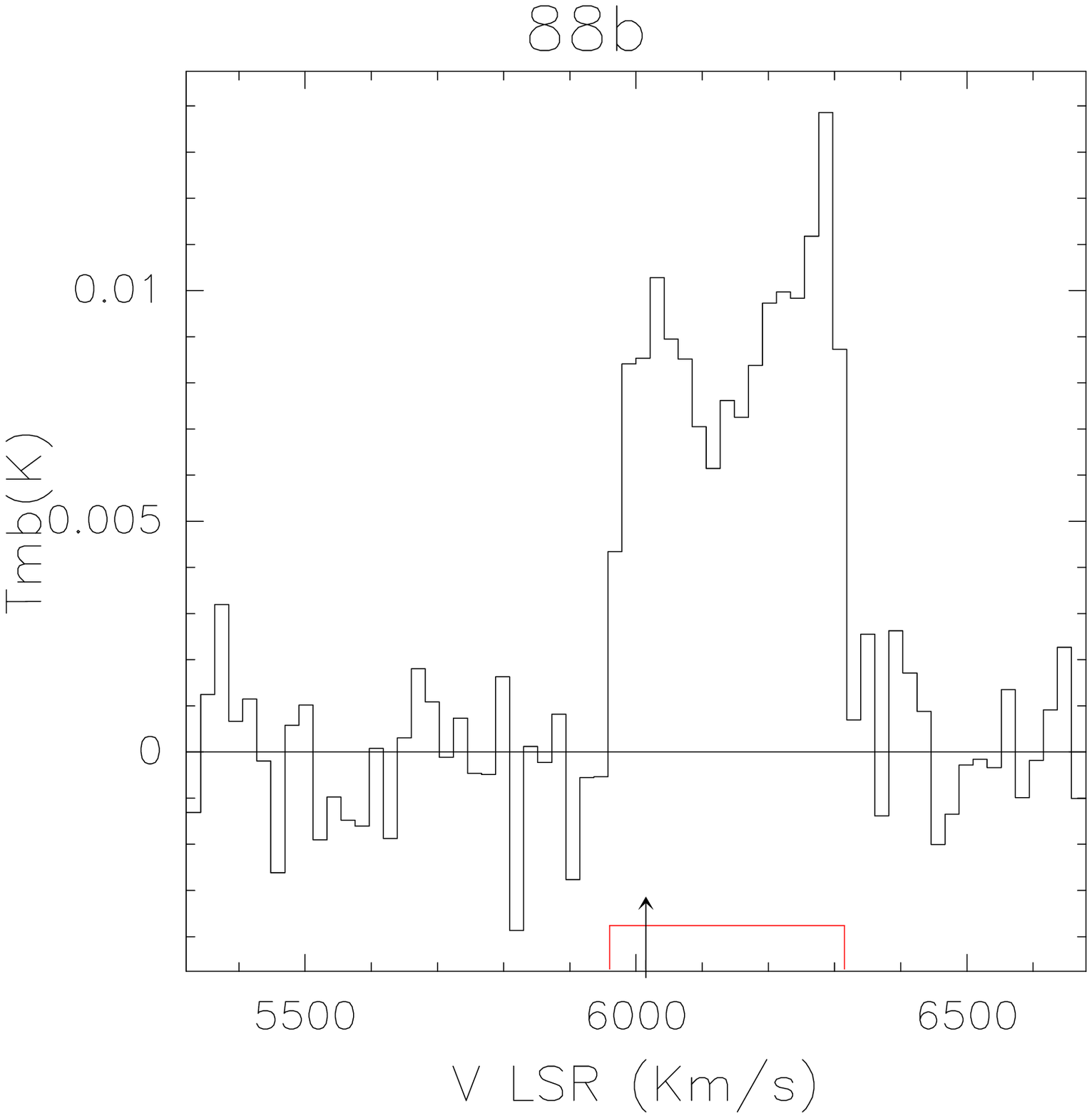,width=3.cm}
}

\quad
\centerline{
\psfig{file=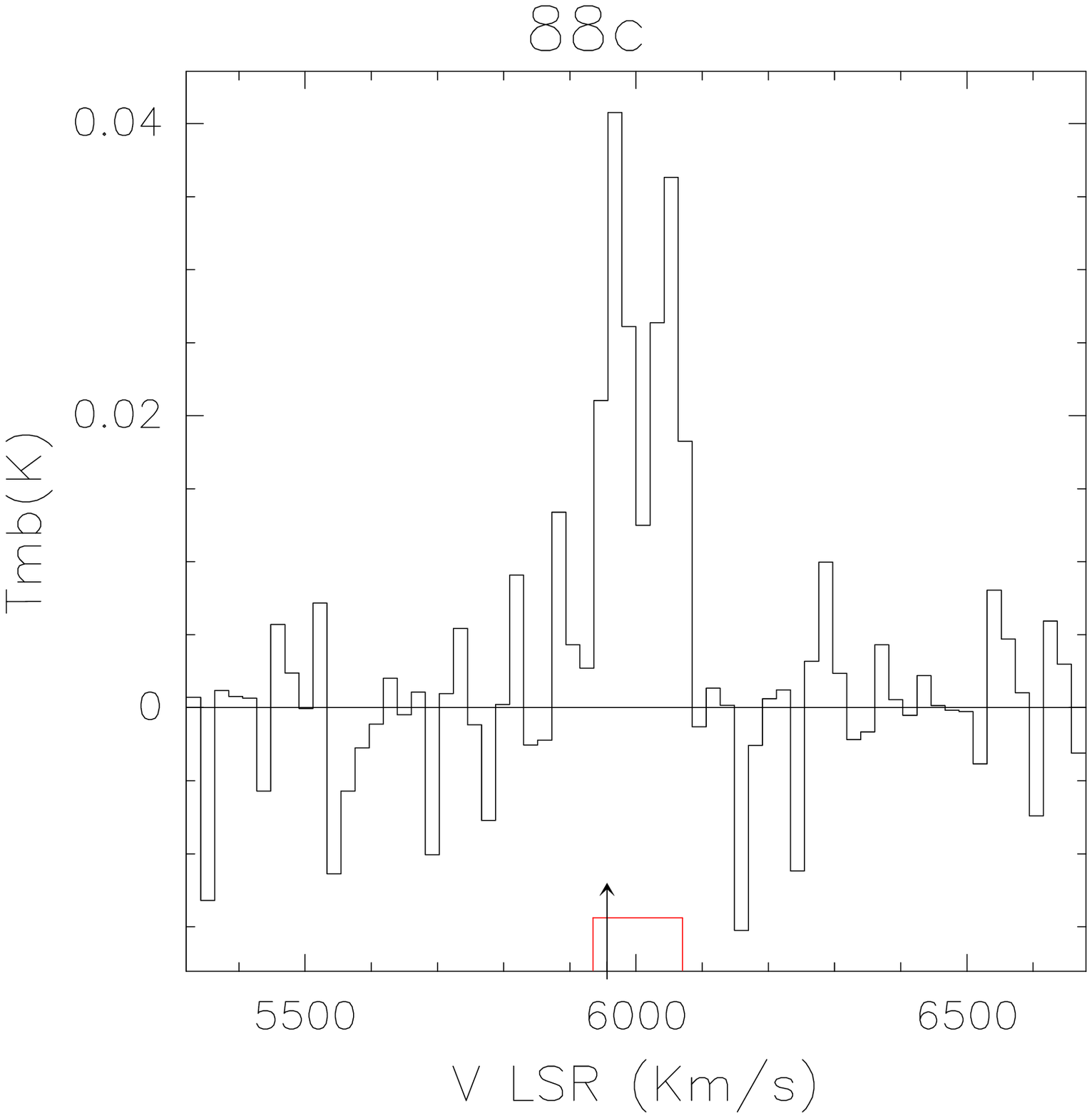,width=3.cm}
\quad
\psfig{file=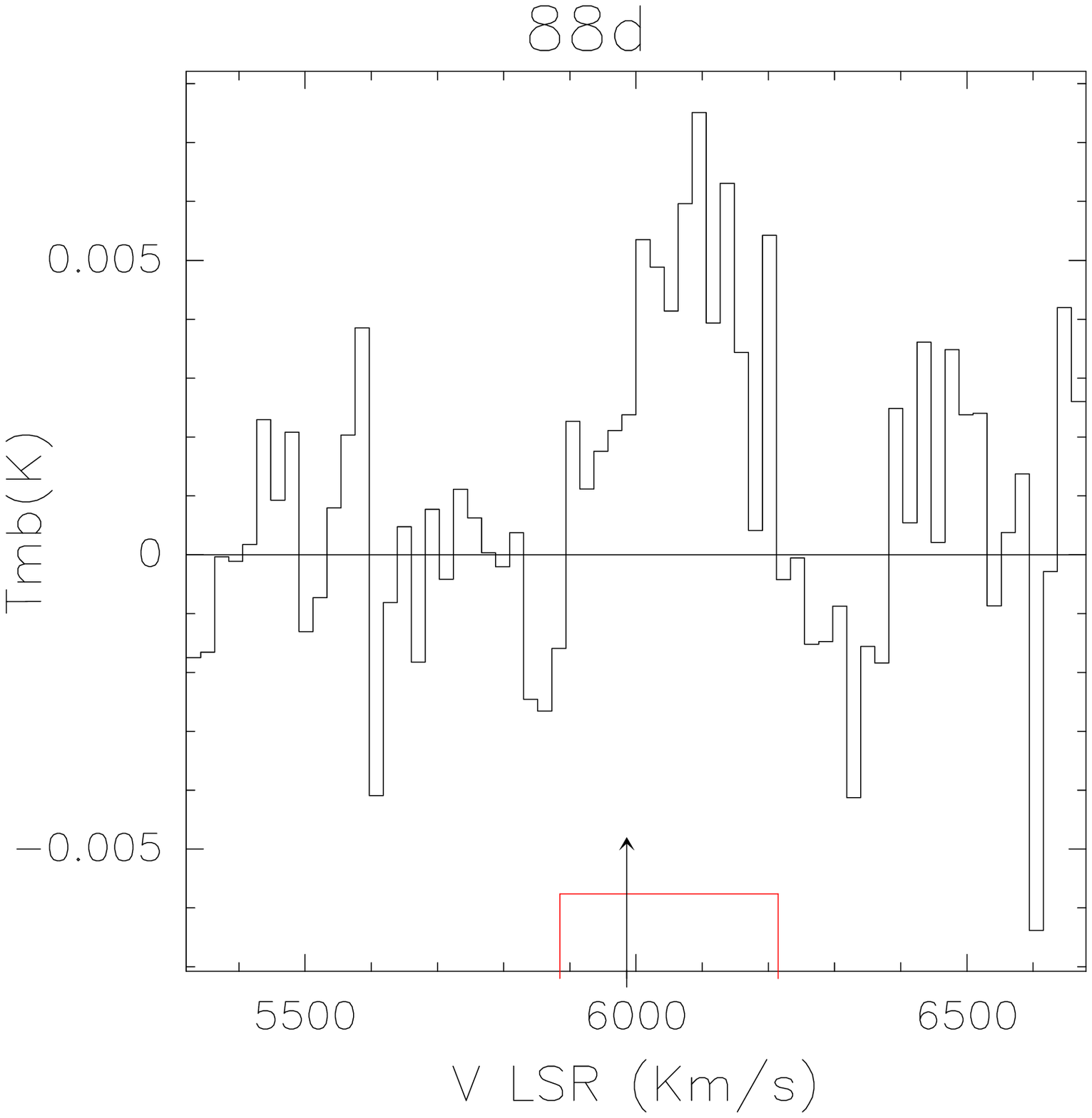,width=3.cm}
\quad
\psfig{file=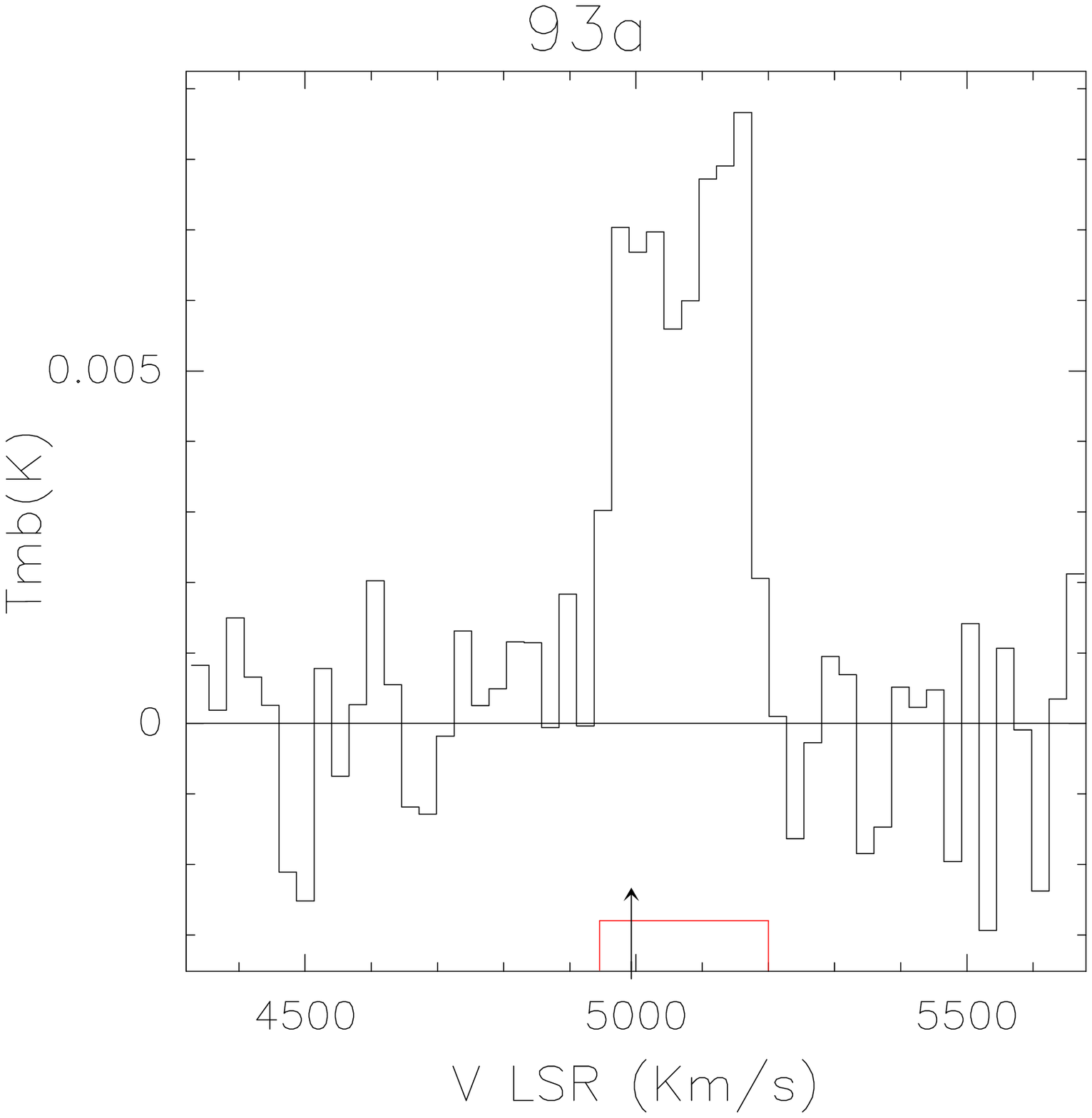,width=3.cm}
\quad
\psfig{file=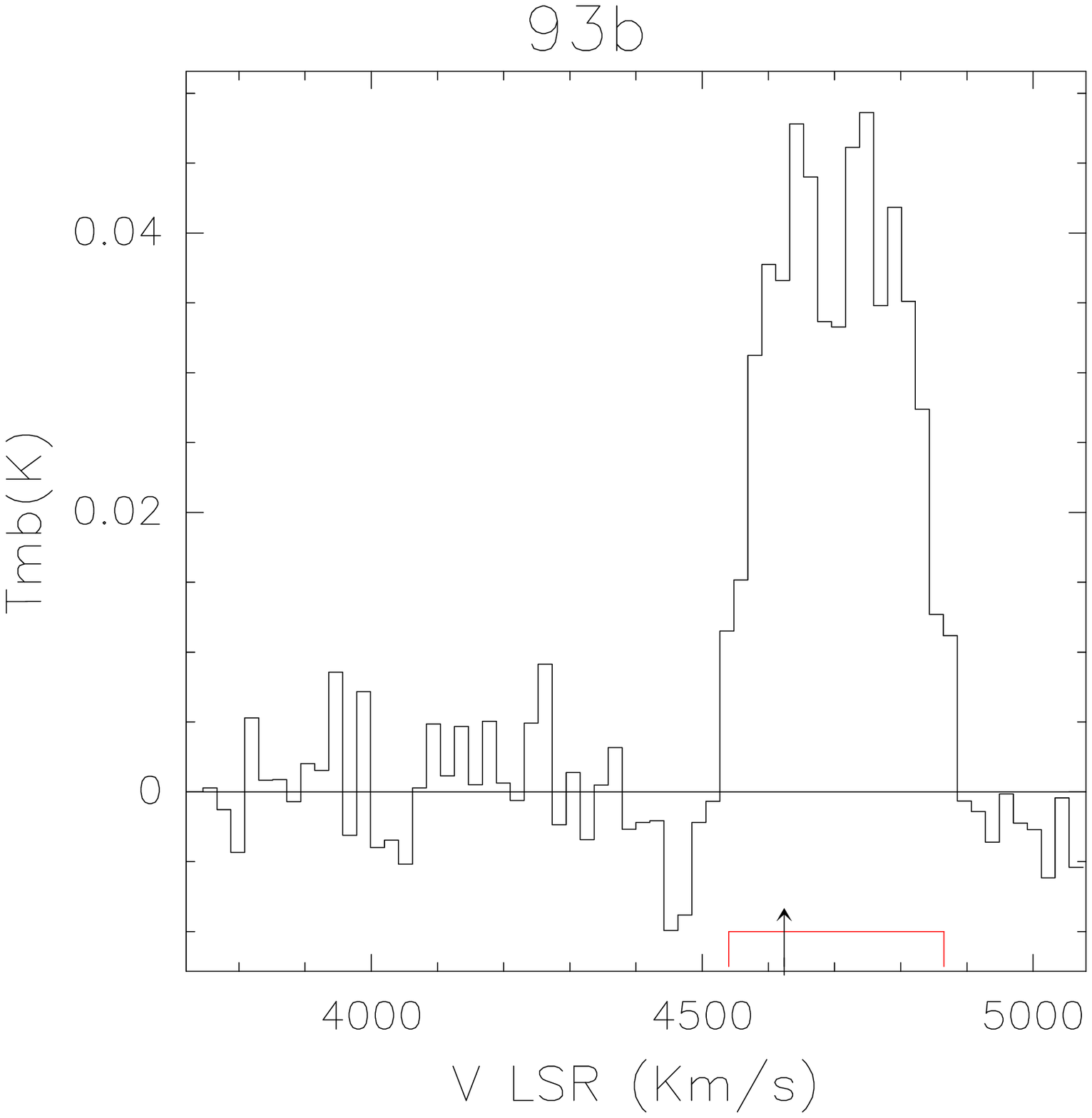,width=3.cm}
\quad
\psfig{file=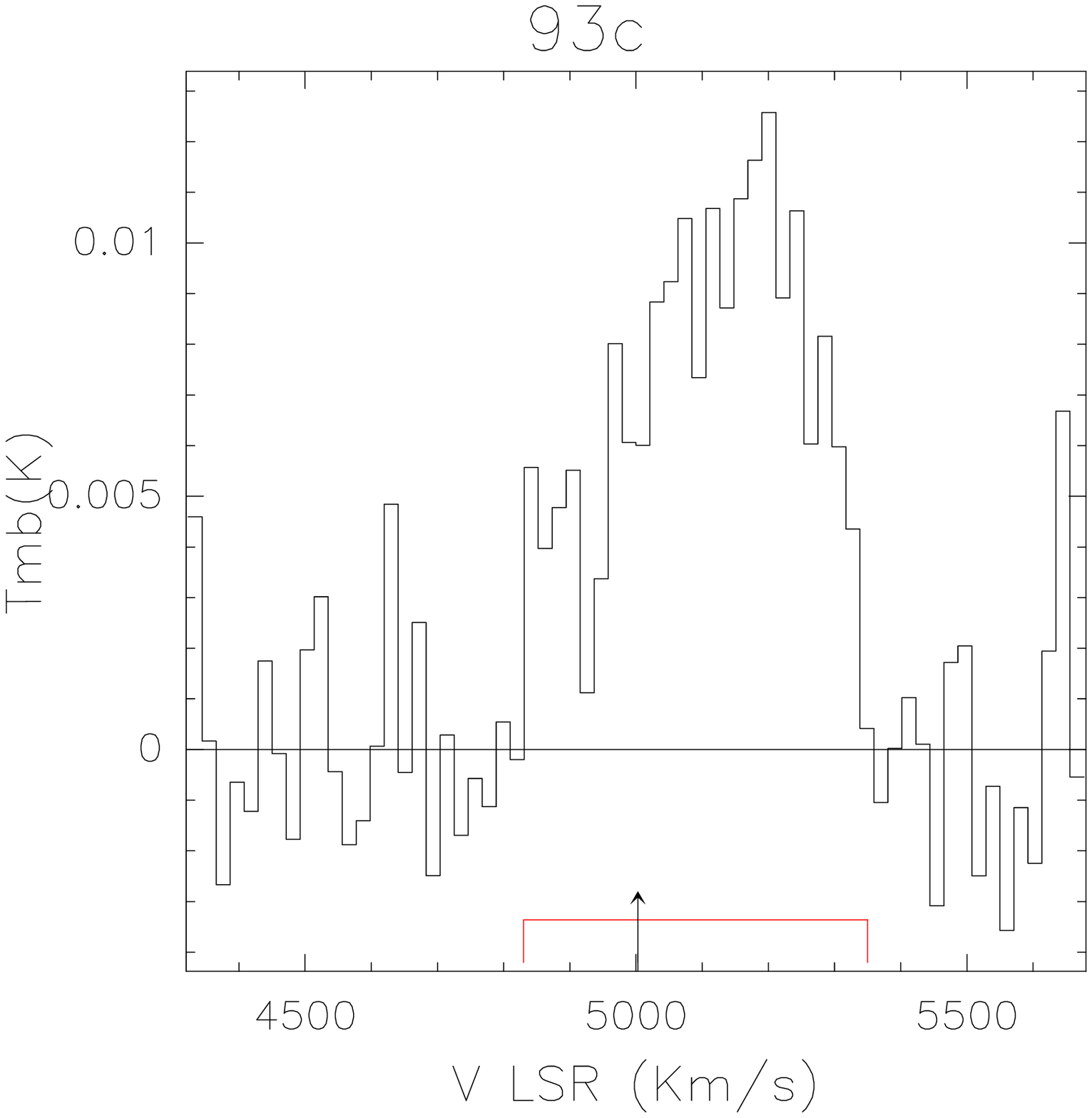,width=3.cm}
\quad
\quad
\quad
}
\quad
\centerline{
\psfig{file=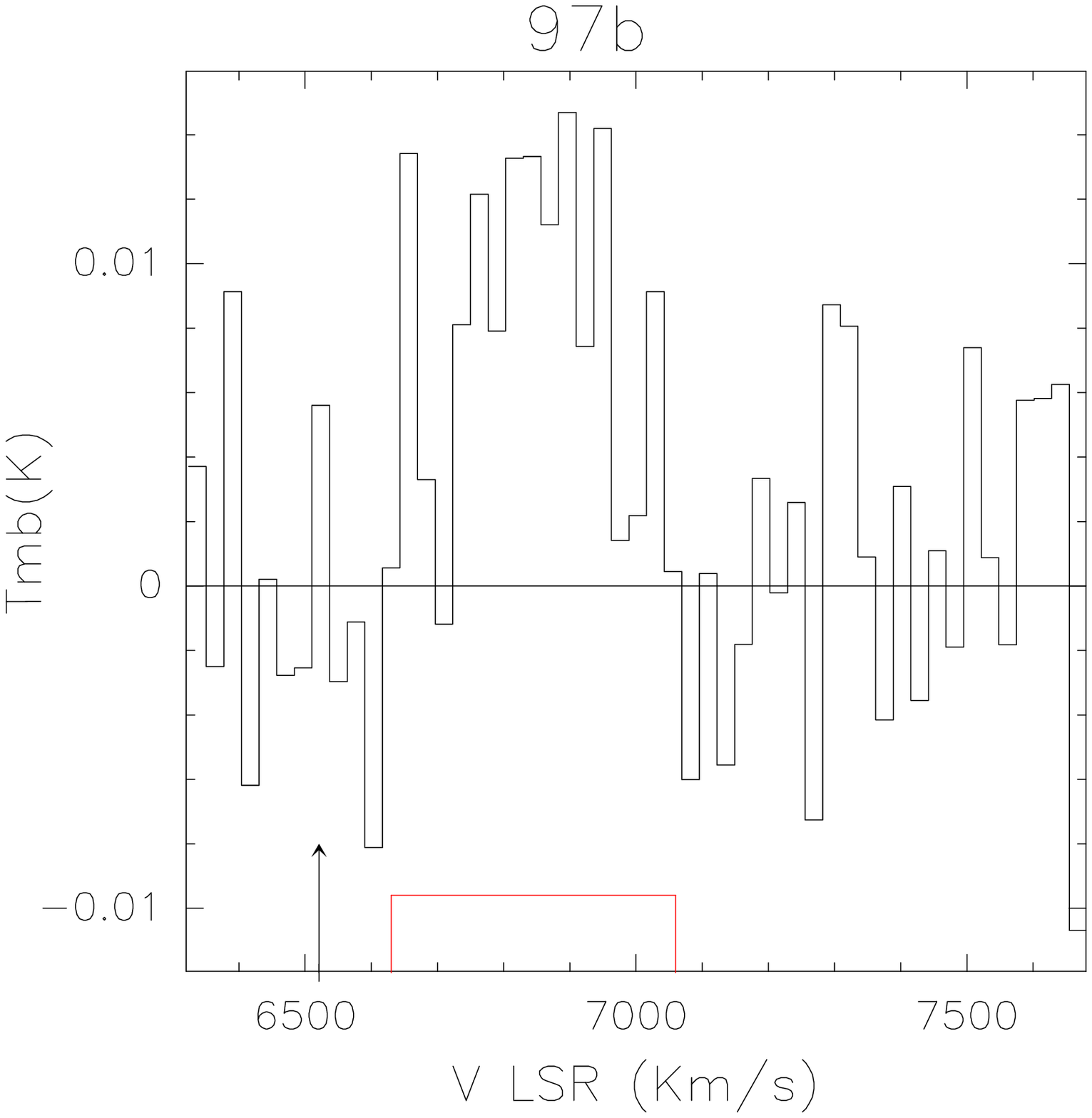,width=3.cm}
\quad
\psfig{file=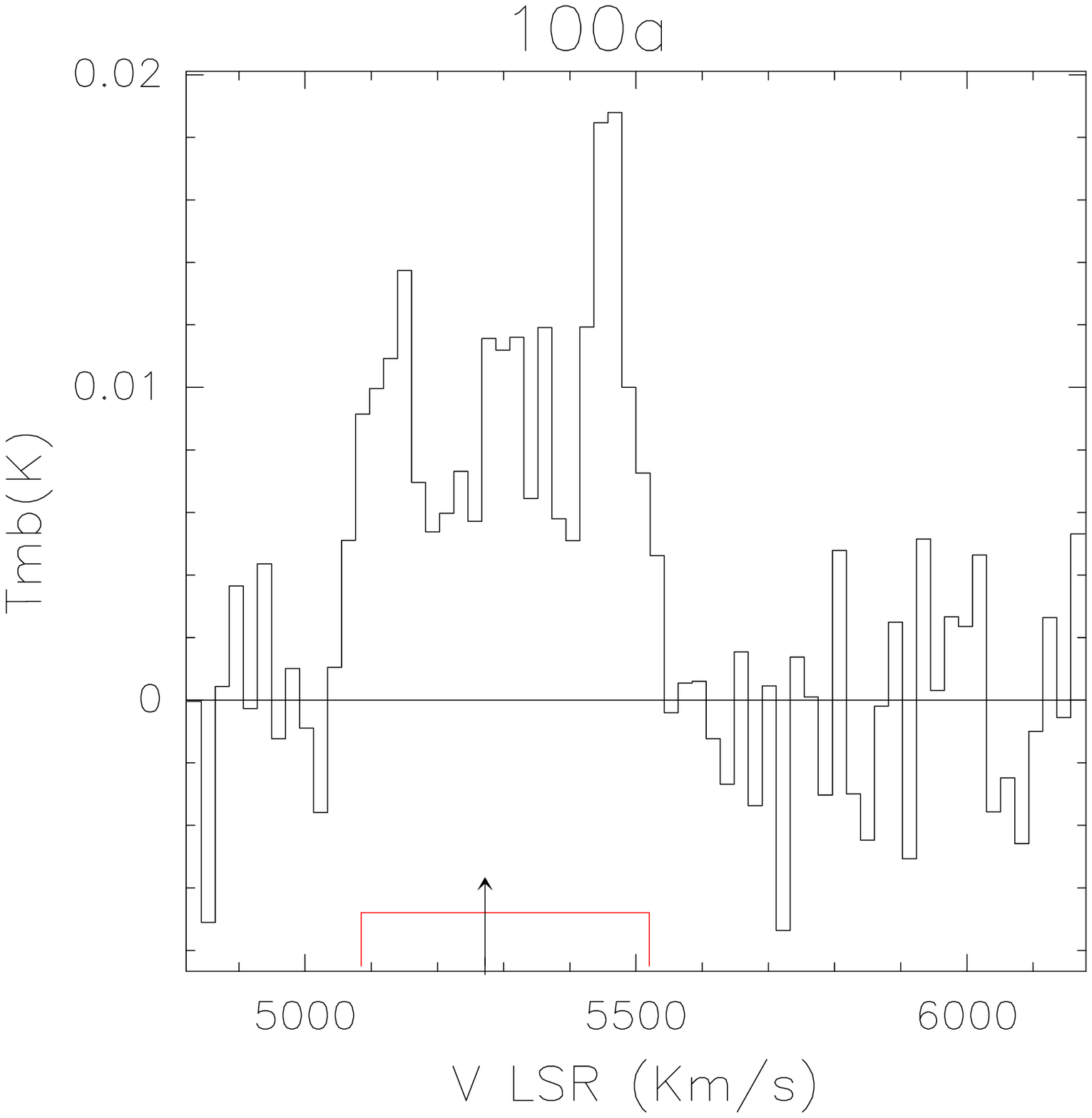,width=3.cm}
\quad
\psfig{file=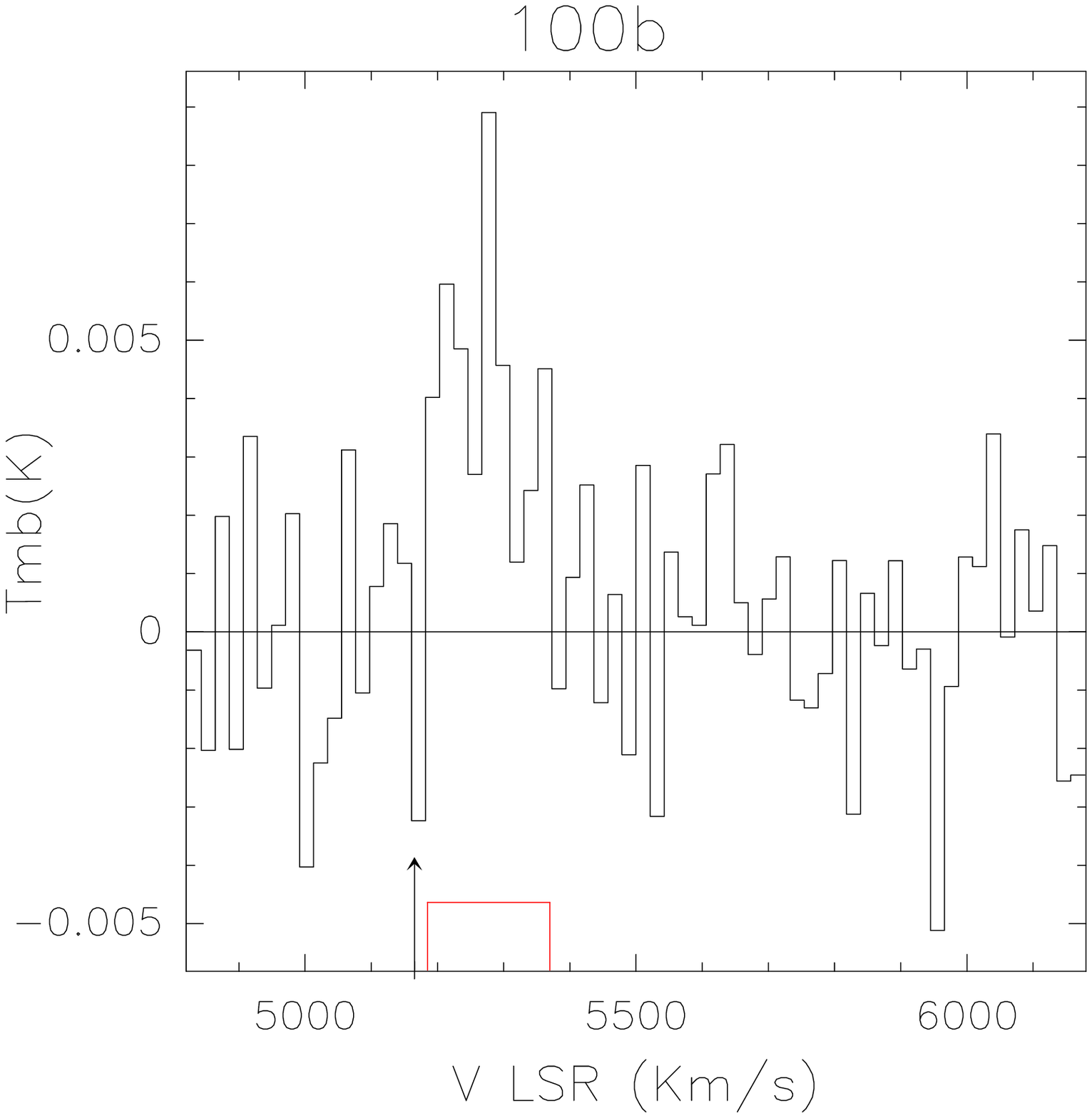,width=3.cm}
\quad
\psfig{file=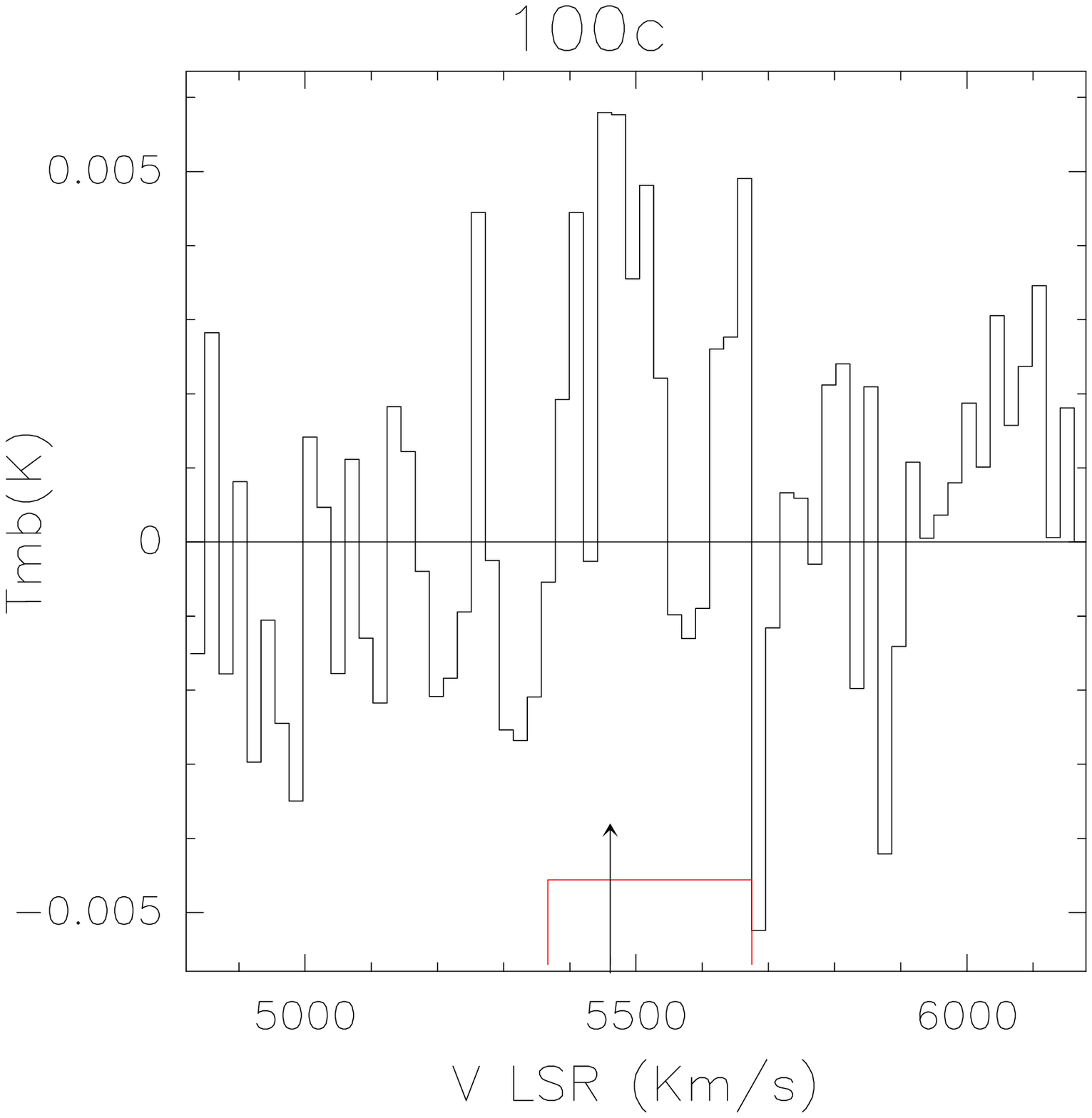,width=3.cm}
}

\caption{CO(1-0) spectra for the detected HCG galaxies. The detection window is shown with a red horizontal line. Main beam temperature ($T_{\rm mb}$, in K) is displayed in the Y axis, and the velocity with respect to LSR in km s$^{-1}$ is displayed in X axis. Velocity resolution is smoothed to 21 or 27 km s$^{-1}$. The optical velocity of the galaxy, converted to the radio definition, is marked with an arrow.}
\label{spec-co10}
\end{figure*}

\begin{figure*}[h!]
\centerline{
\psfig{file=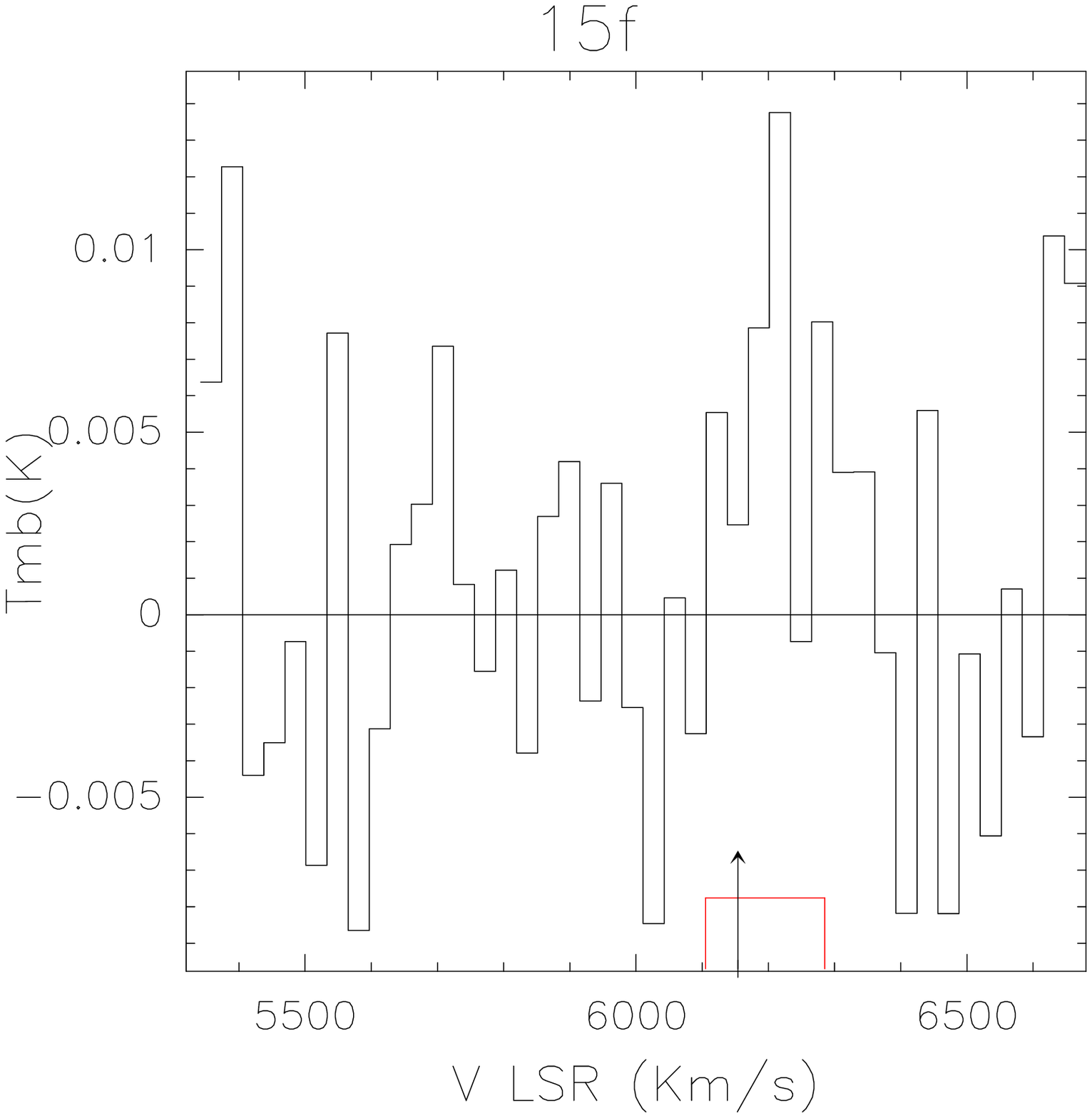,width=3.cm}
\quad
\psfig{file=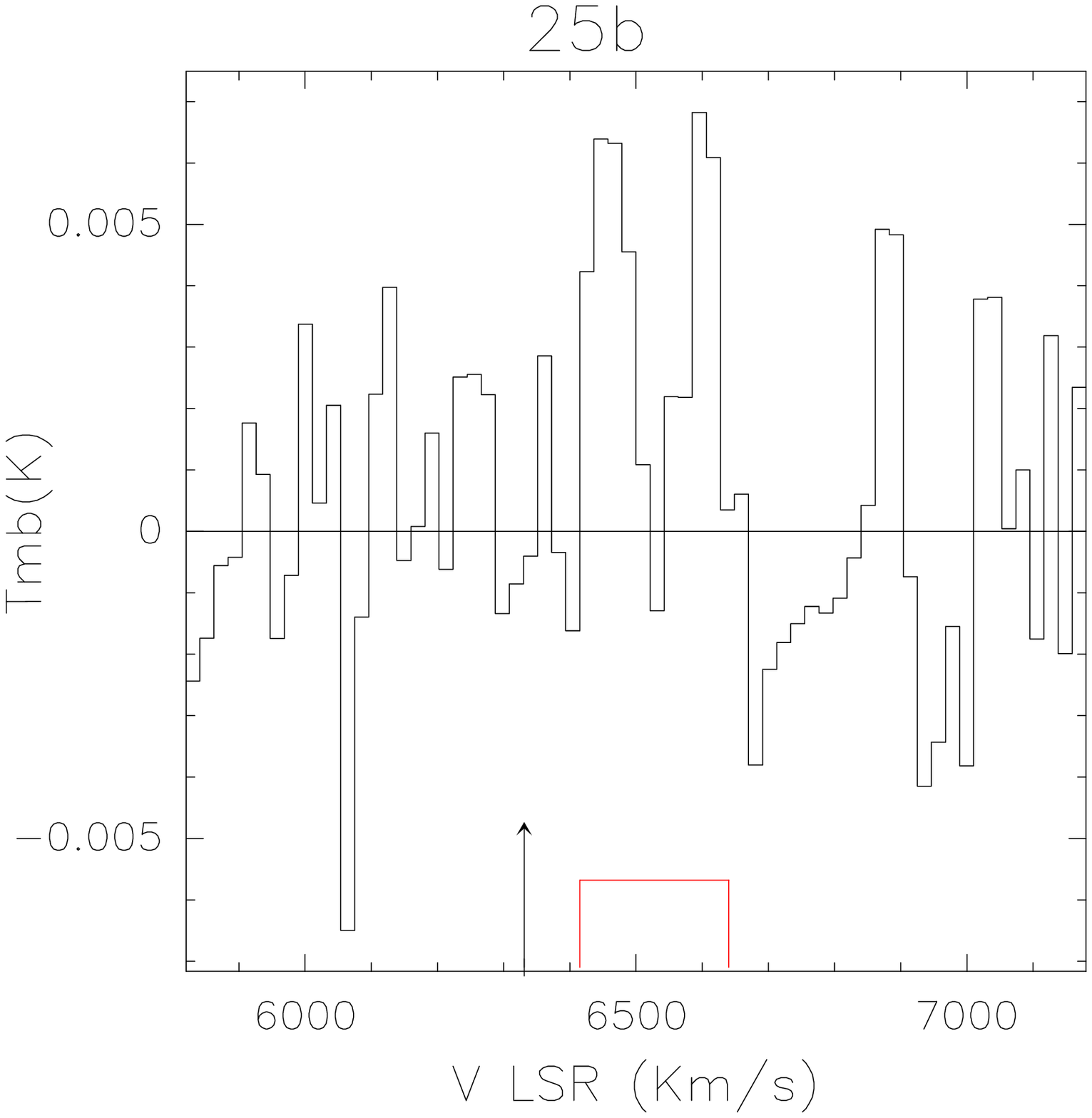,width=3.cm}
\quad
\psfig{file=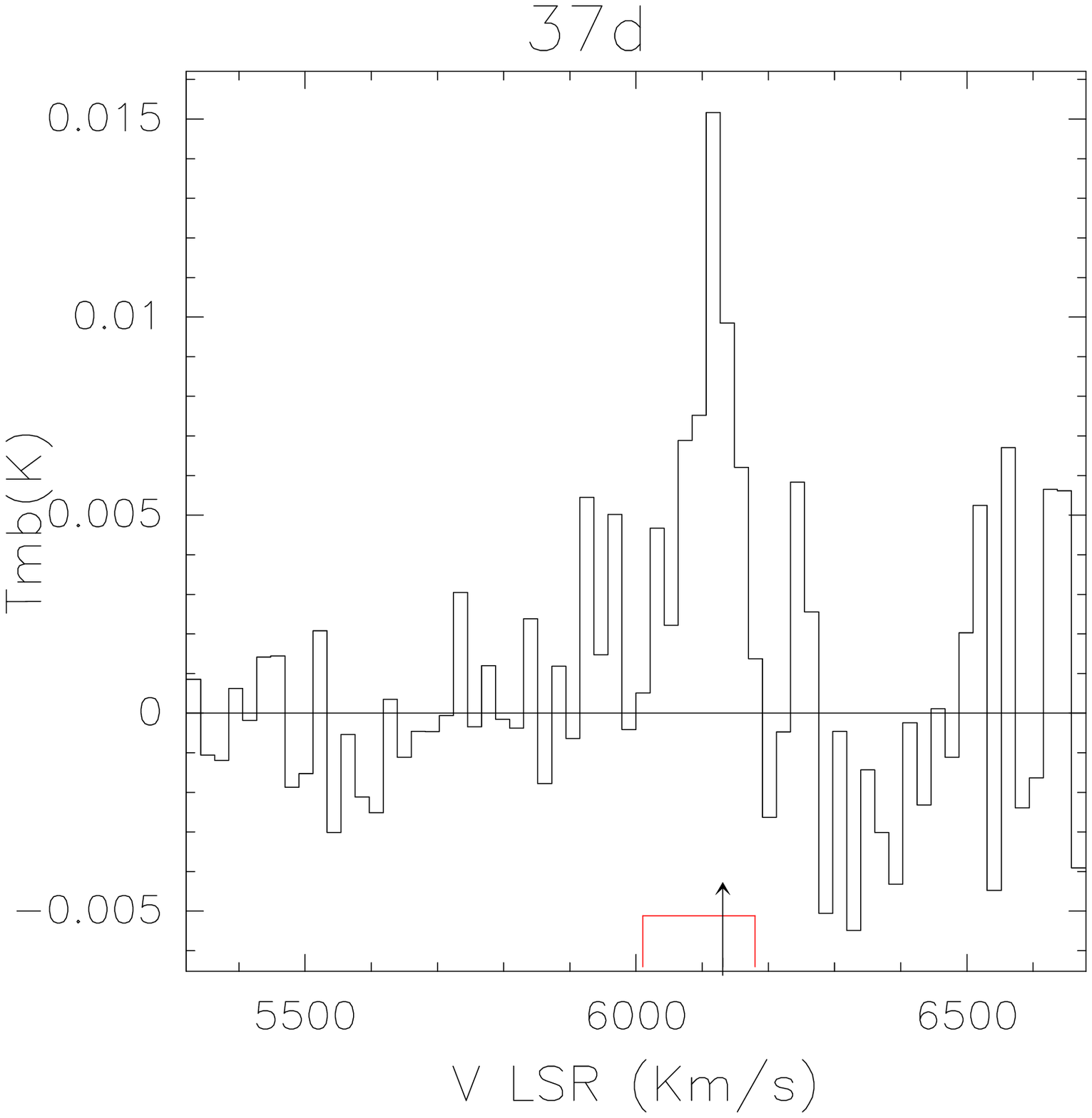,width=3.cm}
\quad
\psfig{file=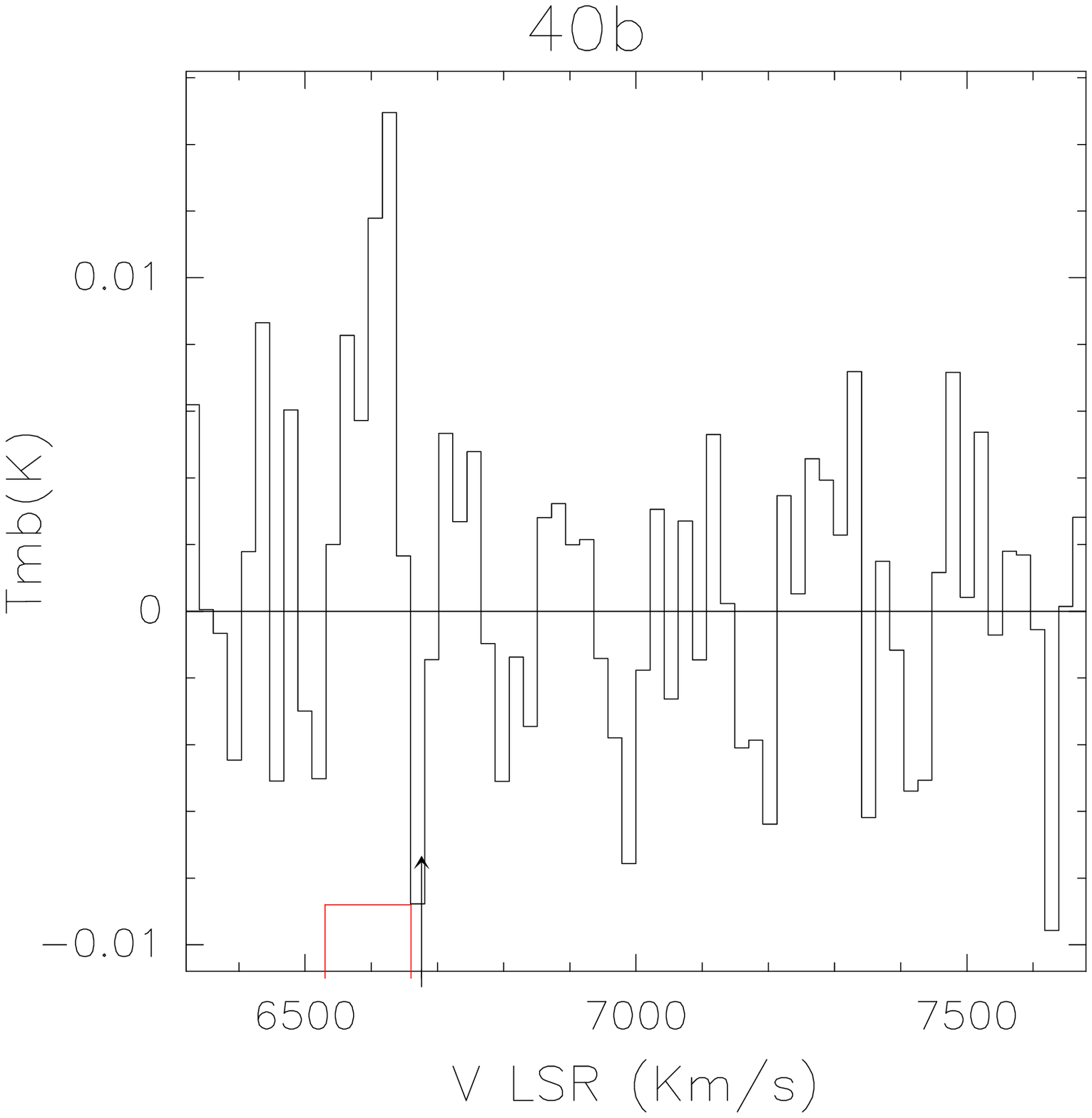,width=3.cm}
\quad
\psfig{file=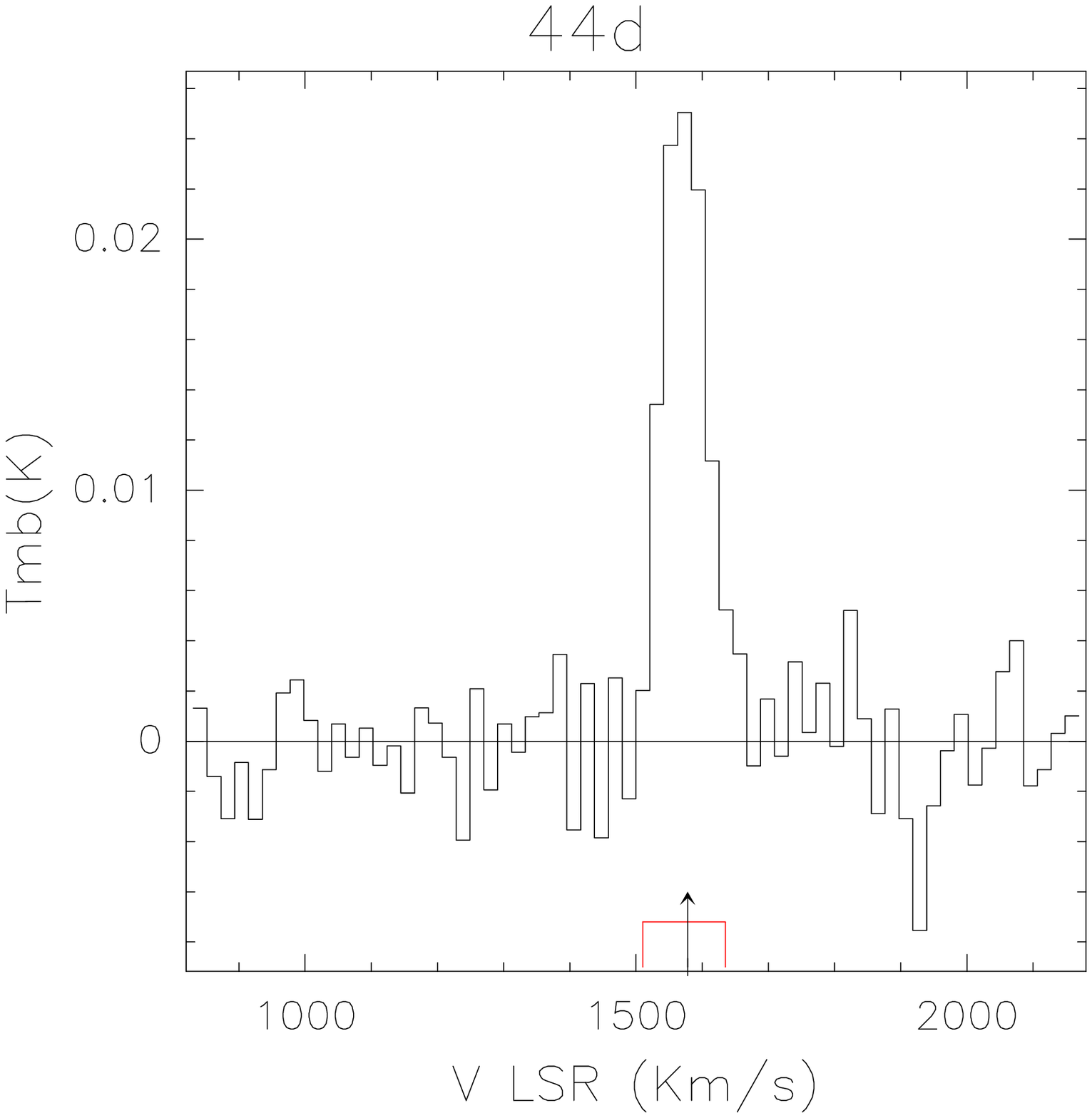,width=3.cm}
}

\centerline{
\psfig{file=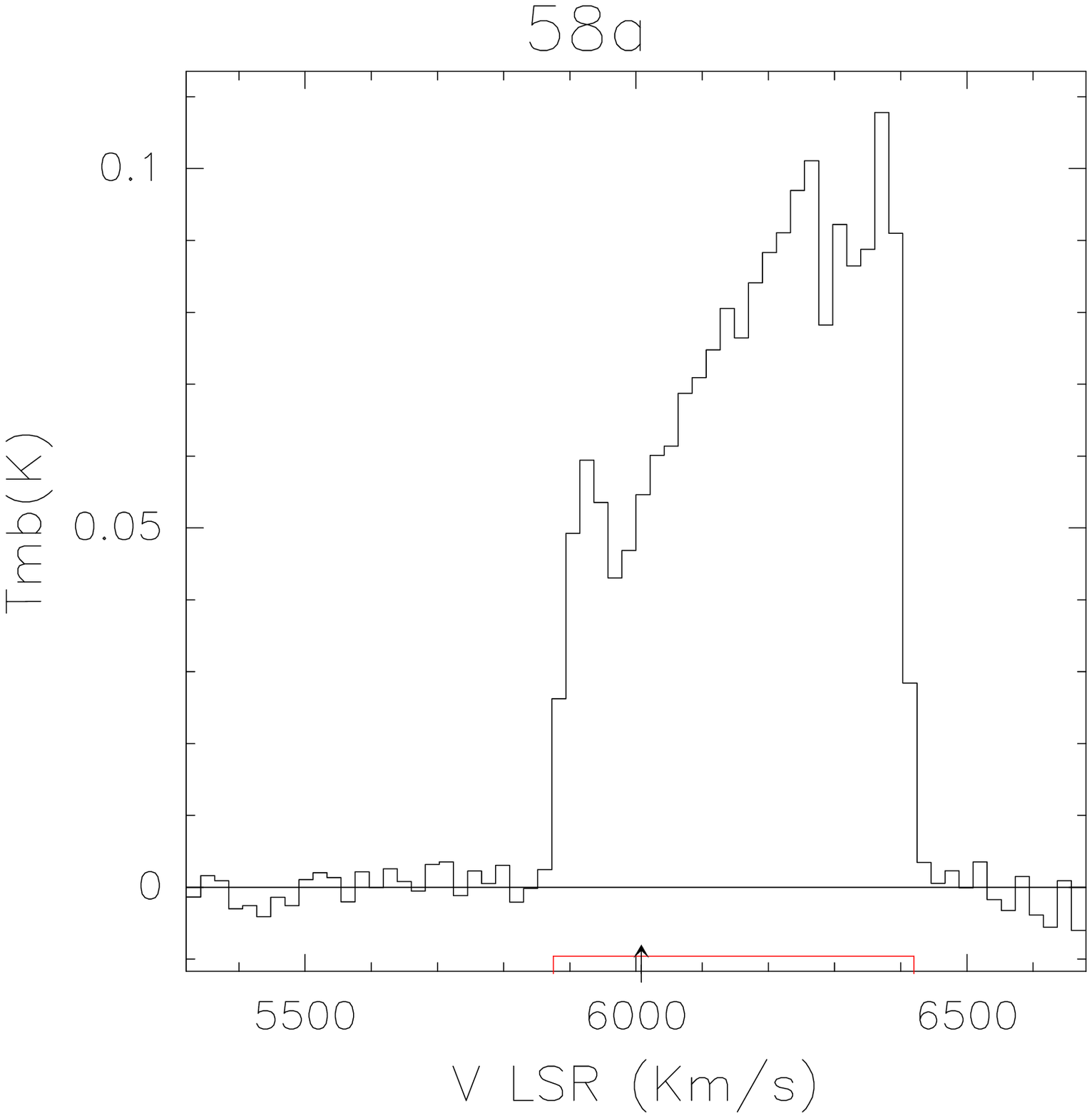,width=3.cm}
\quad
\psfig{file=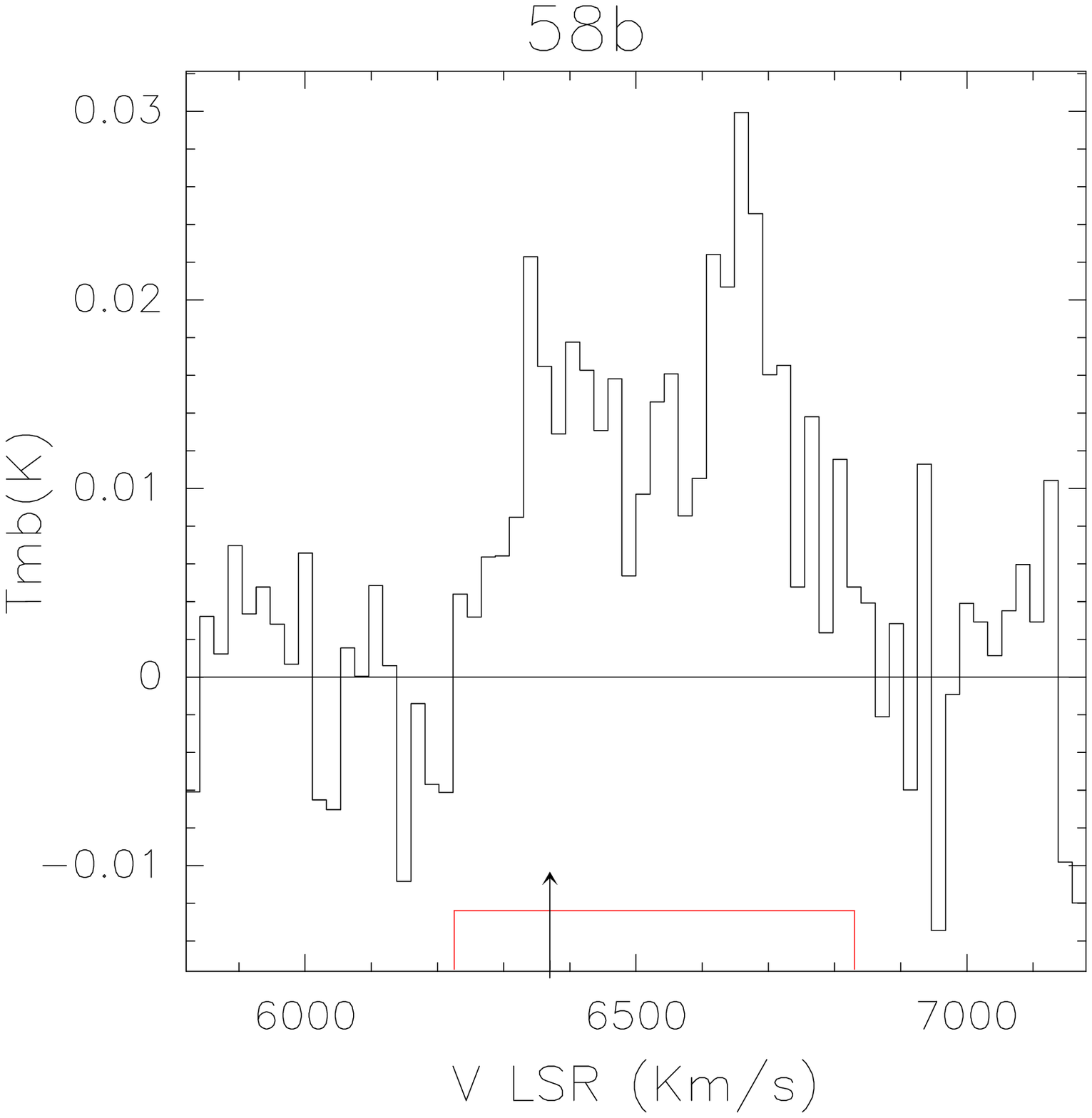,width=3.cm}
\quad
\psfig{file=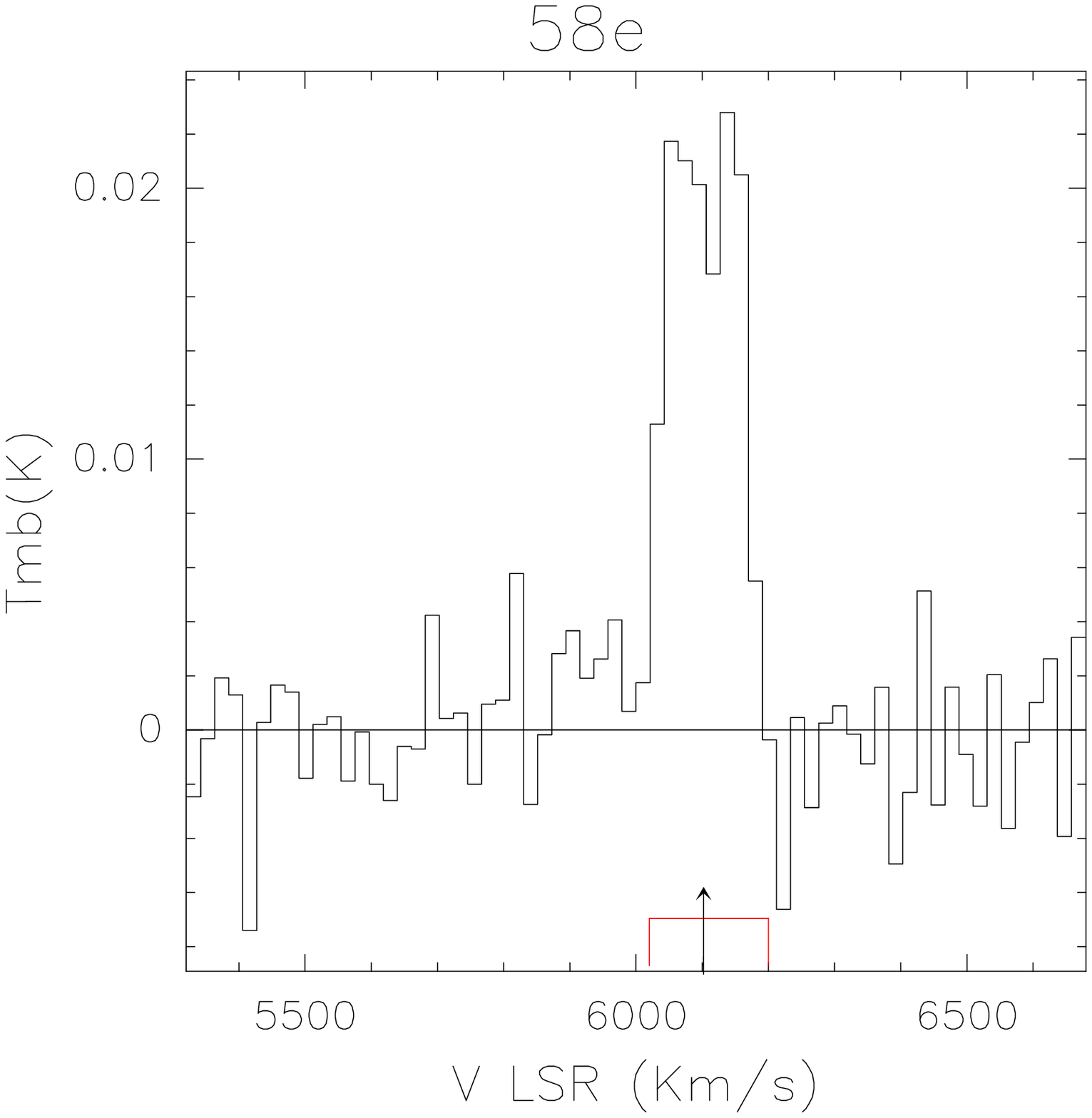,width=3.cm}
\quad
\psfig{file=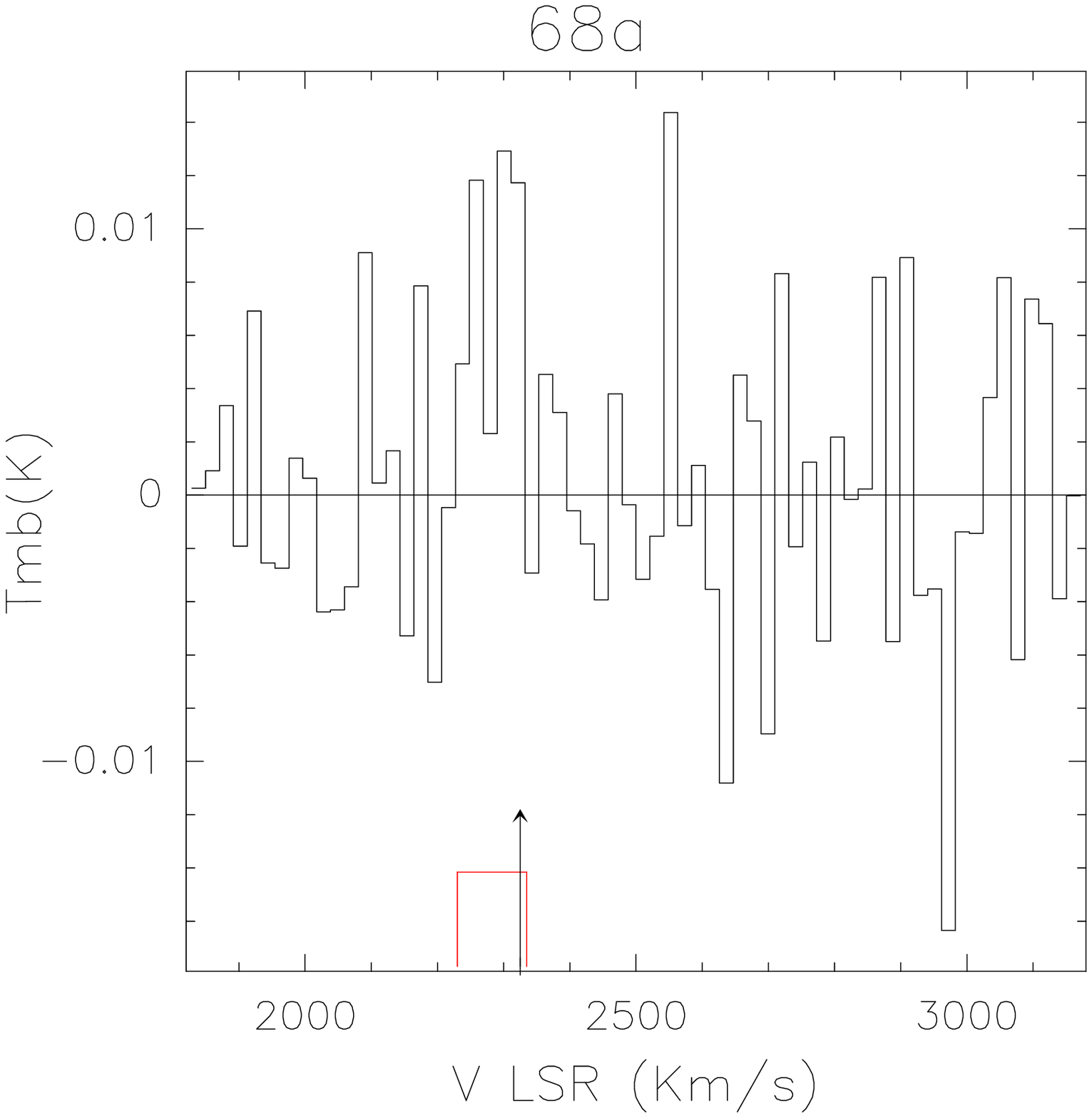,width=3.cm}
\quad
\psfig{file=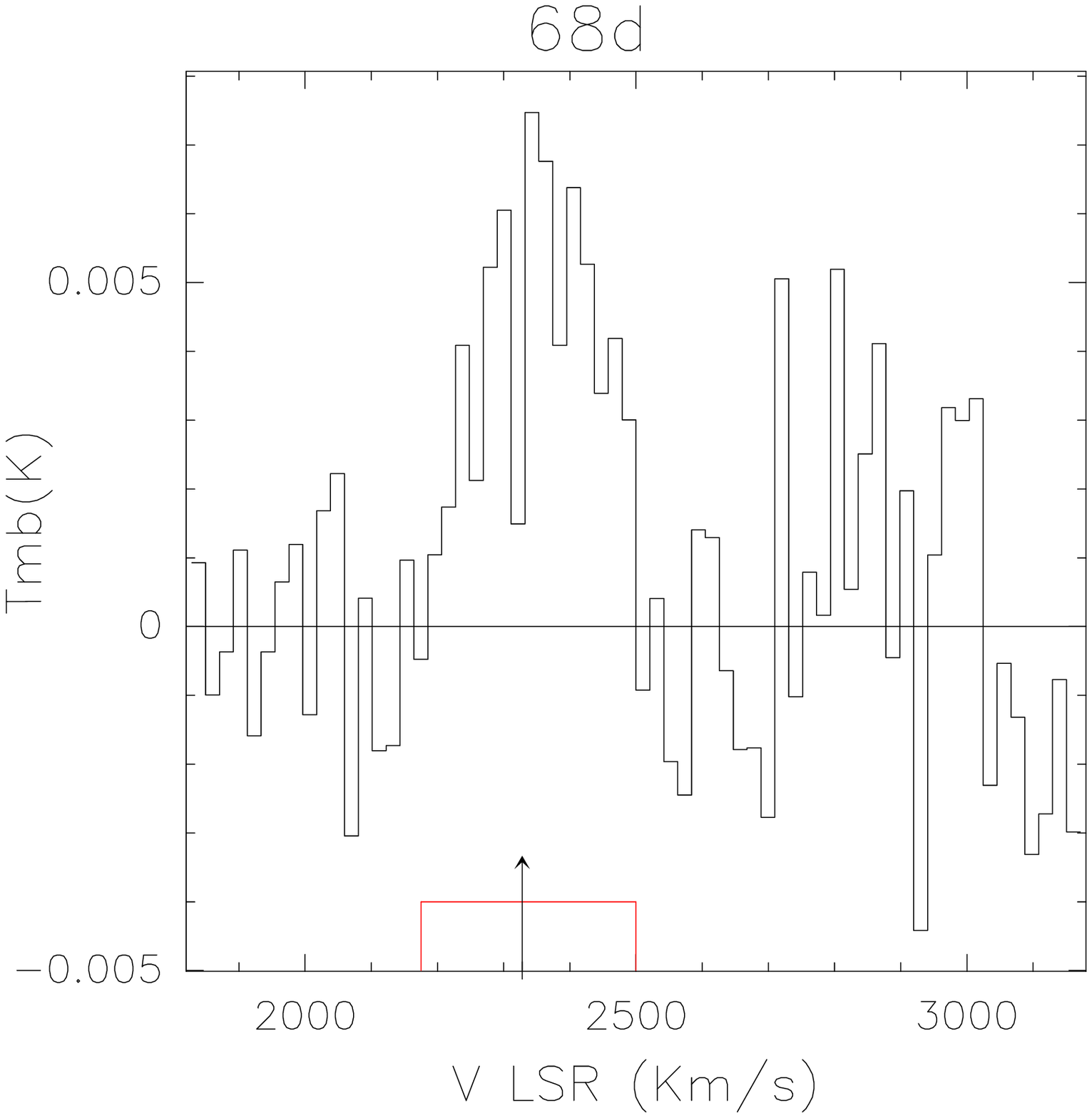,width=3.cm}
}

\centerline{
\psfig{file=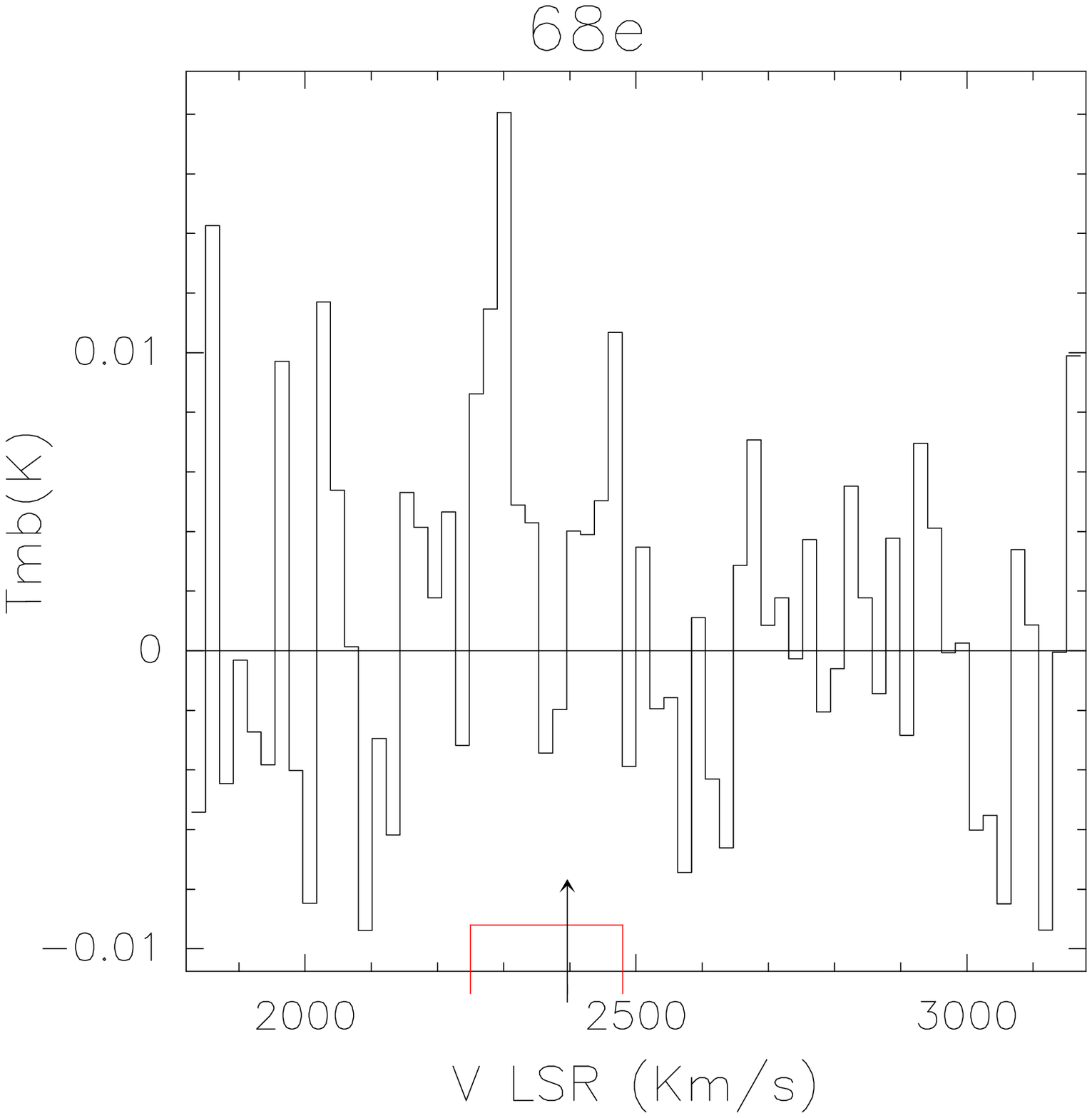,width=3.cm}
\quad
\psfig{file=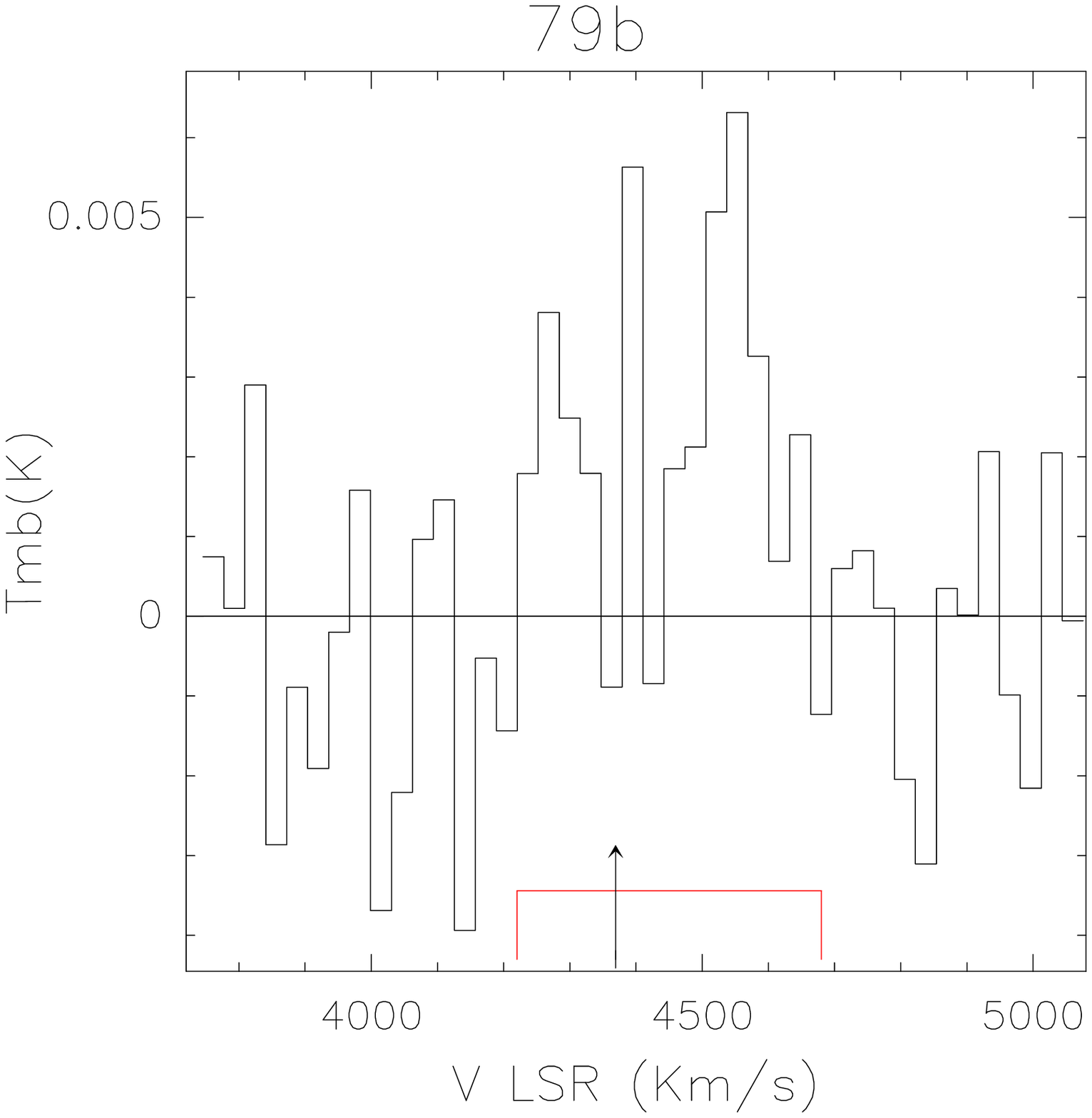,width=3.cm}
\quad
\psfig{file=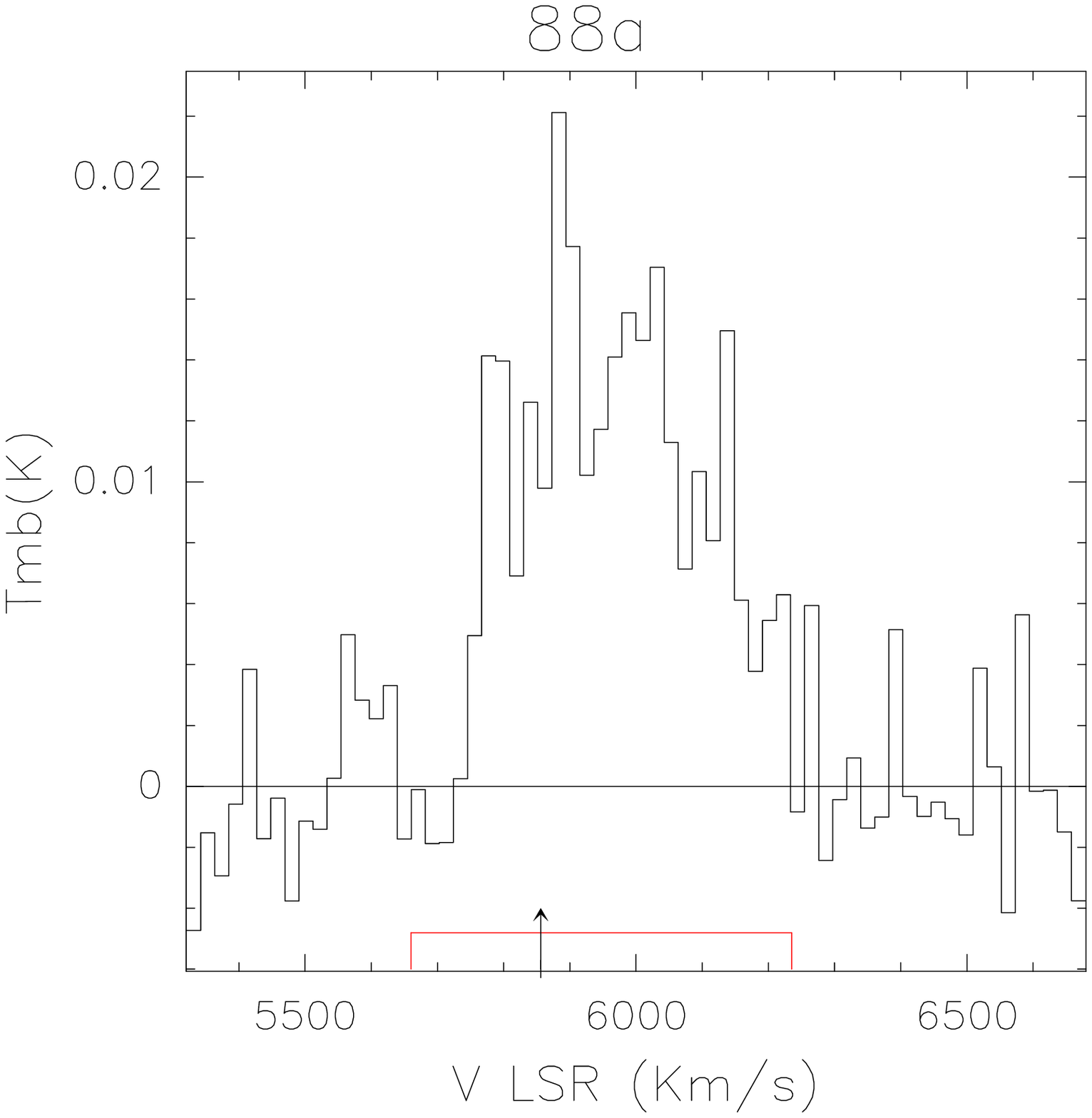,width=3.cm}
\quad
\psfig{file=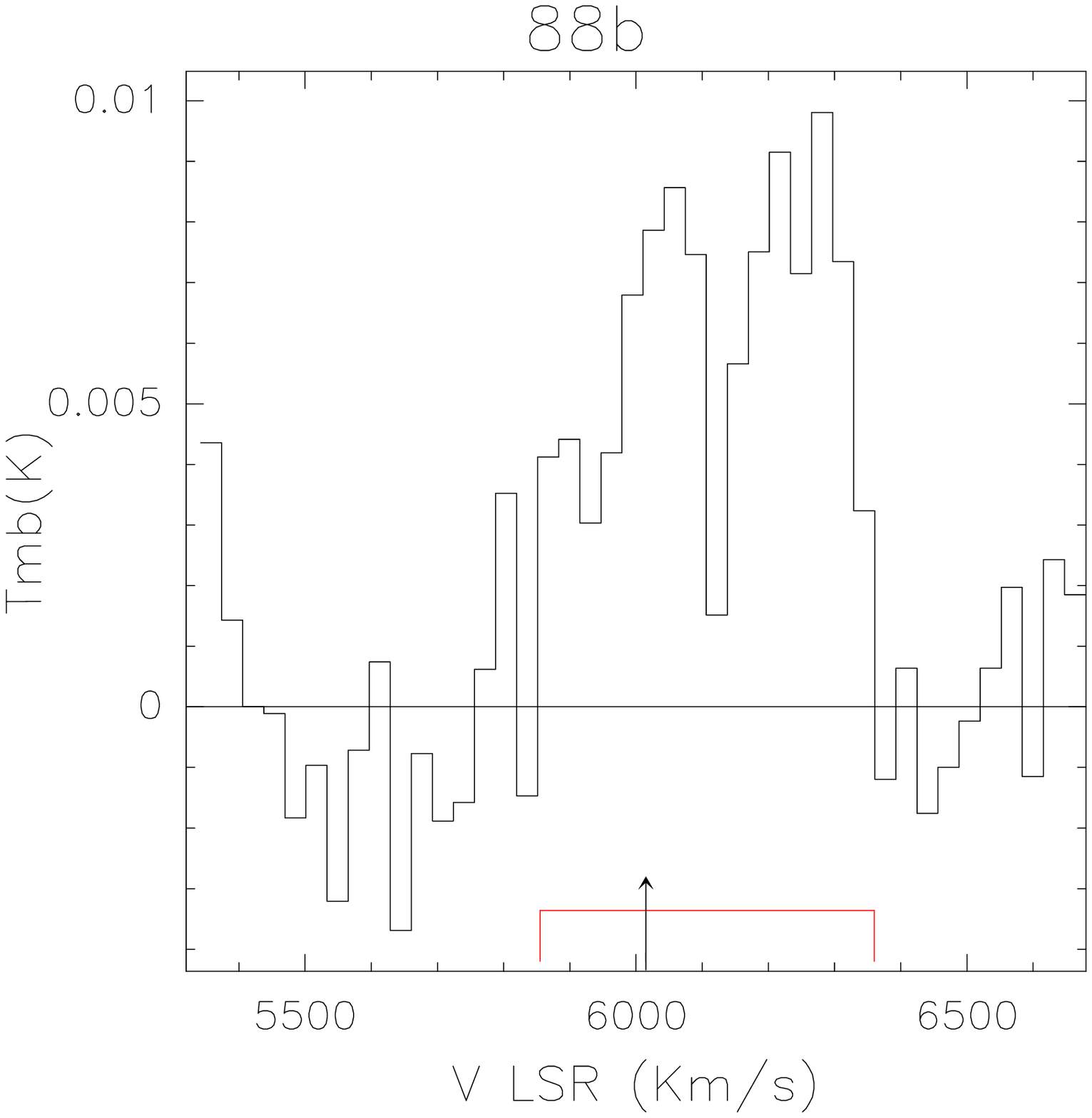,width=3.cm}
\quad
\psfig{file=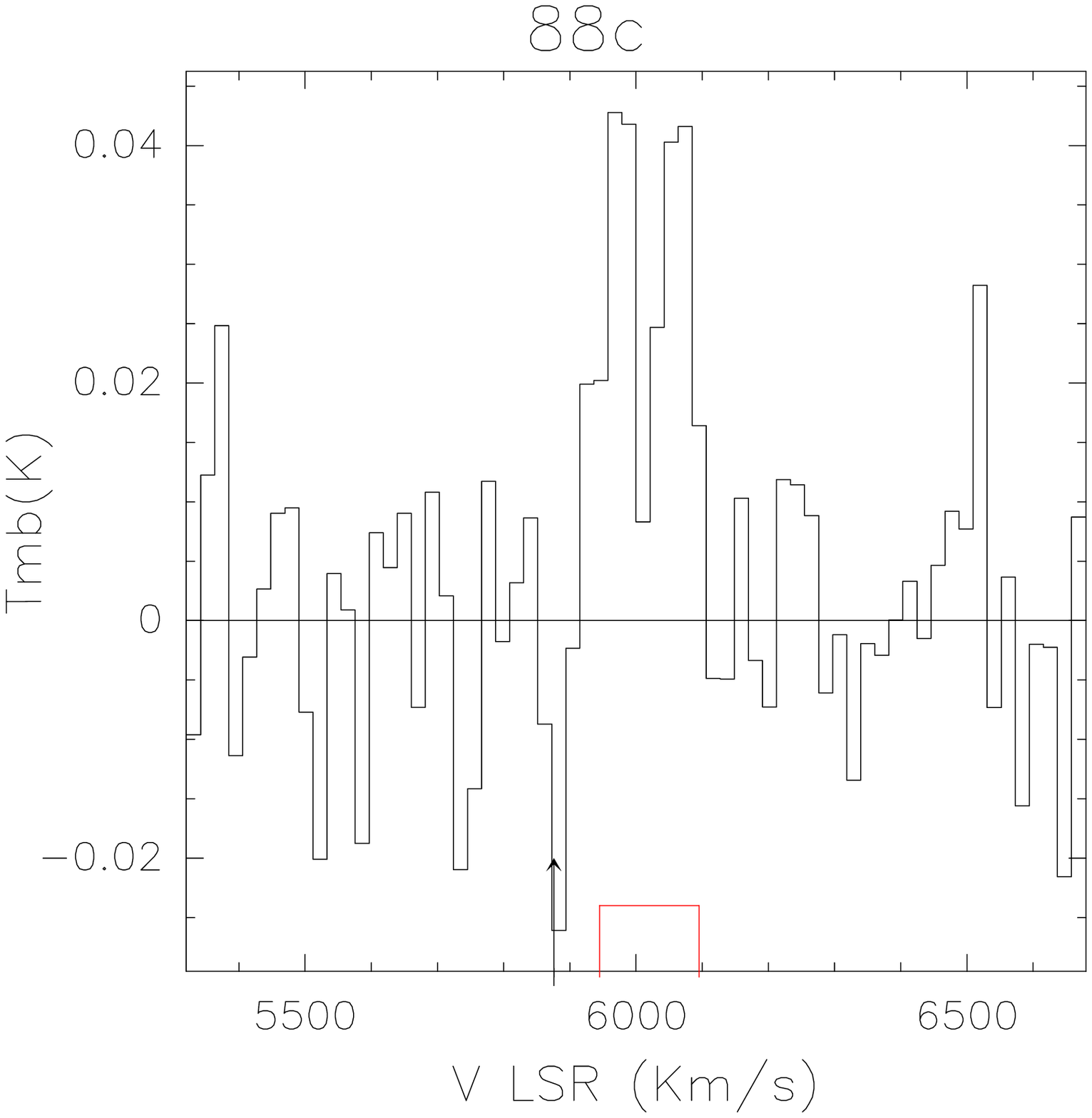,width=3.cm}
}

\centerline{
\psfig{file=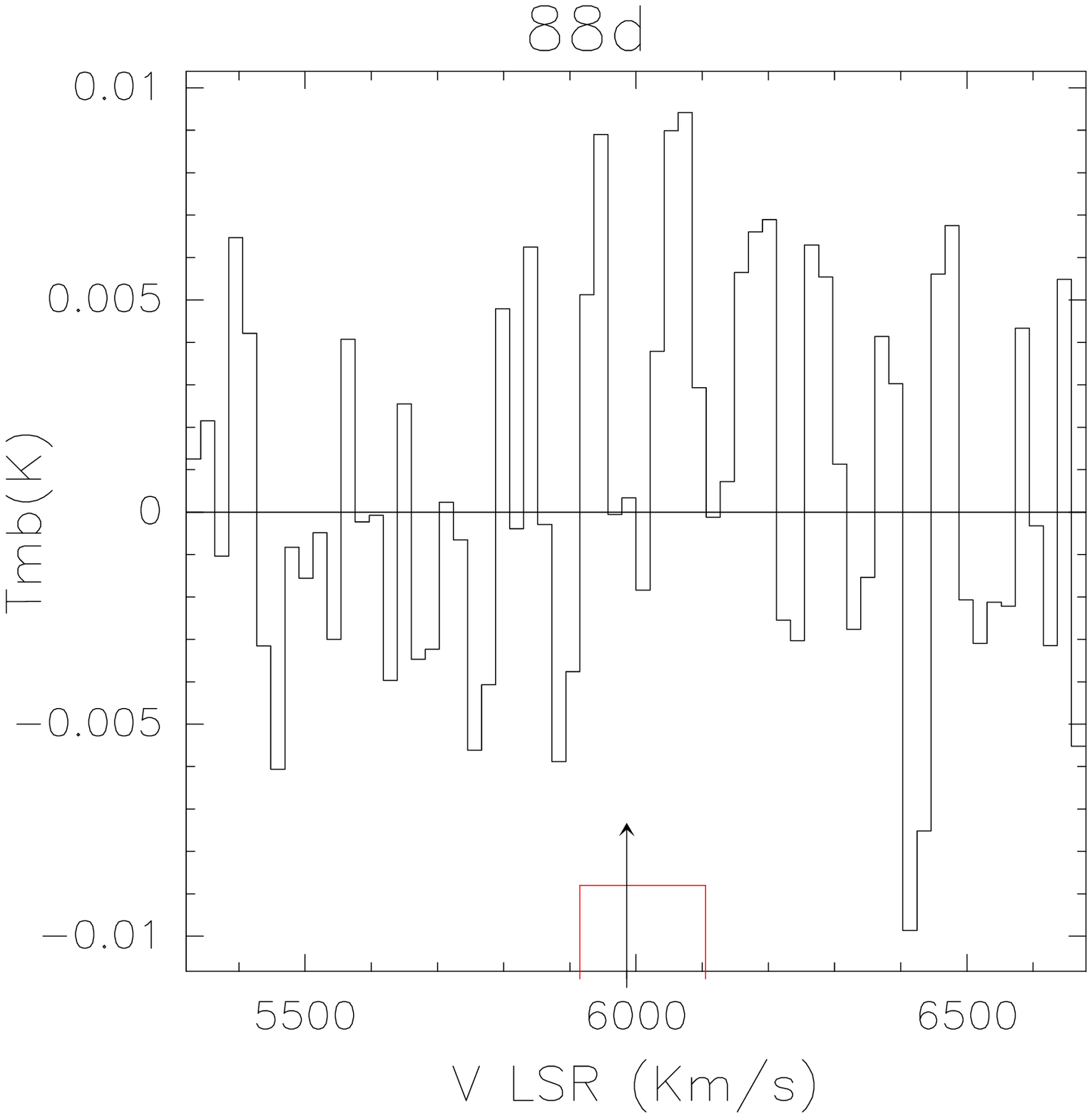,width=3.cm}
\quad
\psfig{file=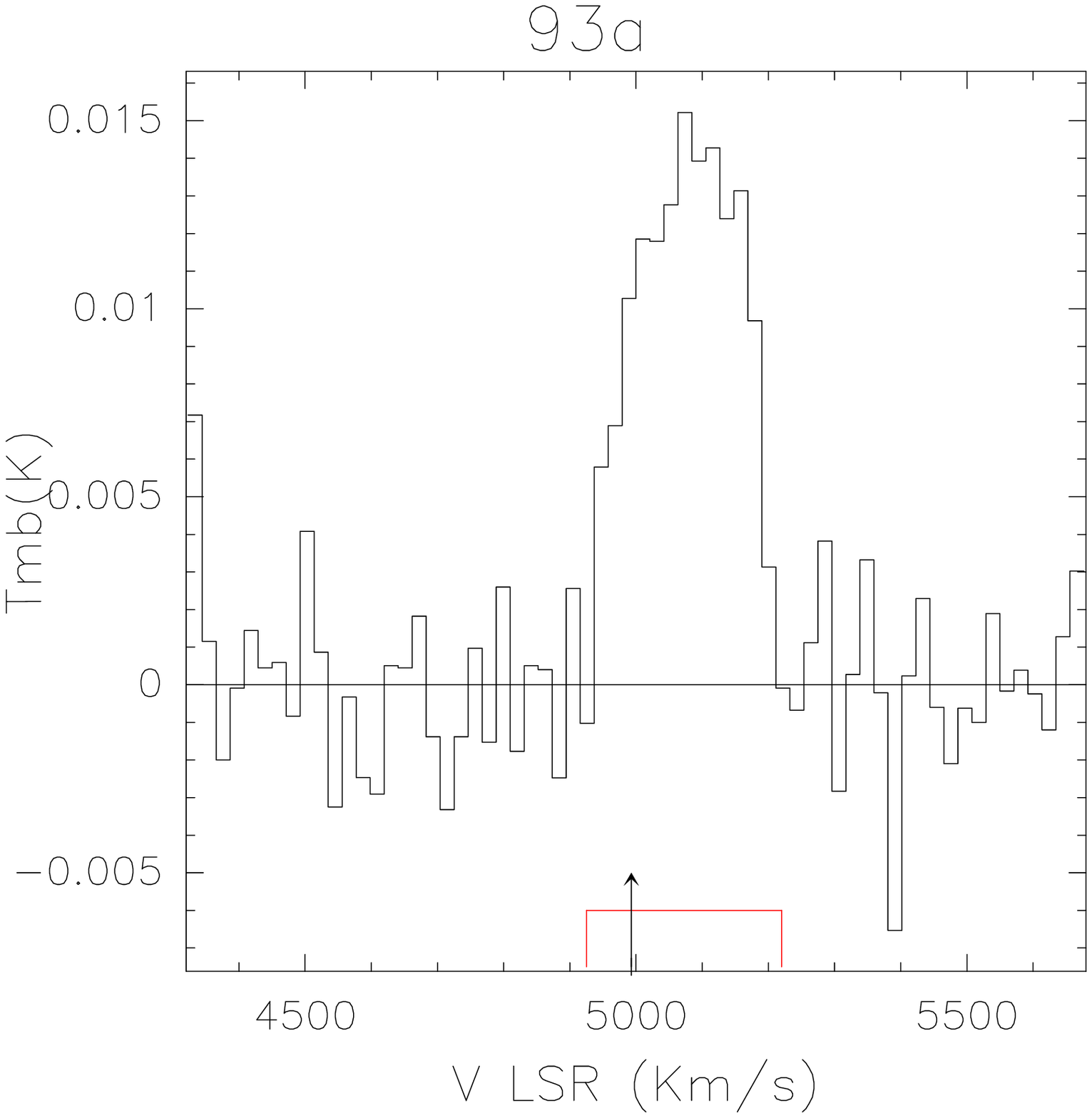,width=3.cm}
\quad
\psfig{file=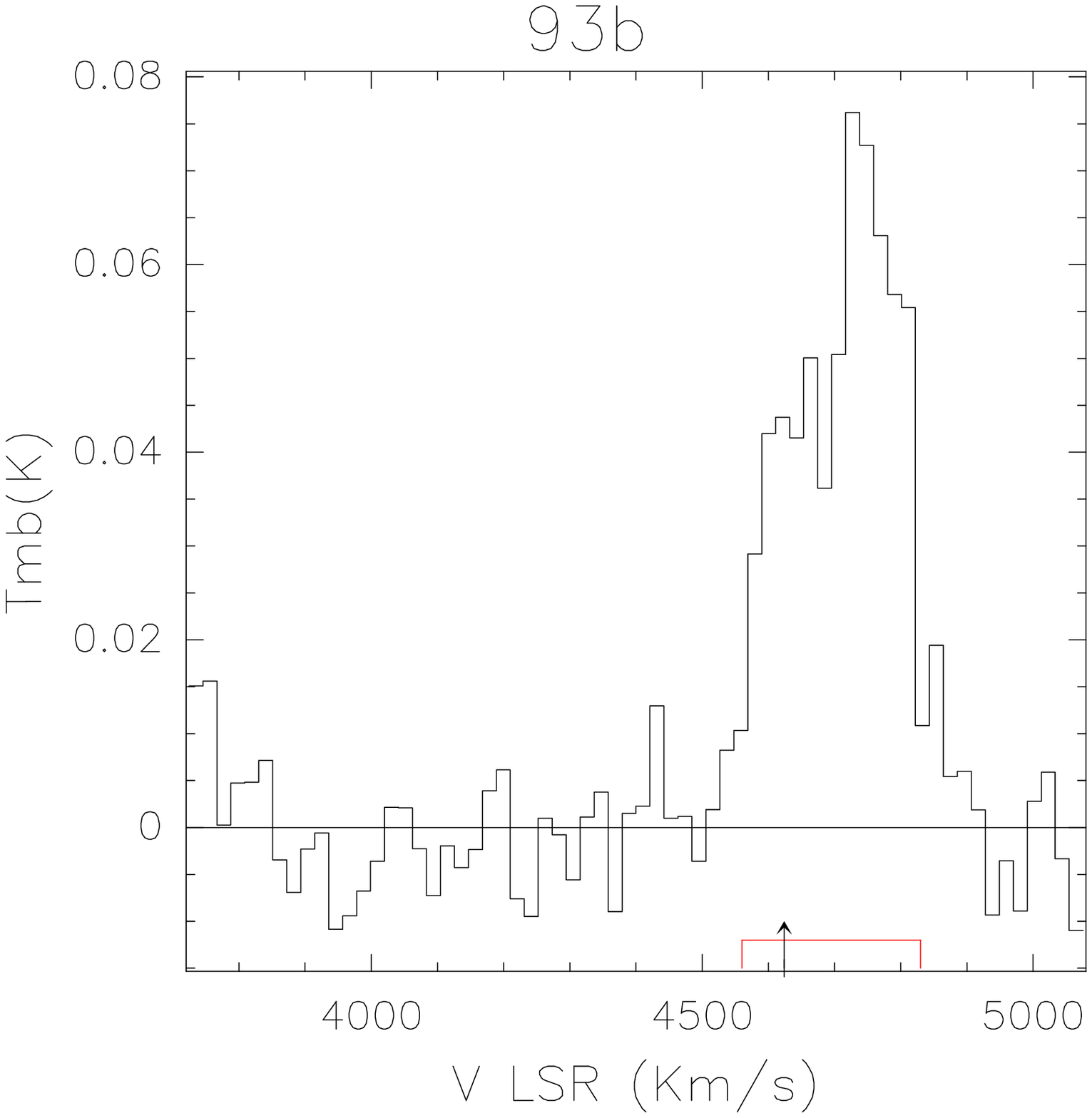,width=3.cm}
\quad
\psfig{file=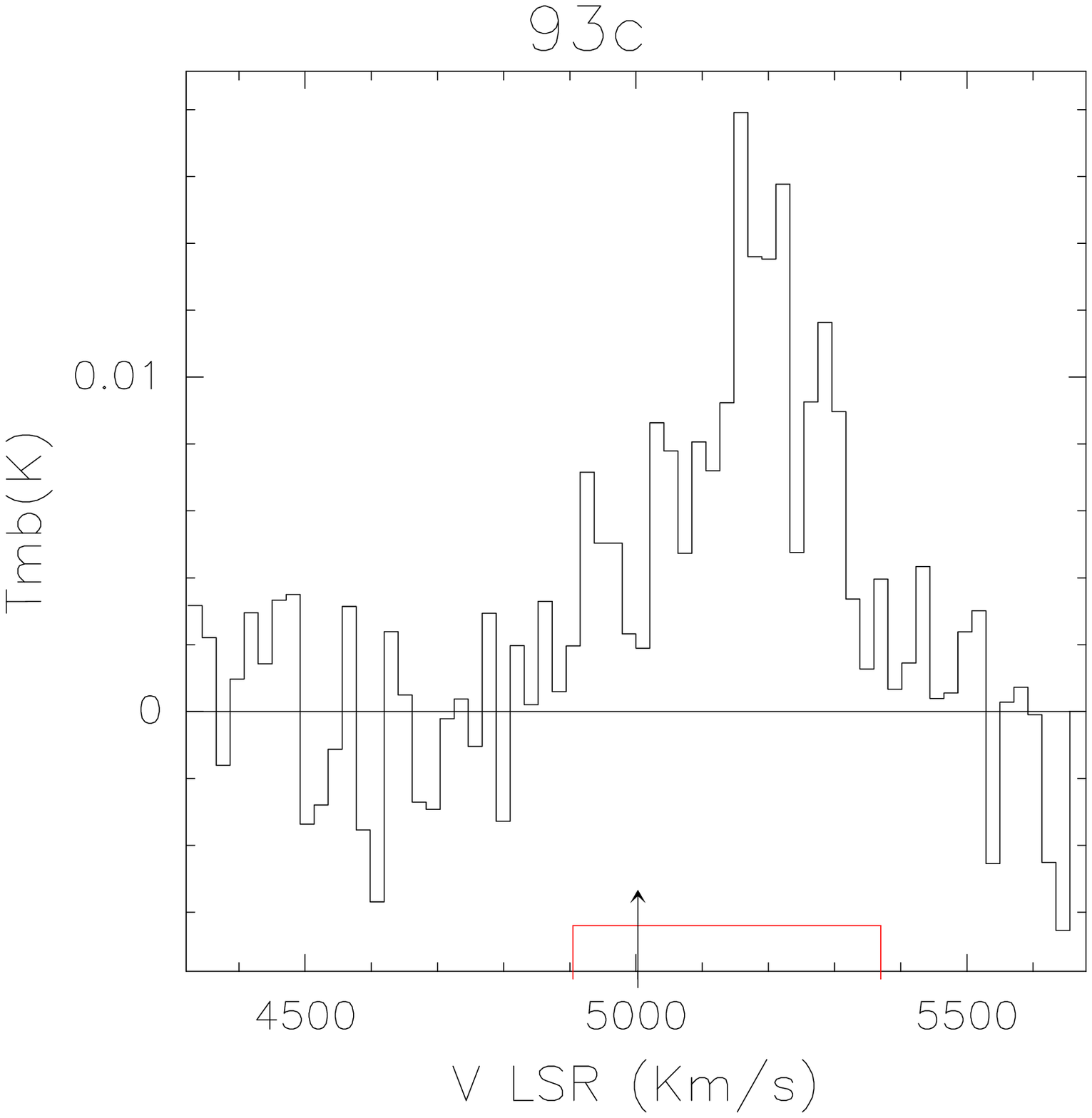,width=3.cm}
\quad
\psfig{file=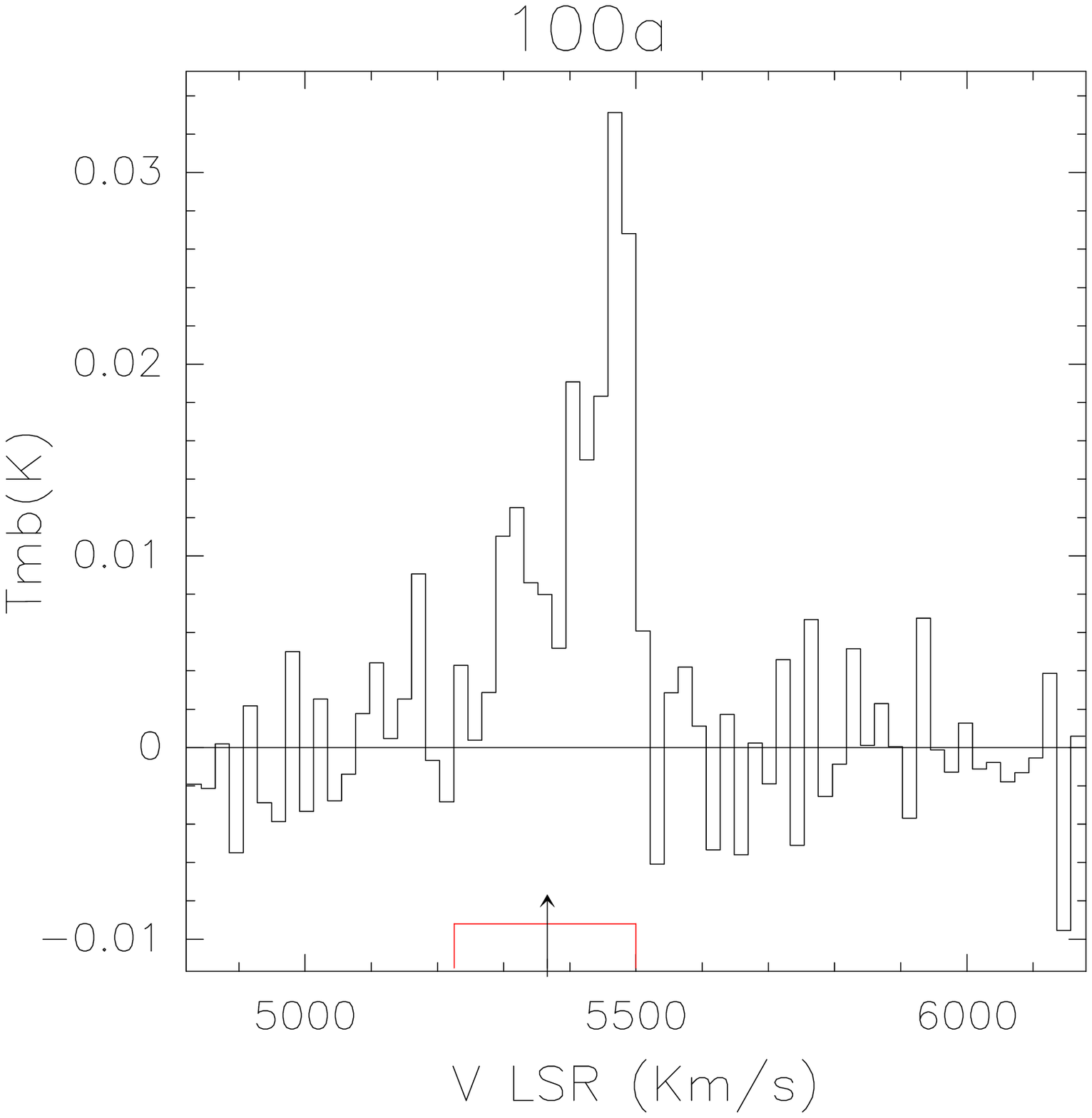,width=3.cm}
}
\centerline{
\quad
\psfig{file=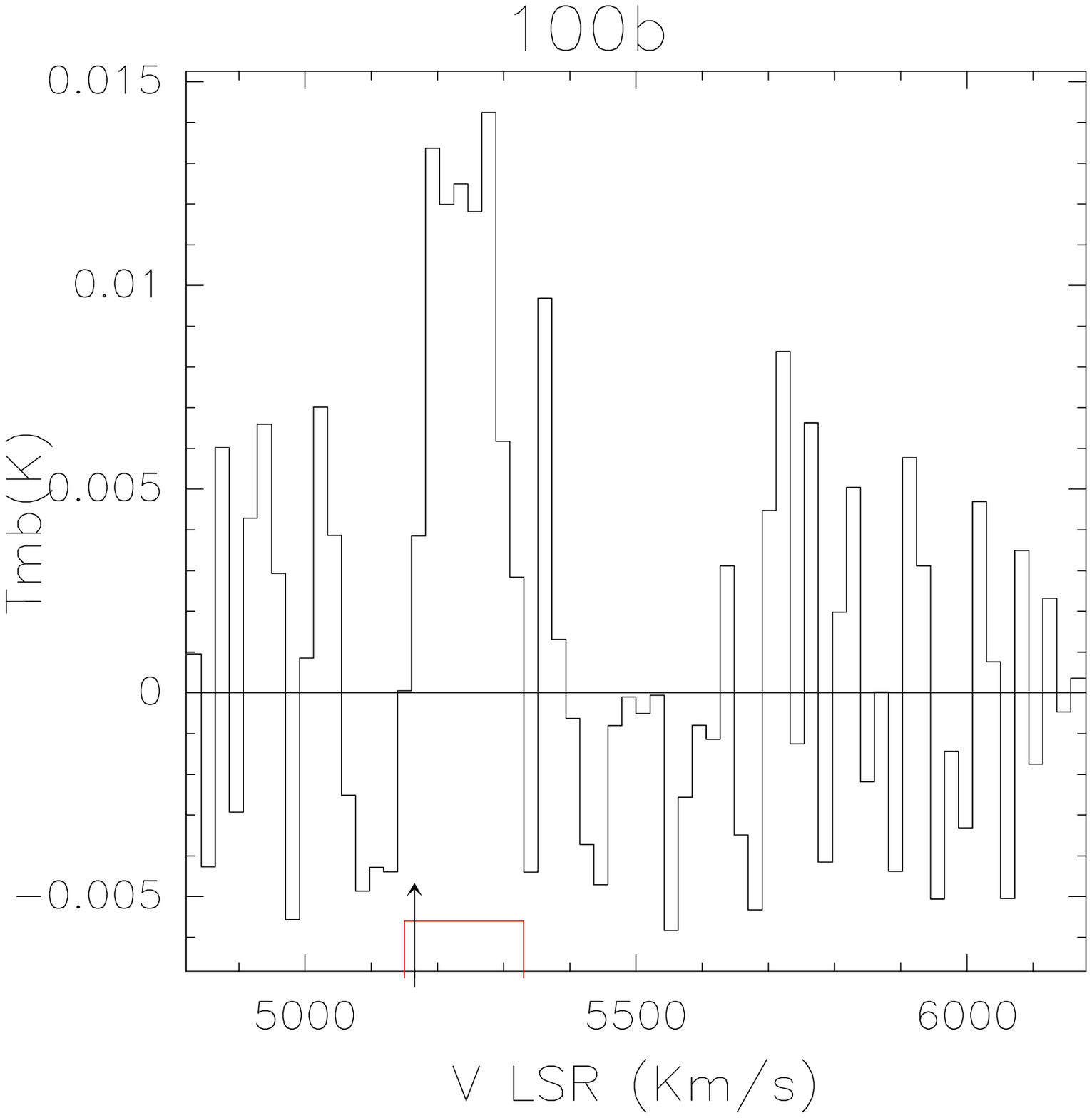,width=3.cm}
\quad
\psfig{file=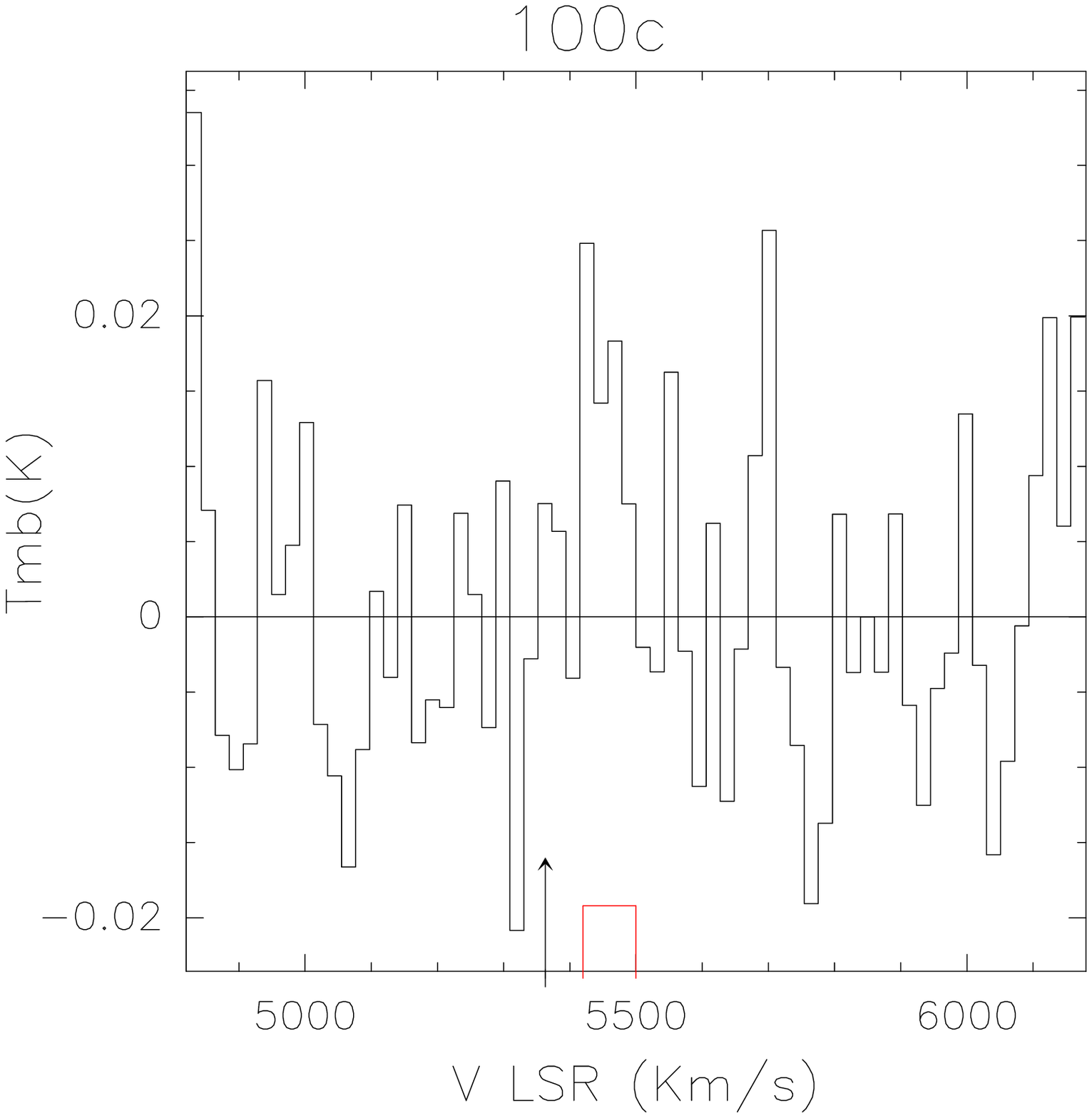,width=3.cm}
}
\caption{CO(2-1) spectra for the detected HCG galaxies. The detection window is shown with a red horizontal line. Main beam temperature  ($T_{\rm mb}$, in K) is displayed in the Y axis, and the velocity with respect to LSR in km s$^{-1}$ is displayed in X axis. Velocity resolution is smoothed to 21 or 27 km s$^{-1}$. The optical velocity of the galaxy, converted to the radio definition, is marked with an arrow.}
\label{spec-co21}
\end{figure*}

\end{document}